\documentclass[preprintnumbers,superscriptaddress,nofootinbib,aps,prd,floatfix]{revtex4}

\usepackage{amsmath,amssymb}
\usepackage{dsfont}
\usepackage{enumitem}
\usepackage{graphicx} 
\usepackage{epstopdf} 
\usepackage{slashed}
\usepackage{caption}
\usepackage{subcaption}
\usepackage{color}
\usepackage{multirow}
\usepackage{pdflscape}
\usepackage{xspace}
\usepackage{verbatim}
\usepackage{hyperref}

\usepackage{footmisc}

\usepackage{array}
\newcolumntype{L}[1]{>{\raggedright\let\newline\\\arraybackslash\hspace{0pt}}m{#1}}
\newcolumntype{C}[1]{>{\centering\let\newline\\\arraybackslash\hspace{0pt}}m{#1}}
\newcolumntype{R}[1]{>{\raggedleft\let\newline\\\arraybackslash\hspace{0pt}}m{#1}}

\newcommand{\be}{\begin{eqnarray*}}
\newcommand{\ee}{\end{eqnarray*}}

\newcommand{\bee}{\begin{eqnarray}}
\newcommand{\eee}{\end{eqnarray}}
\newcommand{\beeq}{\begin{equation}}
\newcommand{\eeeq}{\end{equation}}

\begin{document}

\title{Measuring the signal strength in $t\bar{t}H$ with $H\to b\bar{b}$ }
\begin{abstract} 
A precise measurement of the Higgs boson couplings to
bottom and top quarks is of paramount importance during the upcoming LHC
runs.  We present a comprehensive analysis for the Higgs production process
in association with a semi-leptonically decaying top-quark pair and
subsequent Higgs boson decay into bottom quarks.  Due to 
the 
highly complex
final state and large Standard Model backgrounds, measuring the signal
strength in this process is known to be challenging.  To maximise the
sensitivity, we analyse different, statistically independent, phase space
regions, where one or more of the heavy resonances are boosted.  This allows
us to employ jet substructure techniques, which help to reduce large
$t\bar{t}+\mathrm{X}$ backgrounds. 
We find that combining several 
$t\bar{t}H(b\bar b)$ phase space regions will 
allow one to measure
deviations of the Standard Model signal strength 
of order $20\%$ with 3\,ab$^{-1}$.

\end{abstract}
%
%

\author{Niccolo Moretti} \email{moretti@physik.uzh.ch}
\affiliation{Physik-Institut, Universit\"at Z\"urich,
Winterthurerstrasse 190, 
	CH-8057 Z\"urich,
	Switzerland }

\author{Petar Petrov} \email{p.m.petrov@durham.ac.uk}
\affiliation{Institute for Particle Physics Phenomenology, Department
  of Physics,\\Durham University, DH1 3LE, United Kingdom}

\author{Stefano Pozzorini} \email{pozzorin@physik.uzh.ch}
\affiliation{Physik-Institut, Universit\"at Z\"urich,
Winterthurerstrasse 190, 
	CH-8057 Z\"urich,
	Switzerland }

\author{Michael Spannowsky} \email{michael.spannowsky@durham.ac.uk}
\affiliation{Institute for Particle Physics Phenomenology, Department
  of Physics,\\Durham University, DH1 3LE, United Kingdom}

\preprint{IPPP/15/46, DCPT/15/92,  ZU-TH 34/15}

\maketitle

\section{Introduction}

After the discovery of the Higgs boson \cite{hatlas, hcms} the precise
measurement of its properties is the foremost goal during the upcoming LHC
runs.  All coupling measurements performed so far at 7 and 8 TeV
centre-of-mass energy are in good agreement with Standard Model (SM)
predictions.  However, while Higgs boson couplings to gauge bosons have
already been constrained in a fairly precise way \cite{couplcomb}, Higgs
couplings to fermions are still plagued by large uncertainties.

Measuring Higgs couplings to bottom and top quarks is of particular
importance.  In the Standard Model, the bottom quark coupling drives the
total width $\Gamma_\mathrm{tot}$ of the Higgs boson.  When measuring the
signal strength of any 
Higgs production and decay
process $i$, the
observed number of events depends crucially on the branching ratio,
$\mathrm{BR}_i = \Gamma_i / \Gamma_\mathrm{tot}$, of the Higgs boson into
the process-specific final state.  Hence, if $\Gamma_{\mathrm{tot}}$ is only
weakly constrained due to a large uncertainty on the Higgs-bottom coupling
$c_b$, a precise measurement of any coupling will be hampered~\cite{Lafaye:2009vr}.  
One way of measuring $c_b$ is to 
exploit Higgs boson production in association
with a gauge boson 
in the $H\to b\bar b$ decay channel
\cite{Butterworth:2008iy,Soper:2010xk}.  This process benefits from
mild
combinatorial issues in the reconstruction of the two resonances,
resulting in 
a favourable
signal-to-background ratio.  However, during run 1
ATLAS and CMS were only able to set weak limits on $c_b$, indicating the
need for alternative ways to complement this measurement during the upcoming
runs of the LHC \cite{MSp2009,Bodwin:2013gca,Kagan:2014ila,Koenig:2015pha}.

A direct measurement of the the Higgs-top coupling $c_t$ is desirable for
several reasons: The comparatively large top quark mass, which is not
explained in the Standard Model, is directly proportional to $c_t$ and
contributes in a dominant way to the destabilisation of the electroweak
scale \cite{Degrassi:2012ry, Buttazzo:2013uya}.  An independent measurement
of $c_t$ in addition to a measurement of the Higgs self-coupling can help to
evaluate if the electroweak vacuum is stable or only meta-stable on
cosmological time scales.  Further, many new physics scenarios predict
deviations of $c_t$ from its SM value, e.g.  well-known examples include
generic two Higgs-doublet models, the MSSM or composite Higgs models.  While
the loop-induced 
$gg\to H$ and $H\to\gamma \gamma$ processes are
sensitive to $c_t$, direct access can only
be obtained by measuring the production cross section of the Higgs boson in
association with top quarks, e.g.  $t\bar{t}H$ or $tH$, with the former
having a seven times larger cross section in the Standard Model
\cite{Demartin:2015uha, Frixione:2015zaa,Dawson:2002tg,Beenakker:2001rj}. 
Hence, for the quality of a global fit of Higgs boson properties a precise
measurement of $c_t$ 
through $t\bar{t}H$ production
is indispensable\footnote{Top associated production allows to study the quantum numbers of the scalar particle using differential distributions of the decay products \cite{Buckley:2015vsa,Casolino:2015cza, Li:2015kaa,Santos:2015dja,Kolodziej:2015qsa,Ellis:2013yxa,Yue:2014tya,Farina:2012xp, Degrande:2012gr,Englert:2014pja}.}.

Both of LHC's multipurpose experiments, ATLAS and CMS, have set limits on
$c_t$ in various channels during Run~1 \cite{atlas1,atlas2,cmsTTHall}.  For
a light Higgs boson of $125$ GeV produced 
in association with a $t\bar t $ pair
phenomenological
studies have predicted sensitivity in decays of the Higgs boson to leptonic
taus or $W$ bosons
\cite{Maltoni:2002jr,Belyaev:2002ua,Curtin:2013zua,Craig:2013eta}, photons
\cite{Buttar:2006zd} and bottom quarks \cite{MSp2009,Artoisenet:2013vfa}. 
Decays into leptonic taus and $W$s can give rise to same-sign lepton
signatures which, similarly to di-photon signatures, result in a 
significantly improved
signal-to-background ratio at the expense of a 
very small signal yield. 
However, while same-sign or multi-lepton 
signatures are unlikely to
result in a narrow mass peak after reconstruction, leaving confidence in
having reconstructed the $t \bar{t} H$ final state at stake,
the loop-induced $H\to\gamma\gamma$ decay channel features a nontrivial $c_t$ dependence that does not allow for a completely model-independent measurement of the top--Higgs coupling.

In contrast $t \bar{t} H ( b \bar{b})$ is exposed to large backgrounds but
provides the largest signal yield.  Consequently, the strongest 
constraints on $t\bar t H$ production
during Run 1 were obtained in this channel, 
where ATLAS and CMS have observed exclusion limits between 3.4 and 4.2 times the SM cross section at 95\%  confidence level~\cite{atlasTTHall,cmsTTHMEM,cmsTTHall}.

As it 
allows to access $c_b$ and $c_t$
simultaneously,
$pp\to t\bar{t} H (b\bar{b})$ is
one of the most important processes to
measure during future LHC runs.  
In this paper we present
a detailed study of 
this reaction at 14\,TeV
focusing on exclusive phase space regions that will become accessible 
at high luminosity.
Earlier studies have found that going into the so-called
boosted regime can improve significantly the signal-to-background 
ratio~\cite{MSp2009}, the most limiting factor of $t\bar{t} H (b\bar{b})$ searches\footnote{Predictions for a 100 TeV hadron collider are particularly promising \cite{He:2014xla,Plehn:2015cta}.}.
We extend this approach
using more robust signal and background calculations, 
employing 
experimentally tested
taggers,
and exploiting
a variety of independent event topologies.

We organise this paper as follows: In Sec.~\ref{sec:signal} we describe in
detail the event generation of signal and backgrounds.  Special care is
dedicated
to the validation of high-statistics LO samples for $t\bar t+$multijet
and $t\bar t+b$-jet production---needed for a decent description of the backgrounds in highly
suppressed phase space regions---by means of NLO matched and merged
simulations.
After 
a brief overview of an 
early boosted analysis
of $t\bar{t}H(b\bar{b})$ in Sec.~\ref{sec:mspana}, in
Sec.~\ref{sec:analysis} we present alternative approaches
focusing on several phase space regions in combination with a variety of
reconstruction techniques.  To evaluate how well the signal strength $\mu$
can be measured for each of the aforementioned approaches we perform a
binned profile likelihood test on the resulting distributions in
Sec.~\ref{sec:results}.  Eventually, we offer conclusions in
Sec.~\ref{sec:conc}.

\section{Signal and Background Cross Sections and Event Generation}
\label{se:MC}

\label{sec:signal}  
\def\ttbar{t\bar t}
\def\bbbar{b\bar b}
\def\ttbb{\ttbar\bbbar}
\def\Sherpa{{\sc Sherpa}\xspace}
\def\Amegic{{\sc Amegic}\xspace}
\def\Comix{{\sc Comix}\xspace}
\def\OpenLoops{{\sc OpenLoops}\xspace}
\def\Collier{{\sc Collier}\xspace}
\def\MEPS{{MEPS}\xspace}
\def\MEPSatNLO{{MEPS@NLO}\xspace}
\def\MEPSatLO{{MEPS@LO}\xspace}
\def\SMCatNLO{{S-MC@NLO}\xspace}

The production of $\ttbar$ pairs in association with light and heavy-flavour
jets represents the dominant source of background to $\ttbar H$ searches in the
$H\to\bbbar$ channel, and the accurate simulation of $\ttbar+$\,jet final
states with two light or heavy-flavour jets is a key prerequisite for a
reliable $\ttbar H(\bbbar)$ analysis.  This calls, one the one hand, for Monte Carlo
simulations based on NLO matrix elements.  On the other hand, the highly
exclusive cuts of the boosted analyses presented in this paper
reduce the background by up to a factor $10^{-4}$--$10^{-5}$ with respect to the corresponding inclusive cross section.
In these conditions, the production of NLO Monte Carlo
samples that preserve high statistics in the signal region is very
challenging.
To circumvent this problem we will employ a combination of
NLO and LO simulations.  While the actual analysis will be based on
high-statistics LO samples with appropriate 
dynamical
scale choices and
normalisation factors, NLO samples of lower statistics will be used to
verify that the LO ones 
describe the shapes of all relevant distributions with sufficiently good accuracy.
To generate the LO and NLO samples for signal and backgrounds we employ the
\Sherpa~\cite{Gleisberg:2008ta,Schumann:2007mg} Monte Carlo\footnote{More
precisely we used svn revision 24881 of the \Sherpa\,2.1.1 public release}
and its built-in modules for parton showering, hadronisation, hadron decays
and underlying event.  Tree matrix elements are computed with
\Amegic~\cite{Gleisberg:2007md} and \Comix~\cite{Gleisberg:2008fv}, while
one-loop matrix elements are generated with
\OpenLoops~\cite{Cascioli:2011va,hepforge} in combination with \Collier for
the evaluation of tensor
integrals~\cite{Denner:2014gla,Denner:2005nn,Denner:2010tr}.  Top-quark
decays are treated at LO including spin correlations based on $\ttbar$+jets
Born matrix elements using spin density matrices
\cite{Richardson:2001df,Hoche:2014kca}.  Their kinematics are adjusted a
posteriori according to a Breit-Wigner distribution using the top quark
width as an input.

All LO and NLO samples are generated at $14$\,TeV using CT10 NLO parton
densities~\cite{Lai:2010vv} and the input parameters $m_t=173.2$\,GeV,
$M_Z=91.1876$\,GeV, $M_W=80.385$\,GeV, $M_H=125.0$\,GeV, 
$G_\mu=1.16675\times10^{-5}$\,GeV$^{-2}$,
and
$\alpha=\sqrt{2}G_\mu M_W^2(1-M_W^2/M_Z^2)/\pi$.
Higgs bosons are decayed in the $\bbbar$ channel with branching fraction
\mbox{$\mathrm{BR}(H\to\bbbar)=0.577$~\cite{Heinemeyer:2013tqa}}, while the $\ttbar$ system is decayed into
semi-leptonic final states with NLO branching fractions
\mbox{$\mathrm{BR}(W\to q\bar q')=2\times 0.337303$}
and $\mathrm{BR}(W\to\ell\nu)=2\times 0.108465$, with $\ell=e^\pm,\mu^\pm$.

For the $\ttbar H(\bbbar)$ signal a NLO accurate sample is generated using the
\SMCatNLO method~\cite{Hoeche:2011fd,Hoeche:2012ft}, which
represents the \Sherpa variant of the MC@NLO method~\cite{Frixione:2002ik}.
Since $H\to\bbbar$ decays require a non-zero bottom-quark mass, we adopt the four-flavour (4F)
scheme with $m_b=4.75$\,GeV and we use 4F CT10 parton densities.
For the renormalisation $(\mu_R)$, factorisation $(\mu_F)$ and
resummation $(\mu_Q)$ scales we choose
$\mu_R=\mu_F=\mu_Q=H_T/2=\sum_{i=t,\bar t,H} E_{T,i}/2$, where
$E_T=\sqrt{p_T^2+M^2}$.
The resulting NLO signal cross section (without decays) 
at 14\,TeV
amounts to $\sigma_{\ttbar H}= 558.7(9)$\,fb.

As a precise benchmark for the irreducible $\ttbar\bbbar$ background we
have performed an \SMCatNLO simulation of $\ttbb$ production in the 4F scheme~\cite{Cascioli:2013era}. 
The finite $b$-quark mass in the 4F scheme avoids collinear $g\to\bbbar$ singularities and
permits to cover the full $b$-quark phase space.  Thus it provides a NLO
accurate description of $\ttbar+b$-jet final states 
with $b$-jet multiplicity $N_b\ge 1$ and $N_b\ge 2$.  As in~\cite{Cascioli:2013era}, for the 
factorisation and resummation scale we use
$\mu_F=\mu_Q=(E_{T,t}+E_{T,\bar t})/2$, and the renormalisation scale is
related to the top- and bottom-quark transverse energies by
$\mu^4_R=\Pi_{i=t,\bar t,b,\bar b } E_{T,i}$, where all transverse energies
are defined at parton level.
\def\siLOPS{\sigma_{\mathrm{LO+PS}}}
\def\siSMC{\sigma_{\mathrm{SMC@NLO}}}
\def\siMEPSatLO{\sigma_{\mathrm{MEPS@LO}}}
\def\siMEPSatNLO{\sigma_{\mathrm{MEPS@NLO}}}

\begin{table} 
  \begin{tabular}{l|@{\quad}c@{\quad}|@{\quad}c@{\quad}|@{\quad}c@{\quad}}
   & $ttb$ & $ttbb$ & $ttbb_{100}$ \\
  \hline
  $\siLOPS$[pb] & 8.109 & 1.800 & 0.668 \\[1ex]
  $\siSMC$[pb]  & 15.22 & 2.973 & 1.041 \\[1ex]
  $1.65\times\siLOPS/\siSMC$ & 0.88 & 1.00 & 1.06 \\[1ex]
  \end{tabular}
  \caption{LO+PS and 
\SMCatNLO predictions for $\ttbar+b$-jet production for the
 subsamples with $N_b\ge 1$ ($ttb$), $N_b\ge 2$ ($ttbb$) and $N_b\ge 2$ plus
 an additional $m_{bb}>100$\,GeV cut ($ttbb_{100}$).}
  \label{tab:ttbbXS}
\end{table}

Using the \SMCatNLO sample we have validated a high-statistics LO+PS simulation (based
on the same input parameters and flavour-number scheme) that was used for
the 
analysis in this paper.
In particular, upon application of a constant 
$K$-factor we have checked that LO and NLO predictions are in decent agreement for a
wide range of observables.
In the following we present comparisons of LO and NLO predictions that have
been obtained by switching off top-quark decays, hadronisation and
underlying event.  This approach, allows one to focus on those jets that
originate from QCD interactions and are most sensitive to NLO corrections. 
In Table~\ref{tab:ttbbXS} we compare LO+PS and \SMCatNLO predictions for
$\ttbar+b$-jet cross sections with a different number, $N_b$, of $b$-jets\footnote{For these 
technical comparisons we use the anti-$k_T$ algorithm with $R=0.4$ and we define as $b$-jets 
those jets that contain one or more b-quarks among their constituents.}
with $p_T>25$\,GeV and $|\eta|<2.5$.  The inclusive case with $N_b\ge 1$
(denoted as $ttb$) is compared to more inclusive cross sections with $N_b\ge
2$ ($ttbb$) and $N_b\ge 2$ plus an additional cut, 
$m_{bb}>100$\,GeV, on the invariant mass of the first two $b$-jets
($ttbb_{100}$).  In the actual analysis the LO+PS $\ttbb$ simulation is
improved by a constant $K$-factor of $1.65$.  As can be seen from
Table~\ref{tab:ttbbXS}, this reduces the discrepancy between LO+PS and
\SMCatNLO predictions to 
about 10\% or less
in the three considered subsamples,
while the intrinsic scale uncertainty of the \SMCatNLO prediction is 
around 20--30\%~\cite{Cascioli:2013era}.  We have checked that such good
agreement between rescaled LO+PS and \SMCatNLO predictions holds also for a
number of distributions.  This is
illustrated in Figure~\ref{fig:ttbbmc} for various observables in the
$ttb$ and $ttbb$ subsamples.  The largest discrepancies do not exceed 20\%
and are observed in phase space regions that are not critical for a boosted
$\ttbar H(\bbbar)$ analysis, such as in the region of large $\Delta R_{bb}$. 
In contrast, for the most relevant observables, such as the $p_T$ and
invariant-mass distributions of the $b$-jets, the observed deviations
between LO and NLO predictions are well below the intrinsic \SMCatNLO uncertainty.

\def\stew{.5}
\def\stgw{.8}

\begin{figure} 
\centering
\begin{subfigure}{\stew\textwidth}
  \centering
  \includegraphics[width=\stgw\linewidth]{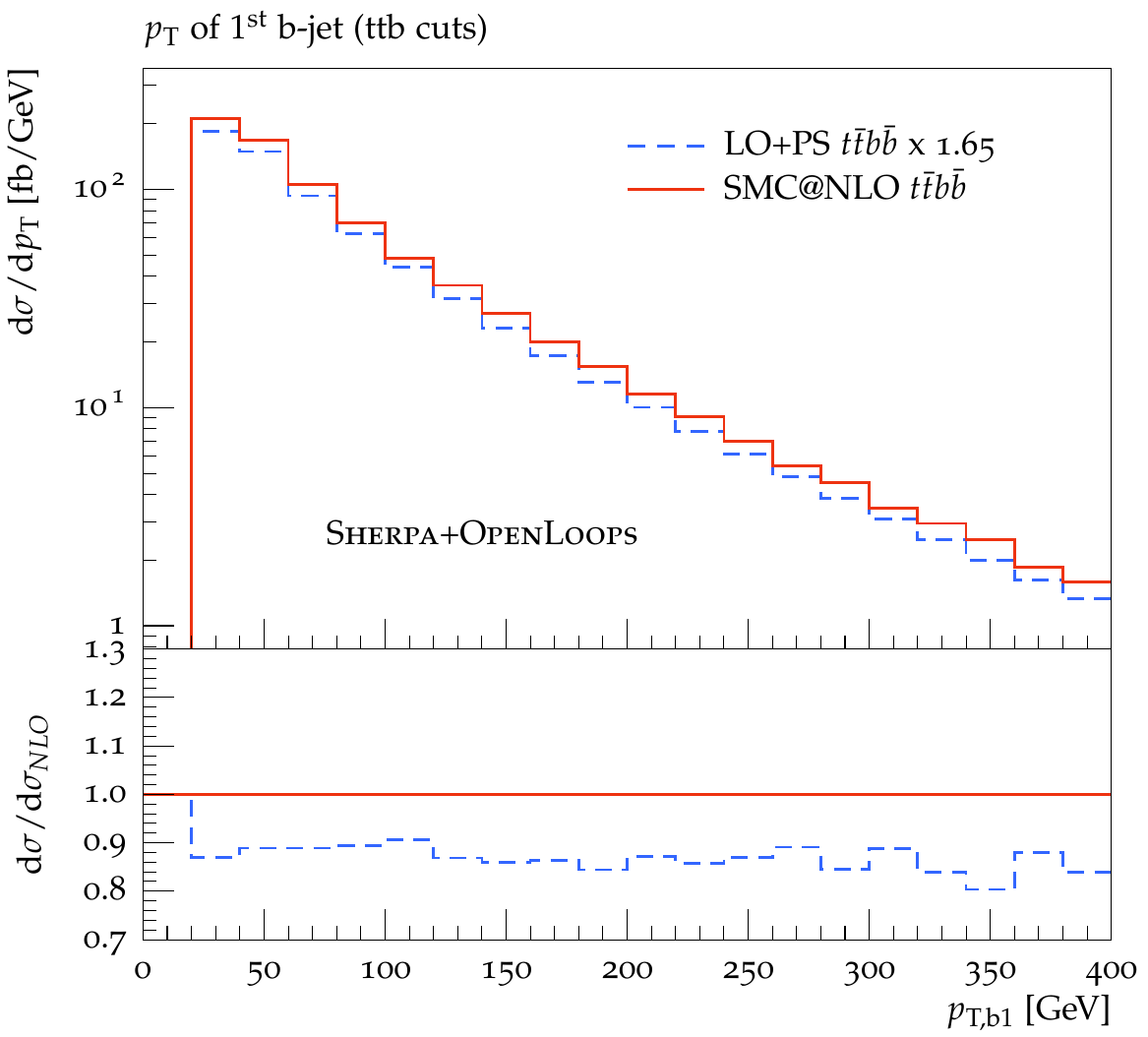}
  \caption{}
  \label{fig:ttbba}
\end{subfigure}
\begin{subfigure}{\stew\textwidth}
  \centering
  \includegraphics[width=\stgw\linewidth]{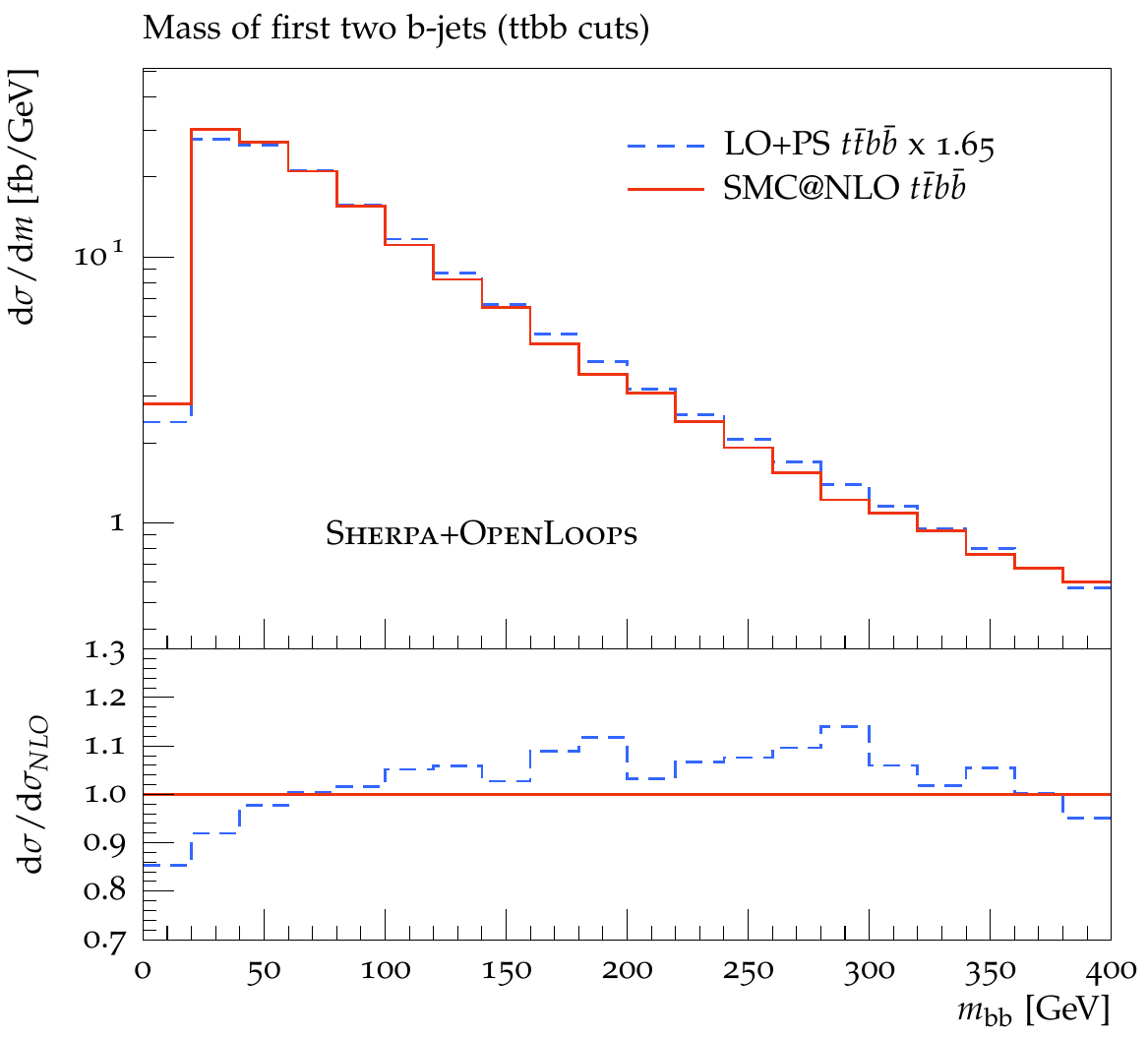}
  \caption{}
  \label{fig:ttbbb}
\end{subfigure}\\[1ex]
\begin{subfigure}{\stew\textwidth}
  \centering
  \includegraphics[width=\stgw\linewidth]{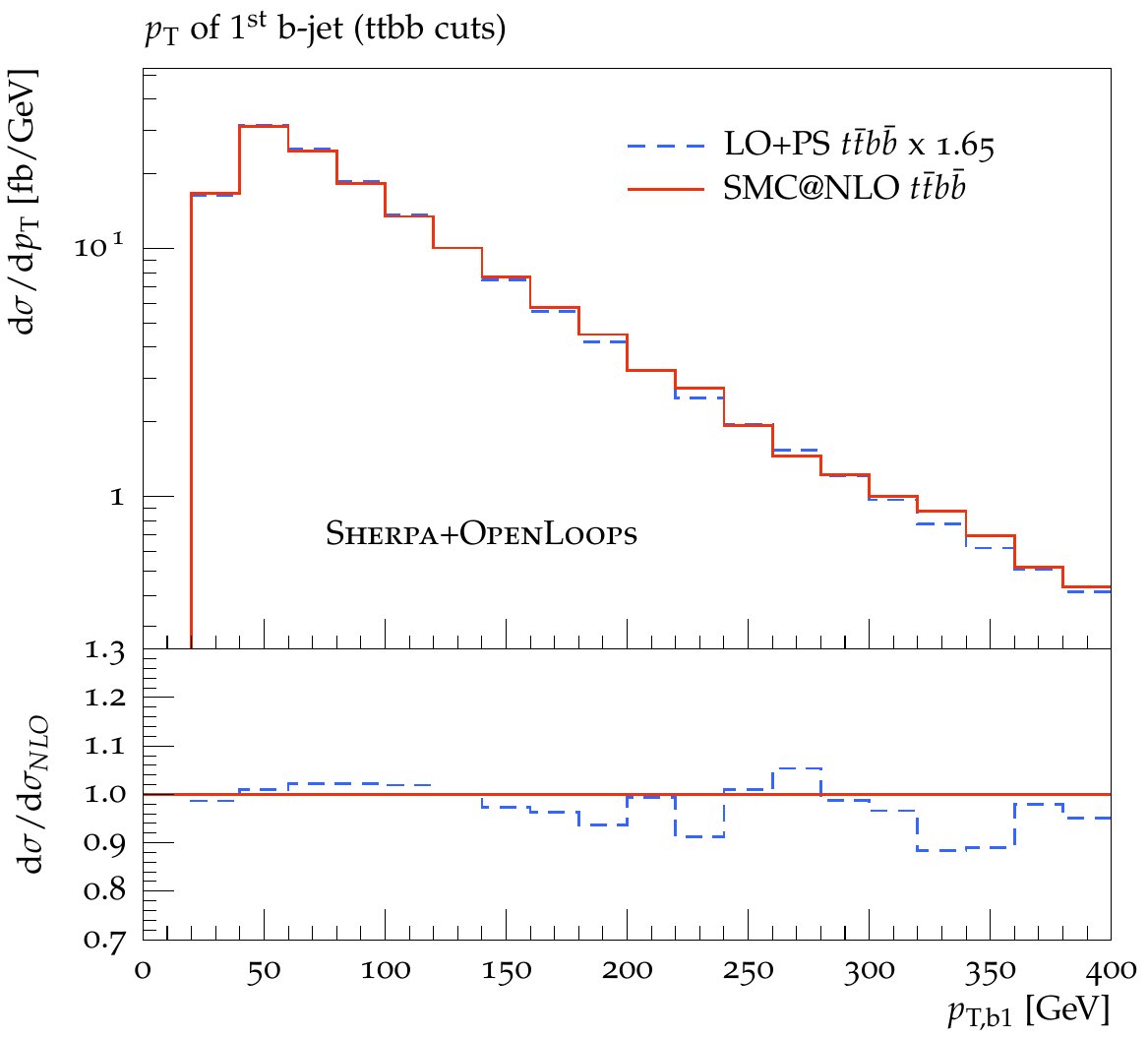}
  \caption{}
  \label{fig:ttbbc}
\end{subfigure}
\begin{subfigure}{\stew\textwidth}
  \centering
  \includegraphics[width=\stgw\linewidth]{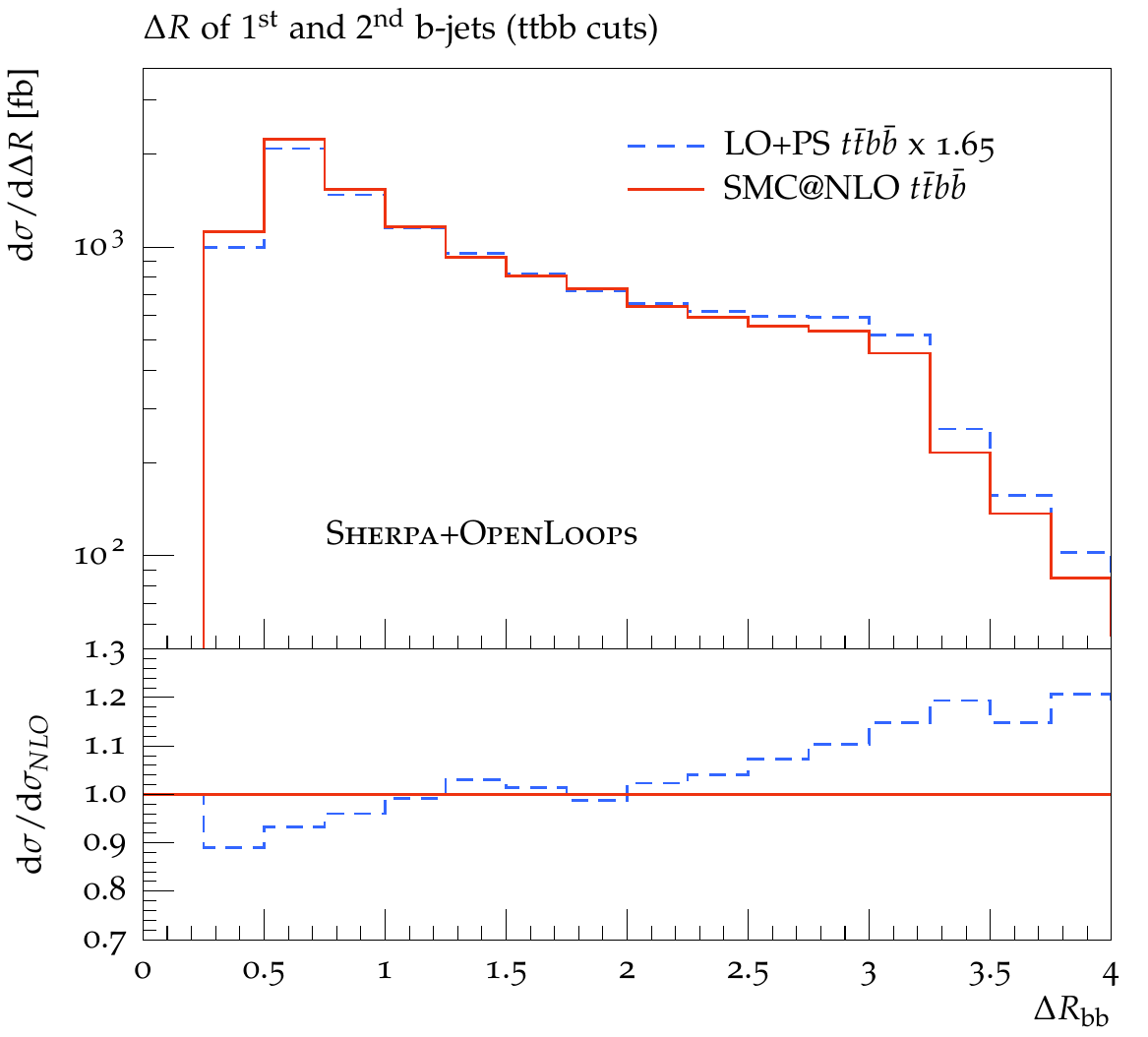}
  \caption{}
  \label{fig:ttbbd}
\end{subfigure}\\[1ex]
\begin{subfigure}{\stew\textwidth}
  \centering
  \includegraphics[width=\stgw\linewidth]{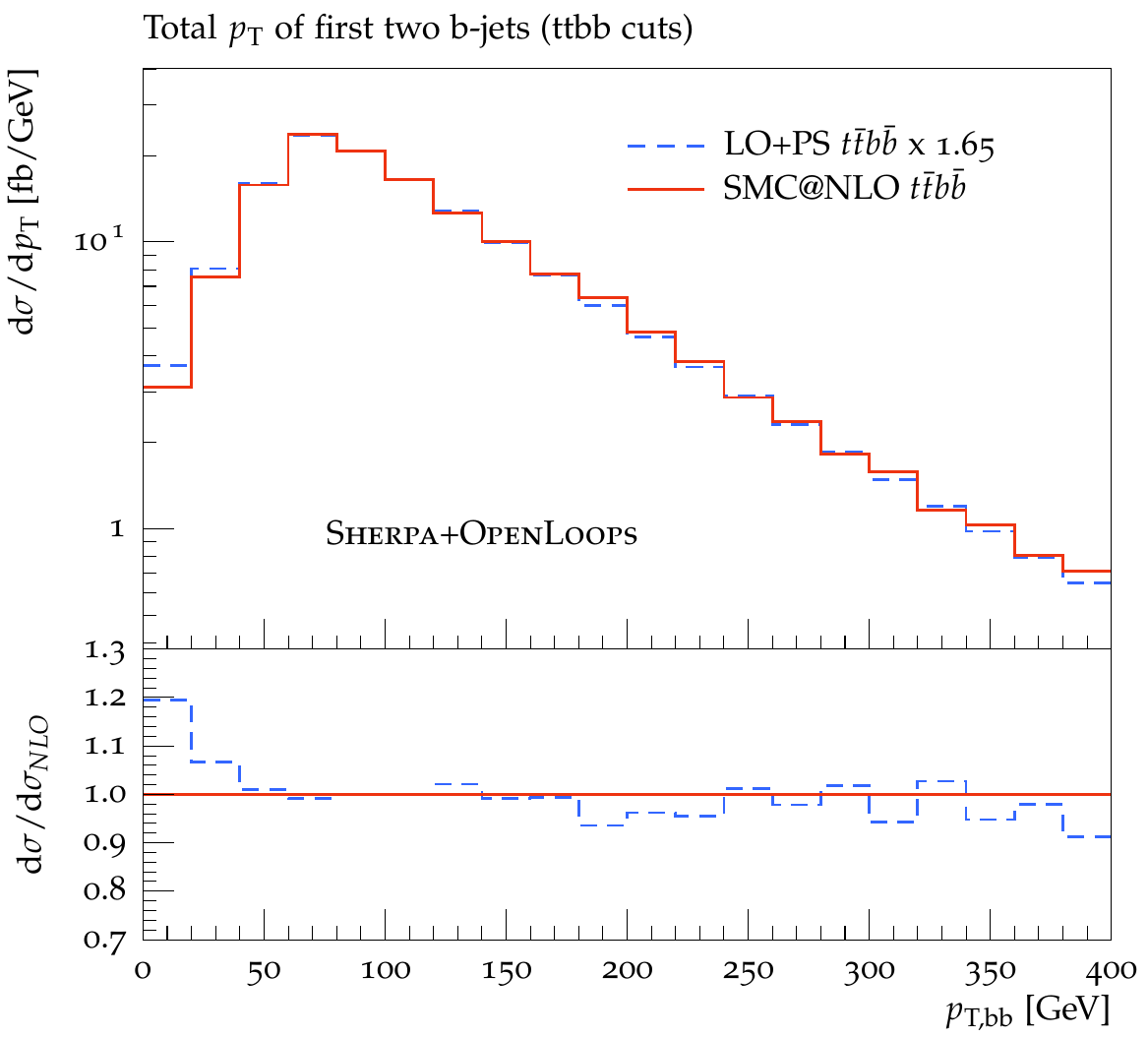}
  \caption{}
  \label{fig:ttbbe}
\end{subfigure}
\begin{subfigure}{\stew\textwidth}
  \centering
  \includegraphics[width=\stgw\linewidth]{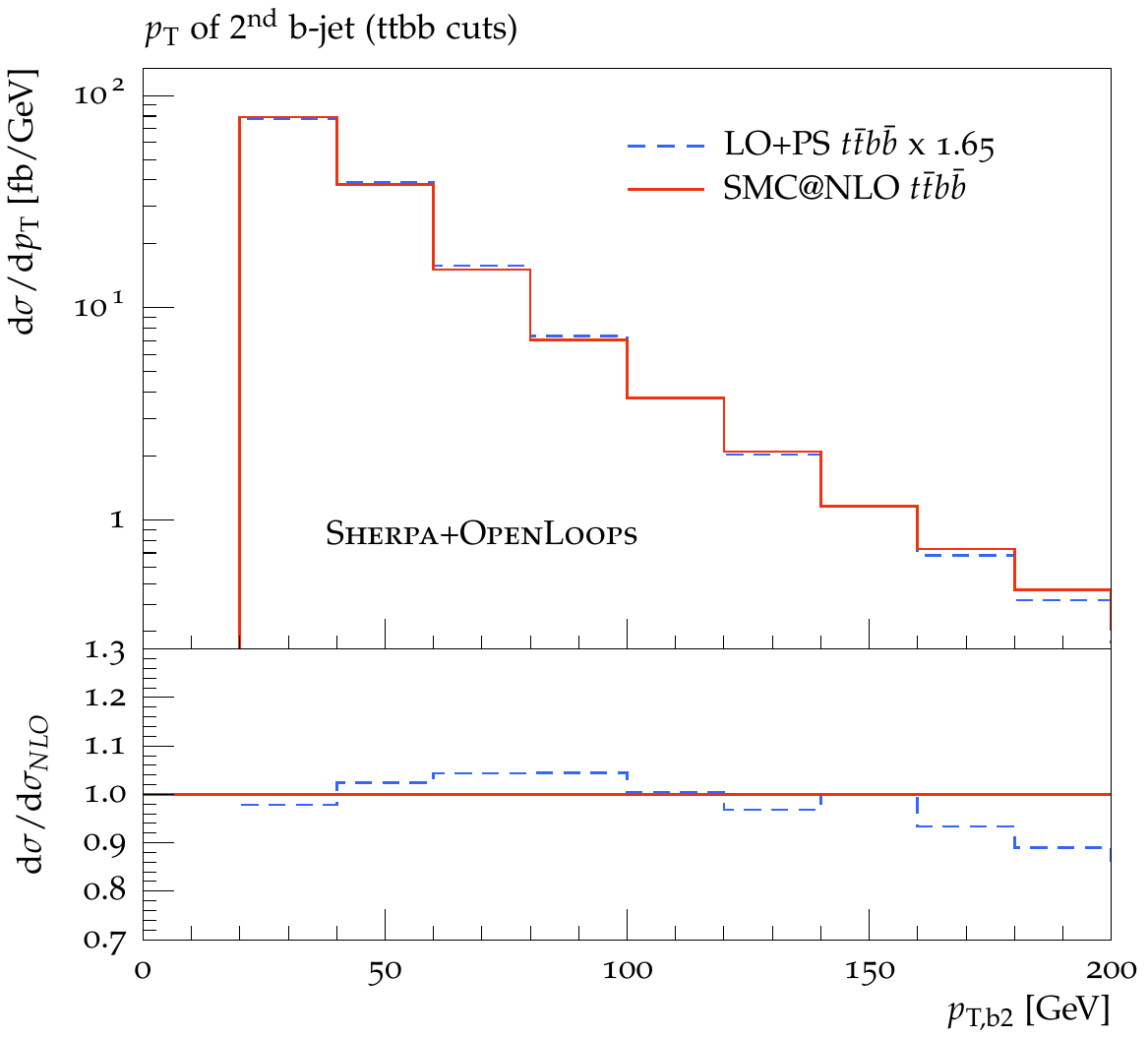}
  \caption{}
  \label{fig:ttbbf}
\end{subfigure}\\[1ex]
\caption{Comparison of LO+PS and \SMCatNLO predictions for 
$p_T$ of first $b$-jet in the inclusive $ttb$ subsample (\ref{fig:ttbba})
and various observables in the $ttbb$ subsample:
invariant mass of first two $b$-jets (\ref{fig:ttbbb}),
$p_T$ of first $b$-jet (\ref{fig:ttbbc}),
$\Delta R$ of first two $b$-jets (\ref{fig:ttbbd}),
total $p_T$ of first two $b$-jets (\ref{fig:ttbbe}) and
$p_T$ of second $b$-jet (\ref{fig:ttbbf}).
In this comparison top decays, hadronisation and underlying event are switched off.
A constant $K$-factor of 1.65 is applied to the LO+PS $\ttbb$ simulation.
}
\label{fig:ttbbmc}
\end{figure}

\def\nmax{n_{\mathrm{max}}} 
\def\mucore{\mu_{\mathrm{core}}}
\def\qcut{Q_{\mathrm{cut}}}

As a benchmark for $\ttbar+$multi-jet
production we have produced an inclusive sample based on the \MEPSatNLO
technique~\cite{Hoeche:2012yf,Gehrmann:2012yg}.  In this approach, \SMCatNLO
 simulations of $\ttbar+n$-jet production  with $n=0,1,\dots,\nmax$ are consistently
merged to form an inclusive sample that provides an NLO accurate description
of any observable involving up to $\nmax$ jets.  First applications of NLO
merging techniques to $\ttbar+$\,multijet production have been presented
in~\cite{Frederix:2012ps,Hoeche:2013mua} for $\nmax=1$ and
in~\cite{Hoeche:2014qda} for $\nmax=2$.  Given the high computational cost
of handling $\ttbar+2$\,jet final states at NLO, in this study we will
restrict ourselves to a \MEPSatNLO simulation for $\ttbar+0,1$\,jets.  As
in~\cite{Hoeche:2014qda}, for the $pp\to\ttbar$ core process we choose the
scales $\mu_R=\mu_F=\mu_Q=\mucore$ with
$1/\mucore^2=1/\hat{s}+1/(m_t^2-\hat{t})+1/(m_t^2-\hat{u})$, while the scale of
the $\alpha_S$ factors associated with additional jet emissions is set equal
to the transverse momentum of the corresponding branchings.  The latter are
determined in a probabilistic way by inverting the \Sherpa parton shower. 
For the separation of the $\ttbar+0$-jet and $\ttbar+1$-jet \SMCatNLO
samples the merging scale $\qcut=30\,$GeV is used.  

Heavy quarks are described in the massless approximation using the
five-flavour (5F) scheme with CT10 5F PDFs.  Double counting with the (N)LO
matched $\ttbb$  sample
in the 4F scheme is avoided by vetoing
final states with one or more $b$-quarks in the (N)LO merged 5F simulation.
In particular, also $t\bar t+b$ configurations with a single $b$-quark need to be 
vetoed since they are already taken into account in the 4F scheme through $gg\to t\bar t b\bar b$ subprocesses where one of the
$b$-quarks remains unresolved in initial-state $g\to b\bar b$
collinear splittings.

Since $b$-production takes
place both through matrix elements and parton showers, the matching of
4F $\ttbb$ production and 5F $\ttbar+$\,jets production needs to be based on a $b$-quark veto
after parton showering. However, $b$-quarks that 
arise from the decay of top quarks (and their subsequent showering)
in $\ttbar+$\,light-jet events are not vetoed since
such configurations cannot arise from $\ttbb$ 4F matrix elements.

\begin{figure} 
\centering
\begin{subfigure}{\stew\textwidth}
  \centering
  \includegraphics[width=\stgw\linewidth]{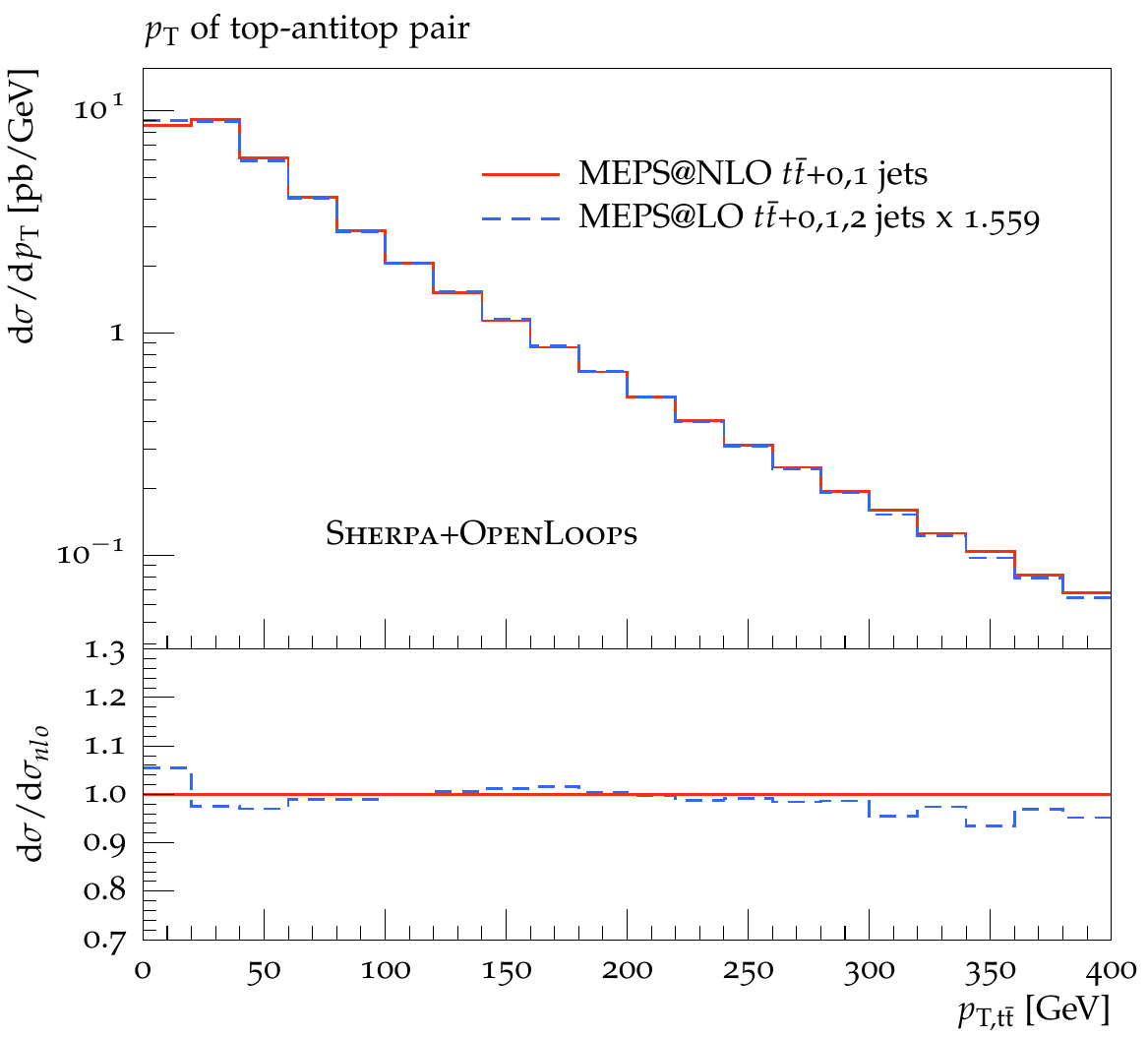}
  \caption{}
  \label{fig:ttjetsa}
\end{subfigure}
\begin{subfigure}{\stew\textwidth}
  \centering
  \includegraphics[width=\stgw\linewidth]{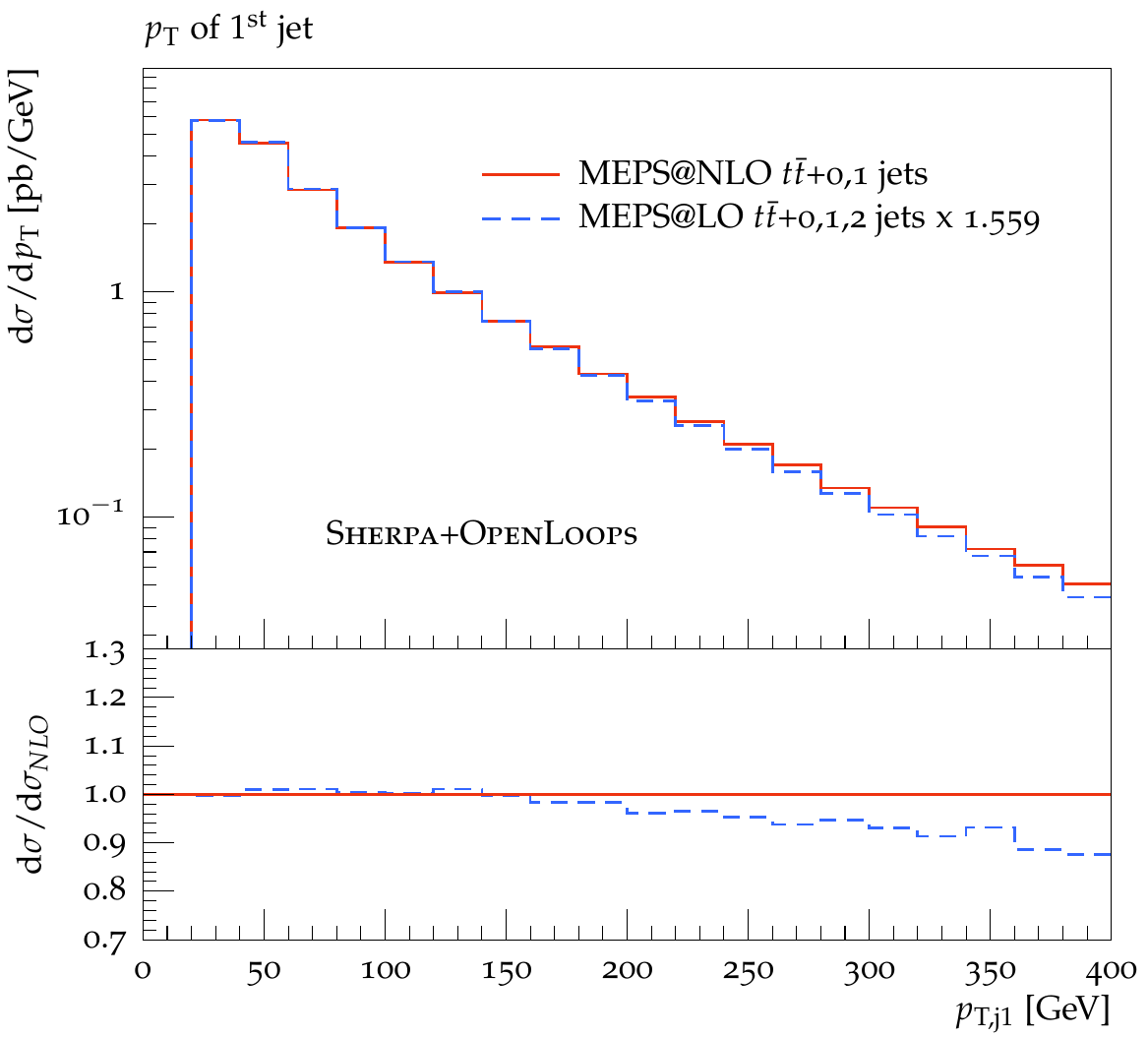}
  \caption{}
  \label{fig:ttjetsb}
\end{subfigure}\\[1ex]
\begin{subfigure}{\stew\textwidth}
  \centering
  \includegraphics[width=\stgw\linewidth]{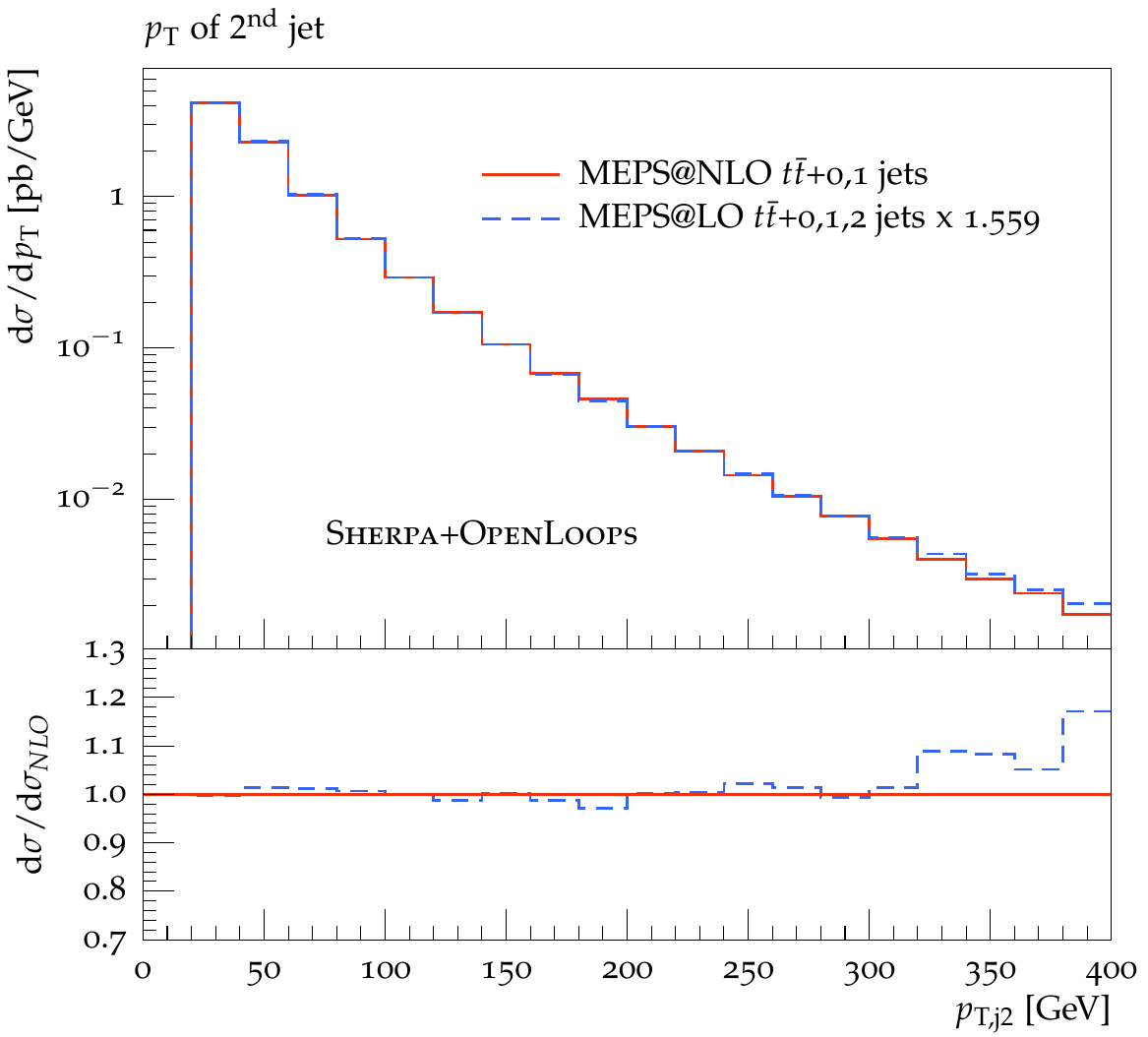}
  \caption{}
  \label{fig:ttjetsc}
\end{subfigure}
\begin{subfigure}{\stew\textwidth}
  \centering
  \includegraphics[width=\stgw\linewidth]{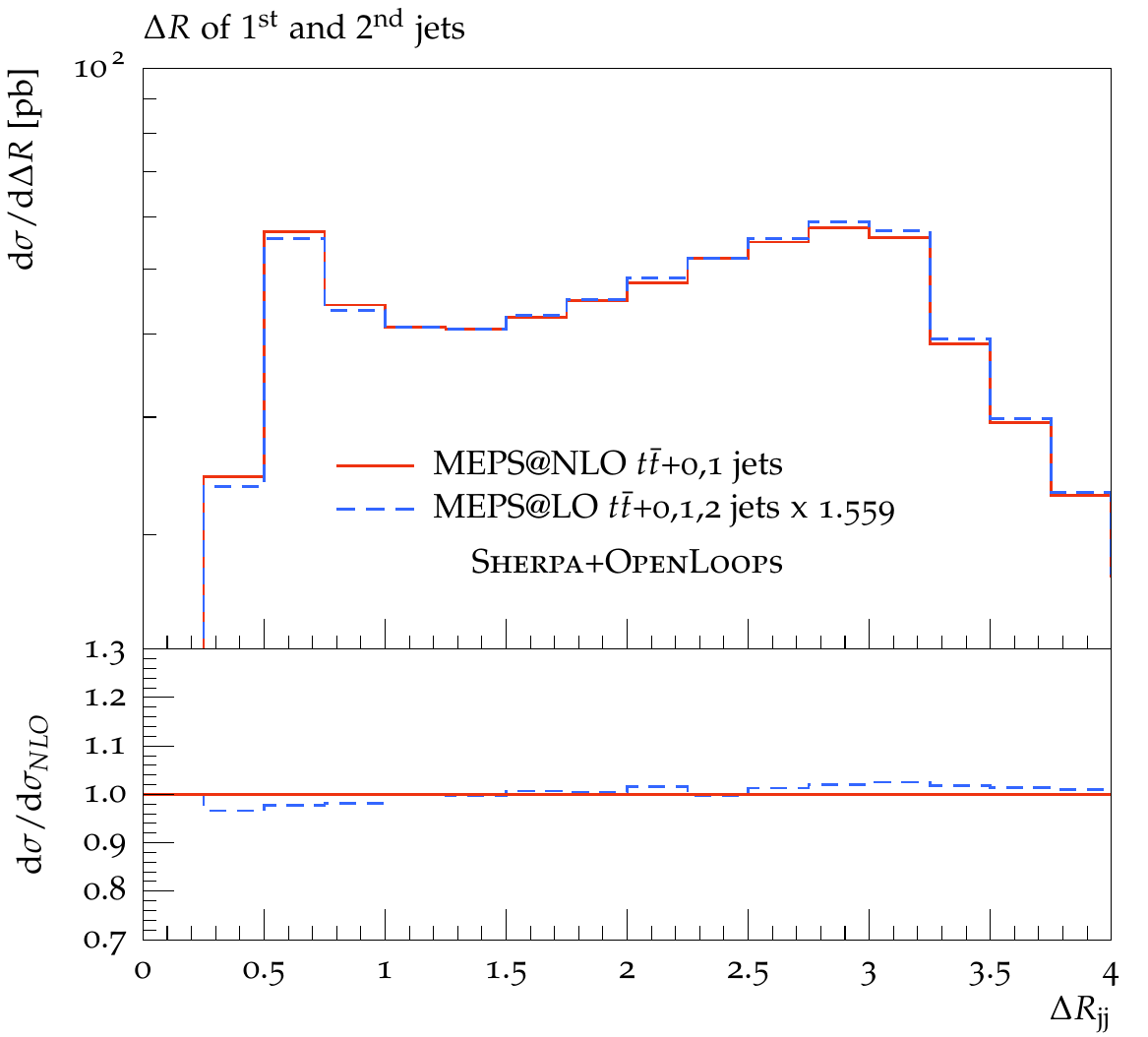}
  \caption{}
  \label{fig:ttjetsd}
\end{subfigure}\\[1ex]
\begin{subfigure}{\stew\textwidth}
  \centering
  \includegraphics[width=\stgw\linewidth]{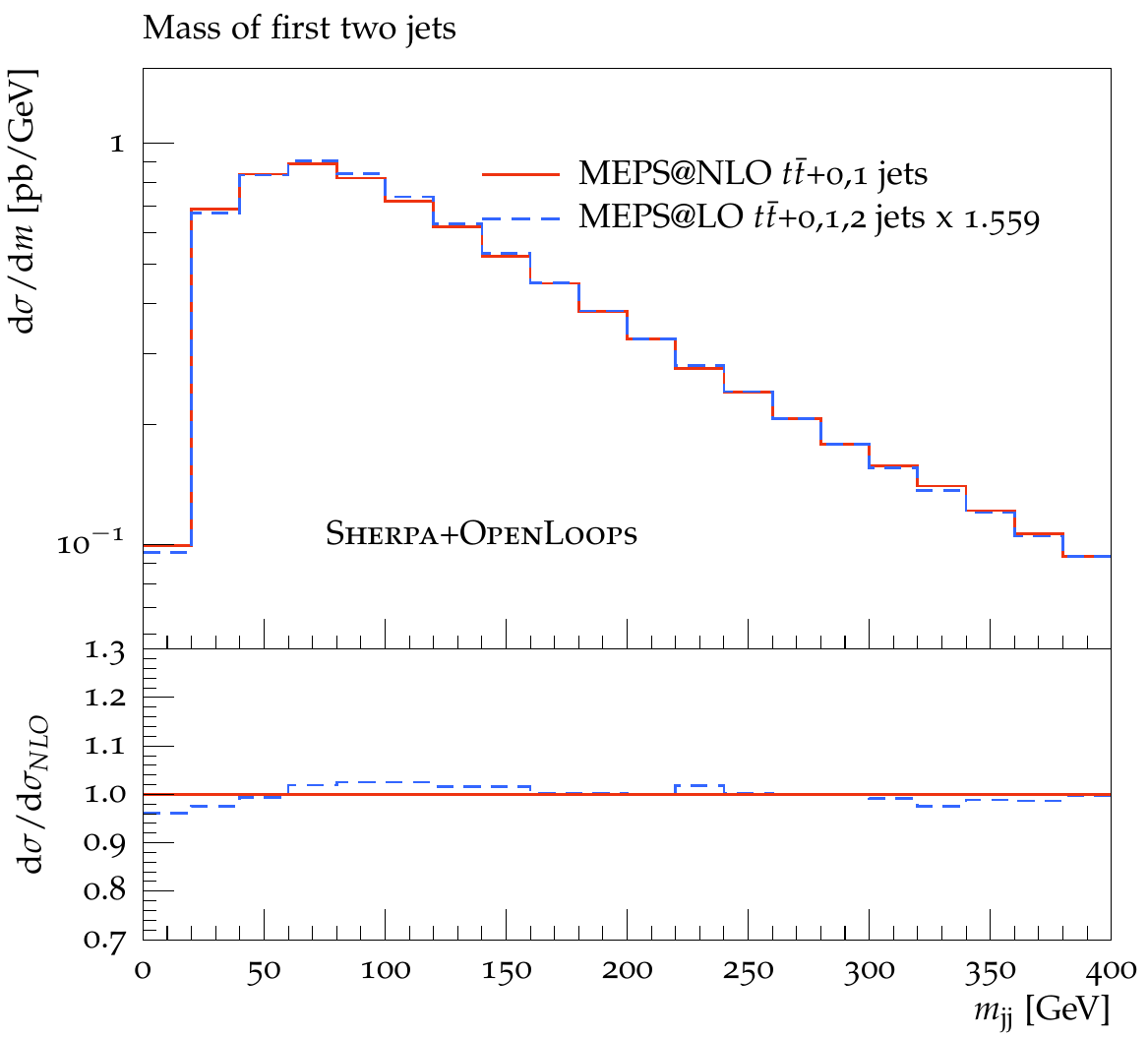}
  \caption{}
  \label{fig:ttjetse}
\end{subfigure}\\[1ex]
\caption{Comparison of \MEPSatLO $\ttbar+0,1,2$\,jets and \MEPSatNLO $\ttbar+0,1$\,jets 
predictions for $p_T$ of $\ttbar$ pair (\ref{fig:ttjetsa}),
$p_T$ of first (\ref{fig:ttjetsb}) and second (\ref{fig:ttjetsc}) jet,
$\Delta R$ of first two jets (\ref{fig:ttjetsd}) and 
invariant mass of first two jets (\ref{fig:ttjetse}).
In this comparison generic (light or heavy-flavour) jets are considered, but $c$- and $b$-jets 
yield only minor contributions.
Top decays, hadronisation and underlying event are switched off.
A constant $K$-factor of 1.559 is applied to the \MEPSatLO simulation.
}
  \label{fig:ttjetsmc}
\end{figure}

Similarly as for the $\ttbb$ simulations, also for the inclusive $\ttbar+$\,jets
background our analysis relies on a high-statistics LO sample.  More
precisely, using the same scale choices as for \MEPSatNLO, we have generated
a \MEPSatLO $\ttbar+0,1,2$\,jets sample, with up to two jets at
matrix-element level.
Again, thanks to the adopted scale choices and an appropriate $K$ factor, LO and NLO predictions
turn out to be in good agreement. Cross sections for $pp\to\ttbar+0,1,2,3$\,jets are compared in 
Table~\ref{tab:ttjets}, using an anti-$k_T$ jet algorithm with $R=0.4$ and counting 
jets with $p_T>25$\,GeV and $|\eta|<2.5$. Applying a constant $K$-factor of 1.559
brings \MEPSatLO and \MEPSatNLO in striking agreement for all considered jet multiplicities.
We also compared the \MEPSatLO simulation rescaled with $K=1.559$ to 
fixed-order NLO calculations for 
$\ttbar+1$\,jet and 
$\ttbar+2$\,jet production~\cite{Melnikov:2010iu,Bevilacqua:2010ve}.
For the respective cross sections with one and two jets with $p_T>50$\,GeV
we found agreement within 5\% or so, i.e.~below the level of uncertainty expected from
NLO scale variations.
In the final analysis the inclusive $\ttbar$ sample is normalised to the NNLO+NNLL 
result $\sigma_{\ttbar}=956.2$\,pb~\cite{Czakon:2013goa} , which corresponds 
to applying an overall $K$-factor of $1.209\times 1.559=1.885$ to the \MEPSatLO sample.
The comparison of differential observables presented in Figure~\ref{fig:ttjetsmc} demonstrates 
that in  \MEPSatLO approximation also the shape of the most important variables is sufficiently 
well described. The observed shape differences between \MEPSatLO and \MEPSatNLO 
predictions approach 10\% only in the hard tails of the jet-$p_T$ distributions.

\begin{table} 
  \begin{tabular}{l|@{\quad}c@{\quad}|@{\quad}c@{\quad}|@{\quad}c@{\quad}|@{\quad}c@{\quad}}
   & $\ttbar+\ge 0$\,jets    & $\ttbar+\ge 1$\,jets    & $\ttbar+\ge 2$\,jets    & $\ttbar+\ge 3$\,jets \\[1ex]
  \hline
  $\siMEPSatLO$[pb] & 506.8    &        268.3 &        113.0 &       42.66 \\[1ex]
  $\siMEPSatNLO$[pb]  & 790.9    &        419.3 &        175.5 &       65.00 \\[1ex]
  $1.559\times\siMEPSatLO/\siMEPSatNLO$ & 1.000    &        1.002 &        0.996 &       0.977 \\[1ex]
  \end{tabular}
  \caption{
\MEPSatLO and \MEPSatNLO predictions for $\ttbar+$multijet production 
in jet bins. Top quarks are kept stable, and 
the jet counting refers to generic (light or heavy-flavour) jets.}
\label{tab:ttjets}
\end{table}

In summary we are going to use
an \SMCatNLO signal sample, a LO+PS $\ttbb$ sample and a 
$t\bar t+0,1,2$\,jet merged sample based on the \MEPS technique at LO. 
As discussed above,  the background samples 
have been rescaled with appropriate $K$-factors
and validated by means of \SMCatNLO and \MEPSatNLO simulations. 
The overlap between the $\ttbb$ and $t\bar t+$multijet samples is
consistently removed by vetoing any $t\bar t+$multijet event containing 
$b$-quarks that do not originate from (showered) top quark-decays.

\section{Standard boosted $t\bar tH$ analysis} \label{sec:mspana}

In this paper we focus on the extraction of the $t\bar t H(b\bar b)$ signal at the
14\,TeV LHC in the semi-leptonic  
channel. As a starting point of our study,  
in this section we will consider a standard boosted analysis along the lines of~\cite{MSp2009}.
While following the general strategy of~\cite{MSp2009},
as detailed below we will introduce a first series of simple improvements,
such as the usage of the HEPTopTagger,
as proposed in \cite{heptoptagger}, updated reconstruction approaches,
lepton isolation requirements, more reliable treatment of $B$-mesons and so
on.  Moreover the accuracy of Monte Carlo simulations for signal and
backgrounds will be upgraded from LO to NLO, with a significant impact on
the expected sensitivity.
A major extension of the boosted approach, including completely new regions of 
phase space, will be introduced in Sec.~\ref{sec:analysis}.

The strategy proposed in~\cite{MSp2009}, in order to 
improve the separation of the $t\bar{t}H(b\bar{b})$ signal from the problematic 
QCD backgrounds of type $t\bar{t}+X$ and $W$+jets, exploits final states where
both the hadronically decaying top, 
$t_\mathrm{had}$, and the Higgs 
boson
are
modestly boosted ($p_T \gtrsim m$) and balance each other's transverse
momenta, so that their decay products 
tend to
occupy different physical regions of
the detector.

The boosted selection processes three types of Monte Carlo (MC)
objects: \textit{hadrons, leptons} and \textit{$B$-mesons}.  
For leptons, $\ell=e^\pm$ or $\mu^\pm$, we require $|\eta_\ell| < 2.5$ and
$p_\mathrm{T \ell} > 25$ GeV, following \cite{atlasTTHall}, and impose the lepton isolation requirement
$\sum_{i \in \Delta R_\mathrm{i\ell}<0.2}~H_\mathrm{Ti} < 0.1\,p_\mathrm{T \ell}$.
All other visible final state particles with $p_\mathrm{T} > 0.5$ GeV and
$|\eta| < 4.5$ are treated as hadrons. 
While $B$-mesons were kept stable in~\cite{MSp2009}, the 
present analysis includes $B$-decays and resulting $b$-jet energy losses in case of
semi-leptonic $B$-decays (see more details in Sec.~\ref{sec:bjet}).
However a simplified $b$-tagging modelling 
based on $B$-mesons before decays is used.
For a jet or subjet to be labelled as b-jet, a $B$-meson with $p_T>10$ GeV
and $|\eta|<2.5$ before decay has to fall within the jet radius ($\Delta
R_\mathrm{B,jet}< R_\mathrm{jet}$).  When all jets and subjets in a
configuration are MC-tagged as $b$-jets or light jets, a $b$-tag weight is
given to the configuration as a whole by calculating the probability of a
specific number of $b$-jets and light jets based on 70\% and 1\% tagging
efficiencies for $b$-jets and light (or charm) jets, respectively.  All jets
that lie outside the rapidity region $|y|<2.5$ are tagged as light jets.

Similarly as in~\cite{MSp2009}, two preselection cuts are applied in order to suppress 
the overwhelming QCD background. First, we require
exactly one isolated charged lepton in the event.  Since an isolated lepton
is produced in the hard process and not in subsequent hadron decays, this
condition makes the pure multi-jet background insignificant and also
separates the semi-leptonic $t\bar{t}+X$ decay channels from the fully
leptonic ones.  Second, the \textit{hadrons} are clustered into
Cambridge/Aachen (C/A) \cite{ca_algo} $R=1.5$ fat jets with
$p_\mathrm{Tj}>200$ GeV, excluding events with less than two such fat jets.
The fat jets are handled as hadronic-top and Higgs candidates.

After preselection cuts the selection enters the main stage, which is based
on following jet-substructure analysis: 
\begin{enumerate} 
\item \label{enum:toptagger} Each fat jet is tagged as $t_\mathrm{had}$
(hadronic top) or non-$t_\mathrm{had}$ jet using a top 
tagger, and we require at least one top-tag in the event.
 Specifically,
the HEPTopTagger is used instead of the top-tagging method described 
in~\cite{MSp2009}. 
Although two hadronic top-tags in a semi-leptonic $t\bar t$ event are
unlikely, there is a significant probability to misidentify a Higgs boson as
a top quark 
(see Sec.~\ref{sec:topreco}).
Thus, 
more than one fat jet
can be
top-tagged at this stage.

\item \label{enum:thadbest} In the interest of retaining as much signal as
possible, instead of vetoing events with more than one top-tagged jet,
we identify as unique top candidate the top-tagged jet
that minimises $\Delta m_{\mathrm{tot}} \equiv |m_\mathrm{t,reco} -
m_\mathrm{t}| + \mathrm{min_{ij}}|m_\mathrm{ij} - m_\mathrm{W}|$.  Here
$m_{t,\mathrm{reco}}$ is the mass of the reconstructed top and $m_{ij}$ is
the invariant mass of the pair of subjets closest to the $W$ mass.

\item \label{enum:yhcut} A rapidity cut $|\eta| <2.5$ is applied to all
remaining fat jets, including top-tagged jets that have not been selected as
top candidates in the previous step.

\item \label{enum:newana} 
For each fat jet (except the top candidate) we apply the mass drop filter proposed in~\cite{MSp2009}.  
If the fat jet has less than two subjets after mass drops it is ignored.
Otherwise the pairs of 4-momenta that survive the mass drop represent possible $H(b\bar{b})$ structures.
They are ordered according to a variant of the
Jade distance \cite{JADE},
\begin{equation}
d_\mathrm{ij} = p_\mathrm{Ti}p_\mathrm{Tj}\Delta R^4_\mathrm{ij},
\end{equation}
and only the first three such pairs in descending distance $d_\mathrm{ij}$ are retained. Next, the
constituents of each subjet pair are filtered into C/A jets of radius
$R_\mathrm{filt} = \mathrm{min}(0.3, \Delta R_\mathrm{ij})$ and
$p_\mathrm{T}>20$ GeV.  Only the first 3 filtered jets are kept and combined
into what we refer to as a \textit{Higgs candidate}.  

\item \label{enum:doublebtag} We require exactly two $b$-tags from the
filtered subjets of the Higgs candidate.

\item \label{enum:trippletag} We request exactly one additional $b$-tag in the event. This condition is applied
after removing the reconstructed Higgs and top, 
which are supposed to involve three of the four $b$-quarks of a signal event,
and after clustering the remaining final state objects of
the Higgs fat jet into C/A jets with $R=R_\mathrm{filt}$ and
$p_\mathrm{T}>20$ GeV ({\it inner jets}), and the objects outside the Higgs
fat jet into C/A jets with $R=0.4$ and $p_\mathrm{T}>30$ GeV ({\it outer
jets}).  As the Higgs fat jet was already processed by a mass
drop/grooming procedure, we choose a more aggressive jet definition.

\item \label{enum:lastbasic} We identify a Higgs candidate as tagged if its invariant mass $m_\mathrm{c}$ lies
in the $[100,130]$\,GeV  mass window. 

\end{enumerate}
 
\begin{figure} 
  \centering
  \begin{subfigure}{\stew\textwidth}
    \centering
    \includegraphics[width=\stgw\linewidth]{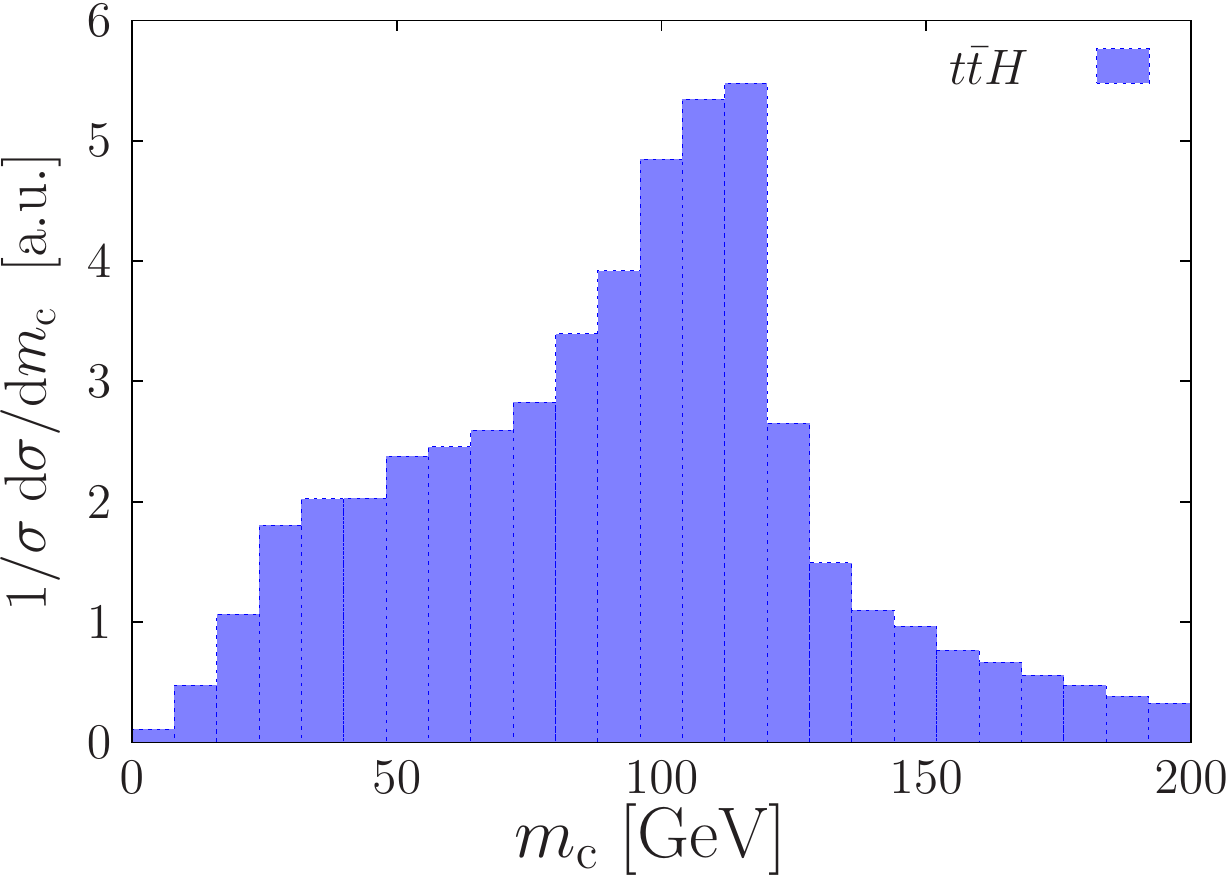}
  \end{subfigure}%
  \begin{subfigure}{\stew\textwidth}
    \centering
    \includegraphics[width=\stgw\linewidth]{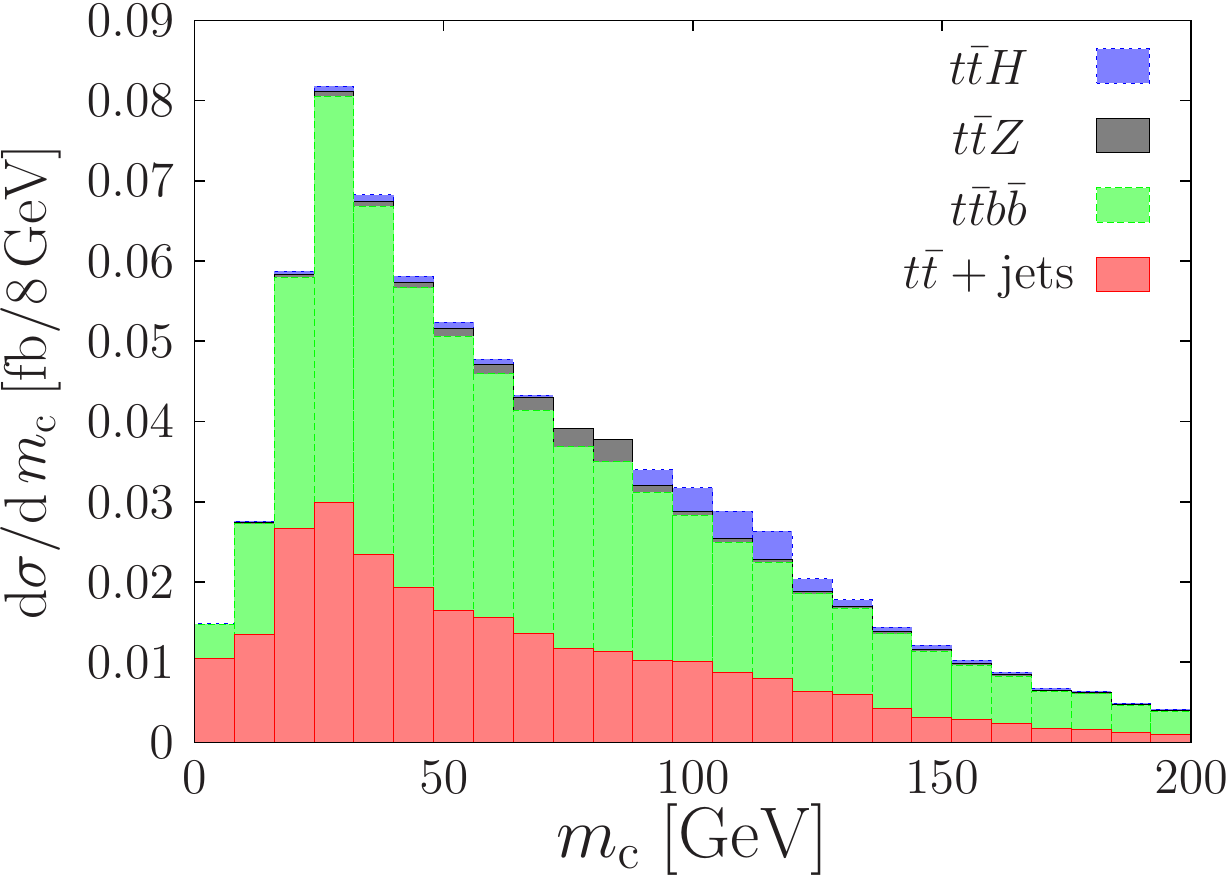}
  \end{subfigure}\\[1ex]
  \caption{Distributions in the Higgs-candidate mass, $m_\mathrm{c}$, 
for signal (left) and signal plus $t\bar t+X$ backgrounds (right)
after 
step~\ref{enum:trippletag} (third $b$-tag) of the standard boosted
analysis of Sec.~\ref{sec:mspana}.}
  \label{fig:basic}
\end{figure}

A first picture of the quality of the Higgs reconstruction in the boosted
analysis
described above
is provided in Fig.~\ref{fig:basic}, which displays the invariant-mass distribution of the
Higgs candidate after step~\ref{enum:trippletag}.  The
normalised $m_\mathrm{c}$ distribution for the $t\bar t H$ signal (left
plot) features a sharp cut-off at large $m_\mathrm{c}$ and a rather long
low-mass tail.  There we observe a bulky structure that points to Higgs
misidentification, i.e.~Higgs candidates that involve $b$-quarks from top
decays.  Moreover, the Higgs peak lies about $15$\,GeV below the true Higgs
mass of $125$ GeV, mainly due to uncorrected energy losses via neutrinos in
$B$-meson decays. 
Superimposing the $\ttbar H(\bbbar)$ signal and the dominant
$t\bar{t}+\mathrm{jets}$ and $t\bar{t}b\bar{b}$ backgrounds (right plot)
illustrates how the latter are dominated by the low mass region.  Nevertheless, 
also due to a certain dilution of the $H\to \bbbar$ peak,
the background contamination of the signal region remains quite serious.
In particular, as discussed in detail in Sec.~\ref{sec:results}, 
when comparing to the analysis in~\cite{MSp2009} we find a 
sizeable reduction of $S/B$ in the signal region,
which can be attributed to the changes in the Monte Carlo simulations of
signal and background processes and to the $b$-jet mistagging in $t\bar t$+jet events.
As a consequence, the systematic (theoretical and
experimental) uncertainty on the background rate and shape may be as large
or even larger than the signal.  
Hence for an optimal signal strength measurement in $t\bar{t}H(b\bar{b})$
a further reduction of the background level through
improved selection strategies, as well as a reduction of the related
uncertainties, are of crucial importance.

\begin{figure}
\centering
  \begin{subfigure}{.35\textwidth}
    \centering
    \includegraphics[width=\stgw\linewidth]{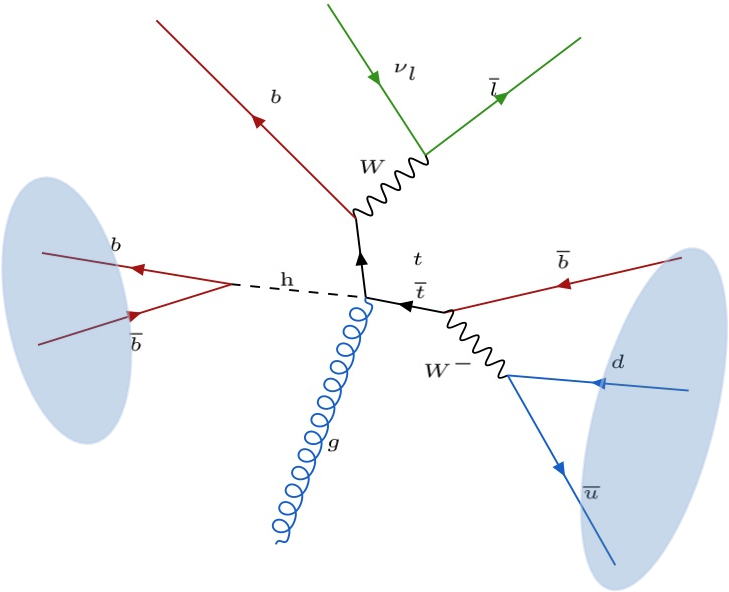}
\caption{}
\label{topolclassa}
  \end{subfigure}%
\hspace{20mm}
  \begin{subfigure}{.35\textwidth}
    \centering
    \includegraphics[width=\stgw\linewidth]{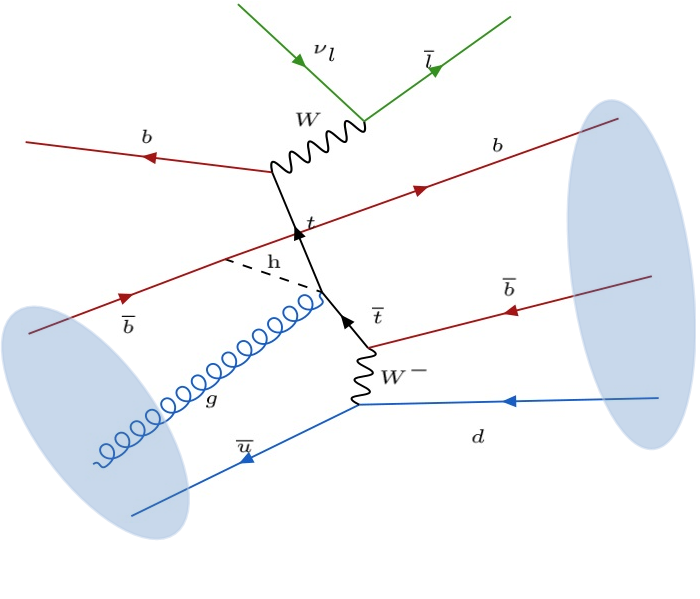}
\caption{}
\label{topolclassb}
  \end{subfigure}\\[5ex]
  \begin{subfigure}{.35\textwidth}
    \centering
    \includegraphics[width=\stgw\linewidth]{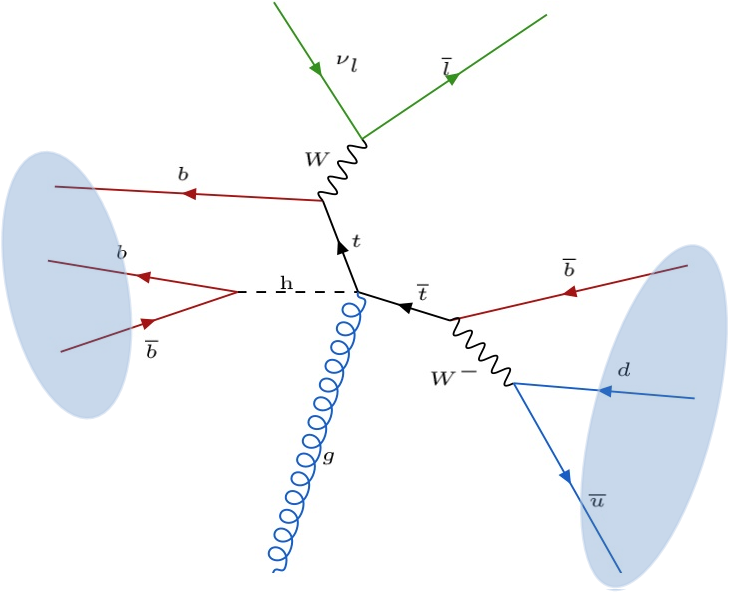}
\caption{}
\label{topolclassc}
  \end{subfigure}%
\hspace{20mm}
  \begin{subfigure}{.35\textwidth}
    \centering
    \includegraphics[width=\stgw\linewidth]{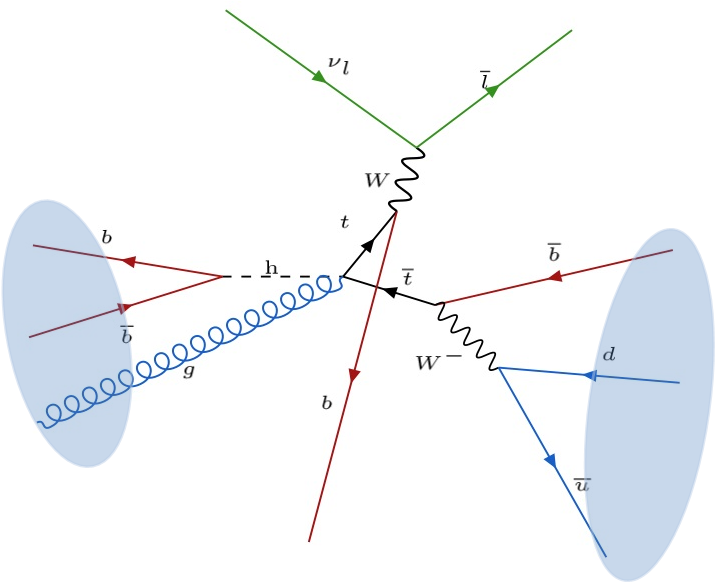}
\caption{}
\label{topolclassd}
  \end{subfigure}\\[1ex]
\caption{
Schematic representation of typical $t\bar{t}H$ event topologies.
The ellipses indicate how partons are clustered to form two fat jets.
Topology~\ref{topolclassa} is the cleanest one: the Higgs products and the
hadronic top products form two separate fat jets without pollution from
other hard particles.  Topology~\ref{topolclassb} features 
misassignements of the Higgs and hadronic top products.
In topology~\ref{topolclassc}
the hadronic top decay products form a fat jet, and the Higgs decay products form another fat
jet with the leptonic top $b$-quark falling within it.  In topology~\ref{topolclassd}
the $b$-quark from the leptonic top decay does not pollute the Higgs fat jet, but there is a
gluon radiation strong enough to form a substructure within the Higgs fat
jet. 
}
\label{fig:topolclass}
\end{figure}

As a preliminary step towards the improved $t\bar t
H(b\bar b)$ selection strategies proposed in Section~\ref{sec:analysis}, 
in the following 
we present a detailed study of the quality of top and Higgs reconstruction
in the 
standard
boosted analysis. 
Specifically,
we attempt to identify the 
patterns that dominate the reconstruction, i.e.~the most probable
ways how $t\bar t H(b\bar b)$ decay products
are grouped into two fat jets. Such configurations will be referred to 
as event topologies, and some typical examples are illustrated in 
Figure~\ref{fig:topolclass}.
The boosted selection is targeted at 
the topology in Fig.~\ref{topolclassa},
where the three quarks from the $t_\mathrm{had}$
decay and the $b\bar{b}$ pair from the Higgs decay form two separate fat
jets, which do not overlap with the extra $b$-jet from the decay of the
leptonic top, $t_\mathrm{lep}$.  However, given the number of final state
objects and the size of the fat jets, the probability is large that the
quarks group up in a different way to form two fat jets with $p_\mathrm{T}>
200$ GeV.  In particular, we are interested in topologies that contribute the
most to Higgs candidate misidentification, resulting in signal dilution and
$t\bar t H$ sensitivity losses. In the following subsection we categorise the signal events according to their quality of Higgs and top reconstruction.

\subsection{Quality of hadronic top reconstruction} \label{sec:topreco}

To define categories that reflect the goodness of fat jets as hadronic top candidates the following
8 binary conditions (true/false) are used: 
\begin{enumerate}
\item $t_\mathrm{had}$: the hadronic top 
quark
is boosted ($p_\mathrm{T,t_{had}} >150$\,GeV)
\item $t_\mathrm{had}$: the hadronic top 
quark
overlaps with the jet ($\Delta R_\mathrm{jet,t_{had}}<R_\mathrm{fat}$)
\item $t_\mathrm{lep}\to b \ell\nu$: the $b$-quark from $t_\mathrm{lep}$ belongs to the jet                      
\item $H\to b\bar b$: the harder $b$ from the Higgs belongs to the jet
\item $H\to b\bar b$: the softer $b$ from the Higgs belongs to the jet
\item $t_\mathrm{had}\to b jj$: the $b$-quark from $t_\mathrm{had}$ belongs to the jet
\item $t_\mathrm{had}\to b jj$: the harder light quark from $t_\mathrm{had}$ belongs to the jet
\item $t_\mathrm{had}\to b jj$: the softer light quark from $t_\mathrm{had}$ belongs to the jet
\end{enumerate}
This characterisation is applied to the ensemble of fat jets in the $t\bar t
H$ signal sample at two levels of the boosted selection: considering
all fat jets just before top tagging and, alternatively, only for 
the jet that has been successfully top-tagged and selected as top candidate in step~\ref{enum:thadbest}.
In practice a $t\bar t H$ event
corresponds to at least two fat jets before top tagging and exactly one
top-tagged fat jet.  Each one of these fat jets falls into one bin of the
8-dimensional discrete space defined by the above conditions.  Overall this
amounts to 256 fat-jet categories, which will be referred to also as jet
topologies in the following.  It turns out that more than 60\% of the top-tagged fat jets
correspond to one of the six jet topologies presented in Table~\ref{tab:topolT}.

\begin{table} 
  \begin{tabular}{c|c|c|c|c}
    label & bin & before top tag & after top tag & tagging efficiency\\
  \hline
$A_1$ & 11000111 & 0.12 & 0.32 & 0.40 \\
$A_2$ & 11001111 & 0.03 & 0.08 & 0.42 \\
$A_3$ & 10111000 & 0.06 & 0.07 & 0.18 \\
$A_4$ & 11010111 & 0.02 & 0.06 & 0.40 \\
$A_5$ & 11100111 & 0.02 & 0.04 & 0.41 \\
$A_6$ & 11011111 & 0.01 & 0.04 & 0.39 \\ [1ex]
  \end{tabular}
  \caption{
The normalised distributions of fat jets before top tagging (column 2) and top-tagged fat jet (column 3)
in the dominant bins of the 8-dimensional jet-category histogram.
The top-tagging efficiency (column 4) is defined as the probability that a fat jet  
is top-tagged in step~\ref{enum:thadbest} of the boosted selection.
The rows are ordered by decreasing fraction after the top-tag.  The bin is identified by
specifying the conditions that are true (1) and false (0) in the order
listed in the text. The left-most digit corresponds to the first condition.
}
  \label{tab:topolT}
\end{table}

\begin{figure} 
  \centering
  \begin{subfigure}{\stew\textwidth}
    \centering
    \includegraphics[width=\stgw\linewidth]{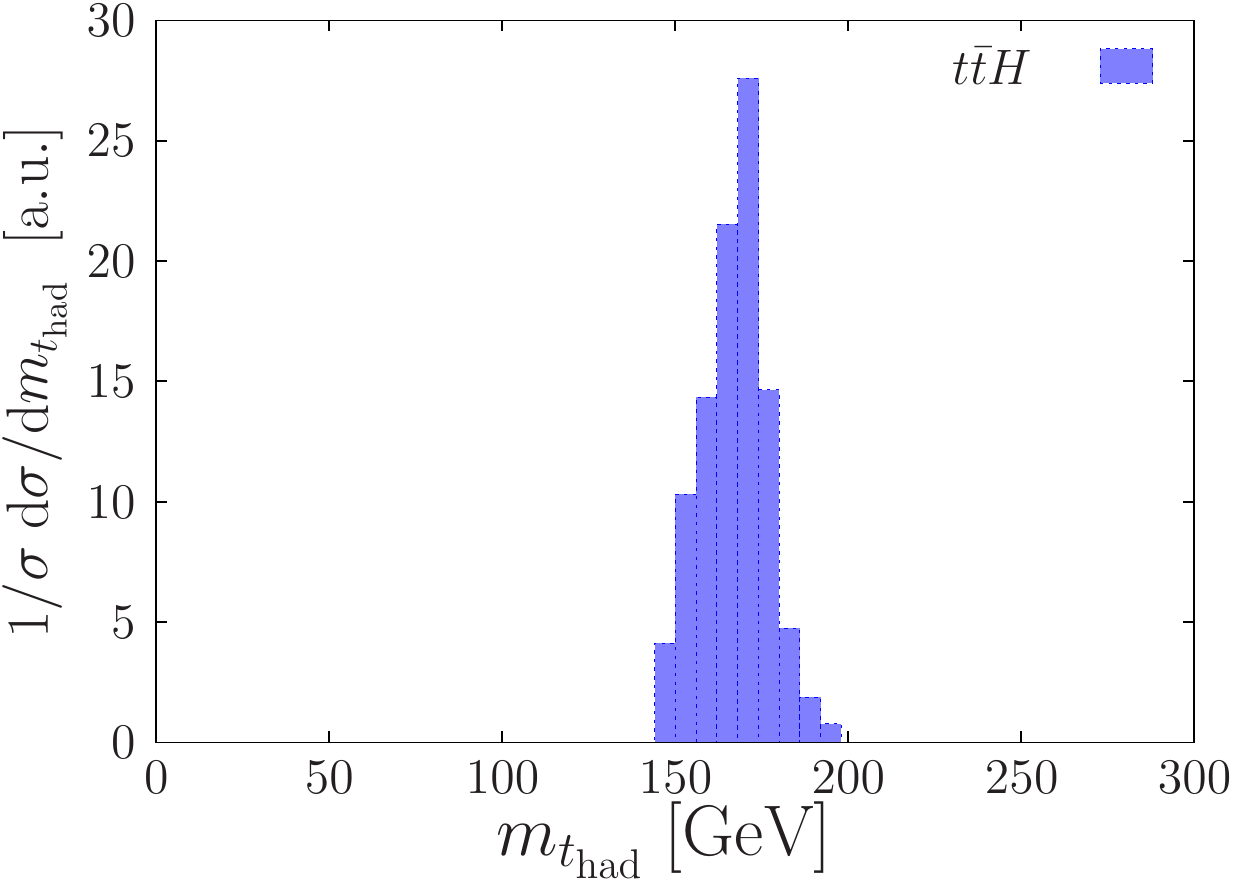}
  \end{subfigure}%
  \begin{subfigure}{\stew\textwidth}
    \centering
    \includegraphics[width=\stgw\linewidth]{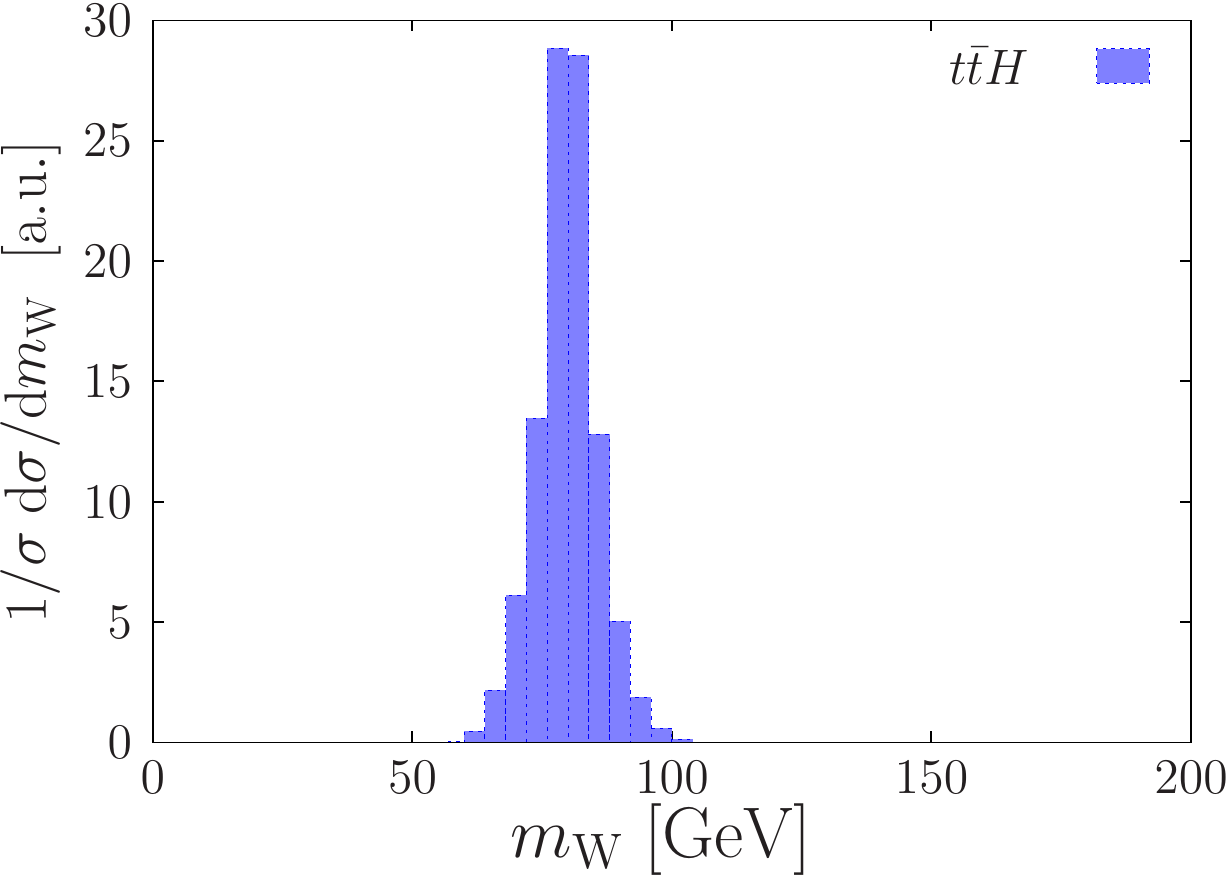}
  \end{subfigure}\\[1ex]
  \caption{Distributions of the
        $m_\mathrm{t_{had}}$ (left) and $m_\mathrm{W}$ (right) invariant masses for the cleanest topology $A_1$ of Table~\ref{tab:topolT}, 
after step~\ref{enum:thadbest}
of the boosted analysis of Sec.~\ref{sec:mspana}.} 
  \label{fig:topolT}
\end{figure}

The $A_1$ topology corresponds to the optimal configuration,
where all $t_\mathrm{had}$ decay products make up 
one fat jet, while the Higgs products and the $t_\mathrm{lep}$ $b$-quark
end up in another direction. As illustrated in Fig.~\ref{fig:topolT},
this topology features an excellent top- and W-mass reconstruction.
However, it corresponds 
to only one third of all events with a top-tag.
For the $A_5$ topology, where the $b$-quark from
$t_\mathrm{lep}$ enters the fat jet of the hadronic top, we find a tagging
efficiency of roughly 40\%, similarly as for $A_1$.
In fact, the top tagger is built in
such a way that a top is recovered  with the same efficiency irrespectively of the
presence of additional structure in the fat jet.
The $A_1$ and $A_5$ topologies allow for a good Higgs identification, since
the Higgs decay products are contained in the remaining fat jet. However
they represent 
less than
40\% of the total signal after top tagging.
There are other configurations, like $A_2$, $A_4$ and $A_6$, where parts of
the Higgs boson as well as the whole top form a fat jet, for which we find
again a tagging efficiency around 40\% as for a fat jet containing only the
top quark.

The $A_3$ topology, where a Higgs fat jet is mistagged as a top,
represents another significant contribution.  The related mistag rate is
around $20\%$, and the corresponding events often involve a second top tag 
associated with the correct hadronic top.  Thus events with more than one top tag 
should not be vetoed, and it is important 
to select the ``best'' top candidate (step~\ref{enum:thadbest} of our selection).  
Obviously topologies where
the reconstructed top contains one or more quarks from the Higgs decay
($A_2$, $A_4$, $A_5$, $A_6$) do not allow for a correct Higgs tag.  Such
configurations amounts to 55\% of the signal after the top-tag stage. 
However, this problem is alleviated by the request of two $b$-tags within the
Higgs candidate fat jet: if one of the Higgs $b$-quarks fall within the top
tagged jet, then the Higgs jet will be unlikely to contain two $b$-quarks,
and such events will be strongly suppressed in the final selection.

\subsection{Quality of Higgs reconstruction} \label{sec:higgsreco}

\begin{table} 
  \begin{tabular}{c|c|c|c|c|c}
    label & bin & before $b$-tags & after $b$-tags & after $m_\mathrm{c}$ cut & tag efficiency\\
  \hline
$B_1$ & 110021 & 0.05 & 0.08 & 0.17 & 0.77 \\
$B_2$ & 110023 & 0.10 & 0.16 & 0.24 & 0.53 \\
$B_3$ & 110123 & 0.09 & 0.40 & 0.38 & 0.32 \\
$B_4$ & 111023 & 0.01 & 0.03 & 0.03 & 0.31 \\ [1ex]
  \end{tabular}
  \caption{
The fraction of the signal cross section at different steps of the analysis
in four of the 144 bins in the 6-dimensional Higgs-jet category histogram. The tag efficiency of the topology is 
reported in the last column, and the bins are ordered by decreasing tag efficiency.  Each row
corresponds to a bin identified by specifying the conditions that are true
and false (or a numerical value if applicable) in the order listed in the
text. The left-most digit corresponds to the first condition.
}
  \label{tab:topolH}
\end{table}

\begin{figure} 
  \centering
  \begin{subfigure}{\stew\textwidth}
    \centering
    \includegraphics[width=\stgw\linewidth]{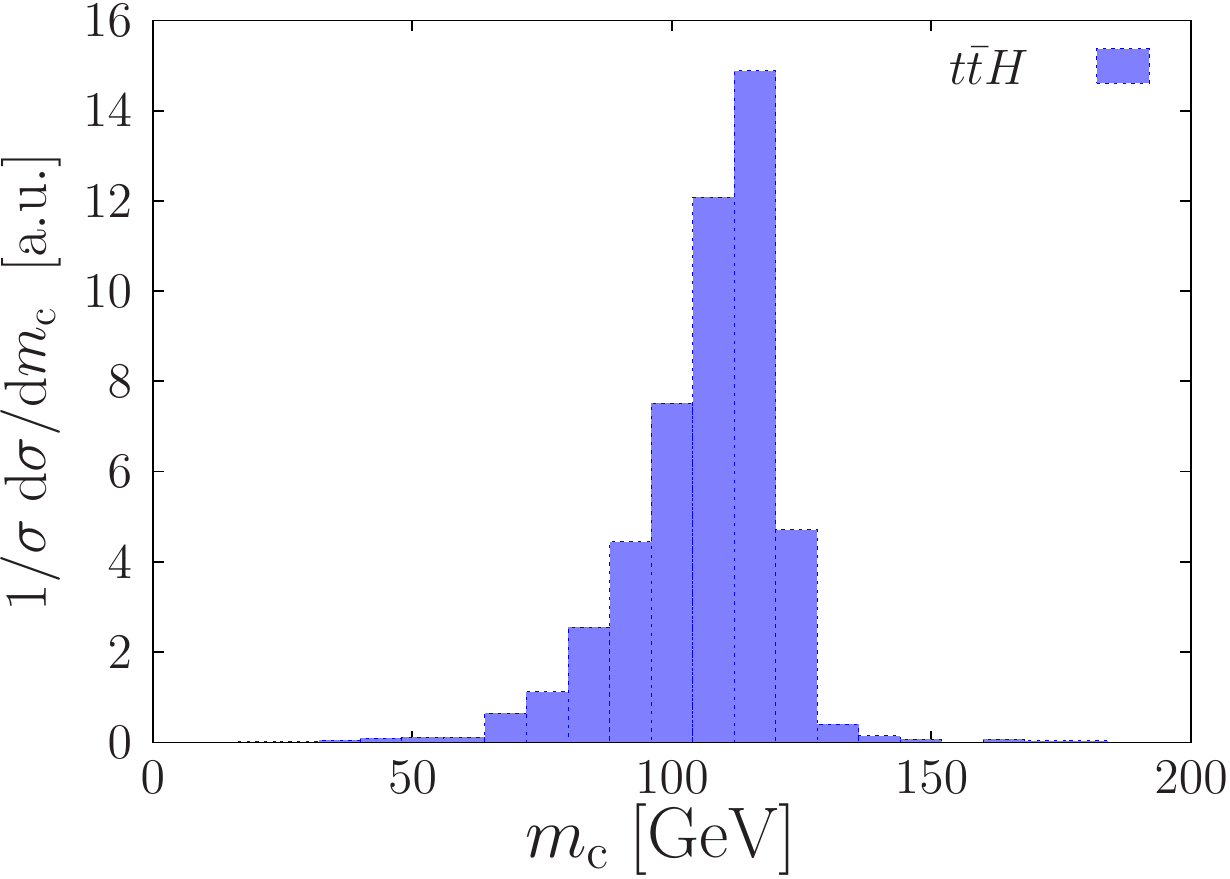}
    \caption{$B_1$\,(110021)}
    \label{fig:topolH2b1pr}
  \end{subfigure}%
  \begin{subfigure}{\stew\textwidth}
    \centering
    \includegraphics[width=\stgw\linewidth]{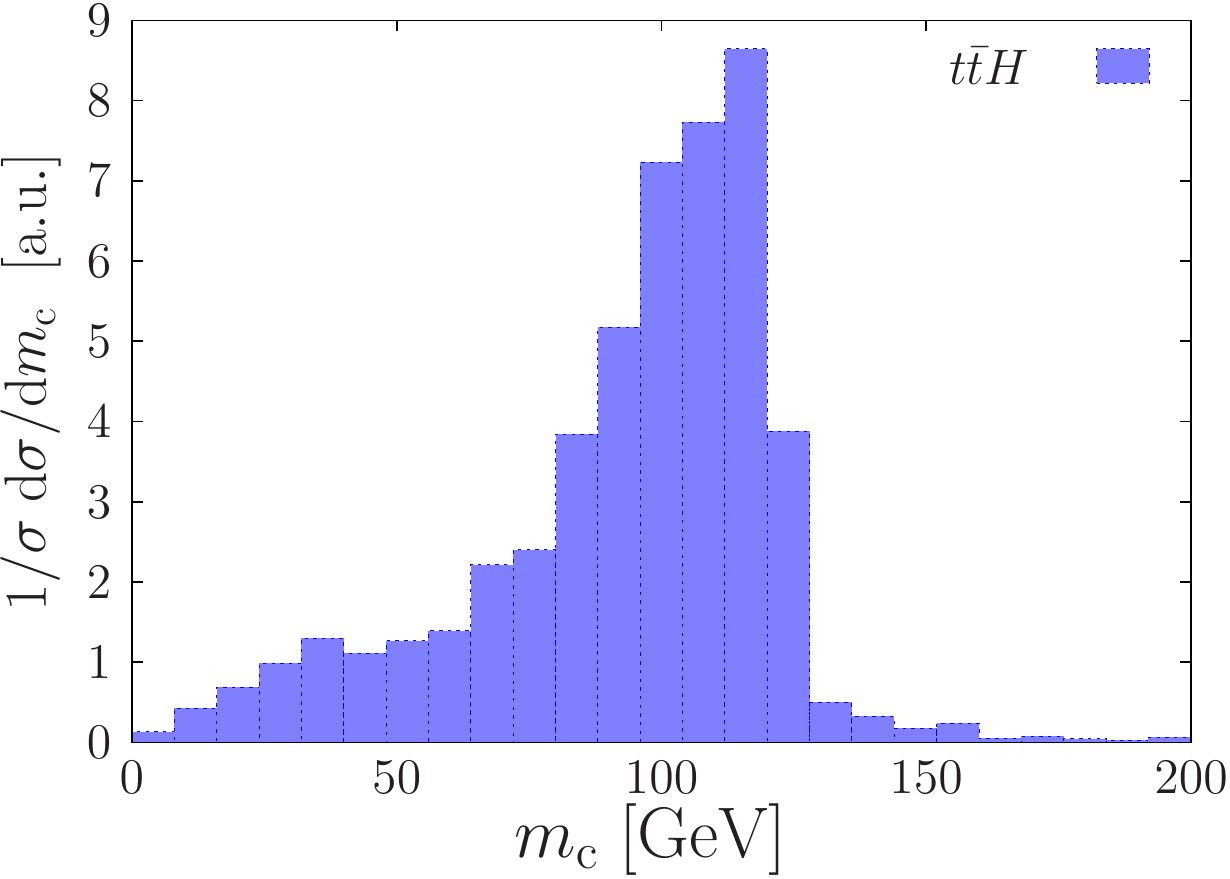}
    \caption{$B_2$\,(110023)}
    \label{fig:topolH2b3pr}
  \end{subfigure}\\[4ex]
    \begin{subfigure}{\stew\textwidth}
    \centering
    \includegraphics[width=\stgw\linewidth]{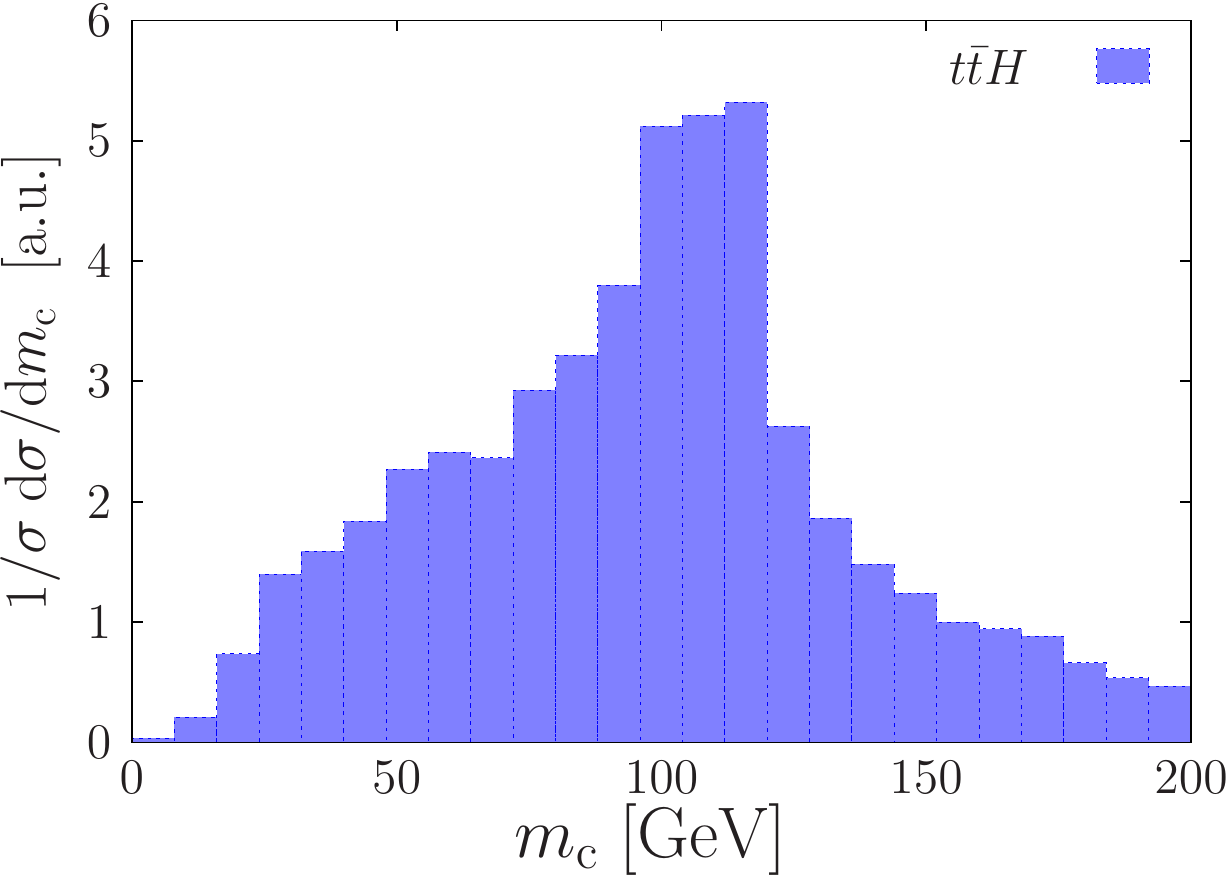}
    \caption{$B_3$\,(110123)}
    \label{fig:topolH3b3pr}
  \end{subfigure}%
  \begin{subfigure}{\stew\textwidth}
    \centering
    \includegraphics[width=\stgw\linewidth]{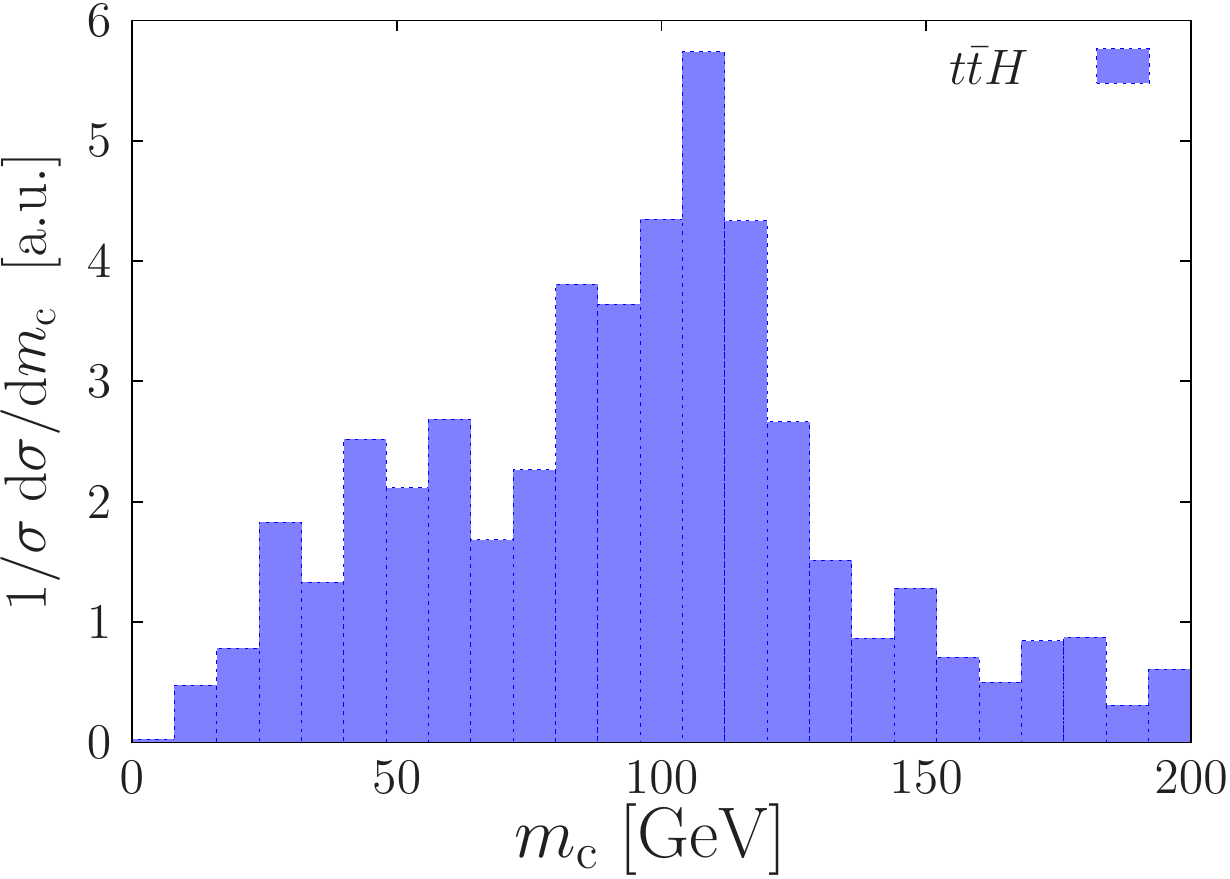}
    \caption{$B_4$\,(111023)}
  \end{subfigure}\\[1ex]
  \caption{Distributions of the Higgs candidate mass, $m_\mathrm{c}$, for different Higgs-jet topologies after requesting three $b$-tags,
i.e.~after 
step~\ref{enum:trippletag}
of the boosted analysis. The figures correspond to the topologies shown in  Table~\ref{tab:topolH}.    }
  \label{fig:topolH}
\end{figure}

To assess the the goodness of fat jets as Higgs candidates we employ categories
based on the following criteria:
 \begin{enumerate}
\item $H$: the Higgs boson is boosted ($p_\mathrm{T,H} > 150$\,GeV) 
\item $H$: the Higgs boson overlaps with the jet ($\Delta R_\mathrm{jet,H}<R_\mathrm{fat}$)
\item $t_\mathrm{had}\to b jj$: the $b$ quark from $t_\mathrm{had}$ belongs to the jet
\item $t_\mathrm{lep}\to b \ell\nu$: the $b$ quark from $t_\mathrm{lep}$ belongs to the jet
\item $H\to b \bar b$: the number of $b$-quarks from the Higgs decay the jet contains is 0/1/2
\item $H\to b \bar b$: the number of $b\bar b$  Higgs candidates in the fat jet  is 0/1/3 
\end{enumerate}
Note that fat jets containing at most one, two, or three $b$-quarks,
can yield zero, one or three $H(b\bar b)$ candidates, respectively.
Conditions number 5 and 6 have three possible outcomes.
This makes a total of
144 categories,  but again only few of them yield significant contributions to 
the accepted cross section.
The relative weight of the four most important topologies
is reported in Table~\ref{tab:topolH} at three levels of the analysis of Sec.~\ref{sec:mspana}: 
before and after $b$-tagging 
(before step~\ref{enum:doublebtag} and after~\ref{enum:trippletag}),
and after the $m_\mathrm{c}$ mass cut (step~\ref{enum:lastbasic}).
These four leading topologies, ordered according to their relative weight after the $m_\mathrm{c}$ cut,
account for 80\% of the $t\bar t H$ signal within the $m_\mathrm{c}$ mass
window. The corresponding distributions in the invariant mass 
of the Higgs candidate are displayed in Figure~\ref{fig:topolH}.  We see that, whenever the Higgs fat
jet contains both Higgs $b$-quarks and no other partons ($B_1$ topology), the $m_\mathrm{c}$ distribution
features a clear peak.  Of course the missing neutrinos from the $B$-meson
decays skew and shift the peak.  If, however, QCD radiation 
produces a third hard structure ($B_2$ topology), the peak
is smeared out due to additional continuum contributions from false Higgs
candidates.  This continuum contribution is greatly diminished by the double
$b$-tag requirement for the Higgs candidate's subjets.

The $B_3$ topology, where the $H\to b\bar b$ products in the fat jet are
contaminated by a third $b$-quark from the leptonic top, is the main
contributor after step~\ref{enum:trippletag} of
the analysis. Since all subjets used to reconstruct the Higgs candidates are b-jets, the continuum distribution from a
false Higgs candidate is comparable to the true peak-shaped distribution in the mass
range of interest.  Moreover, the continuum distribution has a similar shape
as the irreducible $t\bar{t}b\bar{b}$ background. Therefore, as discussed
in the next section, in order to trim this addition to the background from
falsely tagged signal and to sharpen the peak structure in presence of
three-candidate fat jets, we will optimise the Higgs tag by attempting a
reconstruction of the leptonic top.

\section{Improvements and new avenues} \label{sec:analysis}

In this section we propose new selection strategies targeted at a better
reconstruction of topologies that are the major contributors to
misidentified Higgs- or top-candidate fat jets.
In Sec.~\ref{sec:boostAna} we present 
improvements of the standard boosted analysis of Sec.~\ref{sec:mspana}
as well as new boosted analyses that exploit phase space regions with a single fat jet. 
Such boosted selection strategies will be compared to 
a more inclusive multi-variate analysis presented in Sec.~\ref{sec:unboost}.

\subsection{Boosted final state configurations} \label{sec:boostAna}

\begin{figure}
	\centering
	\includegraphics[width=.7\linewidth]{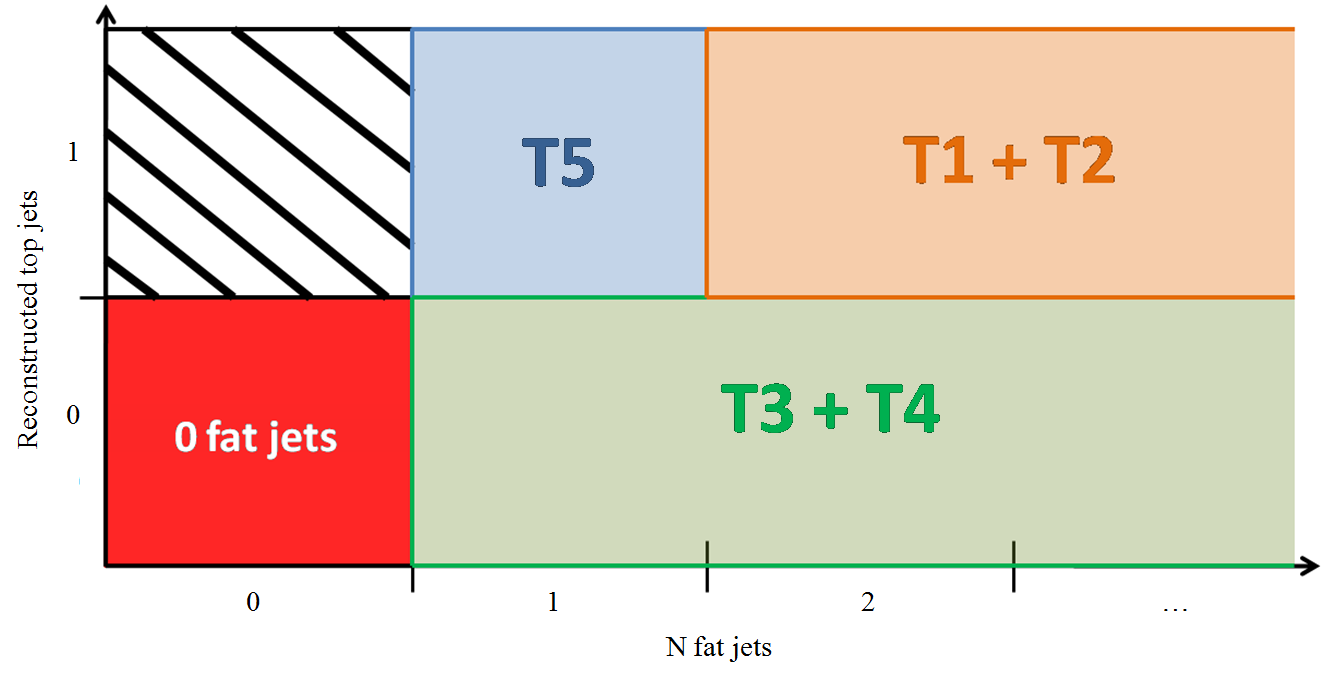}
	\caption{The single-isolated-lepton event phase space with the explored regions labelled as in the text.}
	\label{fig:PhSp}
\end{figure}

The standard boosted analysis of Sec.~\ref{sec:mspana} is targeted at Higgs candidates with 
topology $B_1$, which provides optimal Higgs reconstruction and low mistag rates (see Tab.~\ref{tab:topolH} and
Fig.~\ref{fig:topolH2b1pr}).
The intrinsic difficulty of any reconstruction approach is to maximise the selection efficiency 
for this particular topology and to optimise Higgs reconstruction in fat jet topologies
that feature a less trivial substructure.
To this end, it is useful to perform independent analyses
depending on the number of possible Higgs candidates inside Higgs fat jets.
If the fat jet contains only two hard
substructure objects it can only form one Higgs candidate.
In this case, even though
there are ways to fake
the Higgs candidate (by forming a fat jet from one Higgs decay product and
the leptonic top $b$-quark for example), our results indicate that 
the final selection
is dominated by the desired topology, 
i.e.~by a correctly tagged Higgs boson.
In the following, we will propose improved selection strategies for 
the more challenging configurations with additional hard substructure objects.
Furthermore, in order to increase the statistical
sensitivity to the $t\bar{t}H$ signal, we will 
complement the standard boosted analysis with extra channels
without tagged hadronic tops
or without boosted Higgs candidates.
Thus we will
slice the phase space of 
single-lepton $t\bar t H$ events into the categories illustrated in Fig.~\ref{fig:PhSp}:
\begin{enumerate}
\item[{\bf T1:}] {$\geq 2$ fat jets, 1 tagged boosted top, 1 Higgs candidate}
\item[{\bf T2:}] {$\geq 2$ fat jets, 1 tagged boosted top, 3 Higgs candidates}
\item[{\bf T3:}] {$\geq 1$ fat jets, no tagged boosted tops, 1 Higgs candidate}
\item[{\bf T4:}] {$\geq 1$ fat jets, no tagged boosted tops, 3 Higgs candidates}
\item[{\bf T5:}] {exactly 1 fat jet, 1 tagged boosted top},
unboosted Higgs candidate
\end{enumerate}

Configurations {\bf T1} and {\bf T2}, which will be handled separately here,
cover the entire phase space of the standard boosted analysis of Sec.~\ref{sec:mspana}.
In categories {\bf T3} and {\bf T4} we look for 
a boosted Higgs and an unboosted hadronic top,
and in {\bf T5} we
anticipate  an unboosted Higgs after reconstructing the boosted top.
All in all we examine 
five
statistically independent 
phase space regions
that, when combined,
can enhance the sensitivity to $t\bar{t}H$ events.
Note that here we will not study events without fat jets.

\subsubsection{Topologies {\bf T1} and {\bf T2}: Boosted $\mathbf{t_{had}}$ and boosted $\mathbf{H}$} \label{sec:boostTHana}

In the following we describe dedicated selections for 
event categories with one ({\bf T1})  and more ({\bf T2}) Higgs candidates
after step~\ref{enum:newana} of the standard boosted analysis of Sec.~\ref{sec:mspana}.

As the Higgs mass peak in 
the {\bf T1} channel is already fairly narrow,
one way to
further separate signal from $t\bar{t}+X$ backgrounds is to exploit the
colour singlet nature of the Higgs boson.  The colour dipole, formed by the
$b\bar{b}$ pair, disfavours radiation away from the Higgs decay products,  
while $b\bar{b}$ pairs originating from the QCD background feature 
a different radiation pattern.  In order to take advantage of this distinctive 
signal feature we use the 
ellipticity jet-shape variable $\hat{t}$~\cite{Wshower}
computed in terms of the Higgs candidate's constituents.
\begin{figure} 
  \centering
  \begin{subfigure}{\stew\textwidth}
    \centering
    \includegraphics[width=\stgw\linewidth]{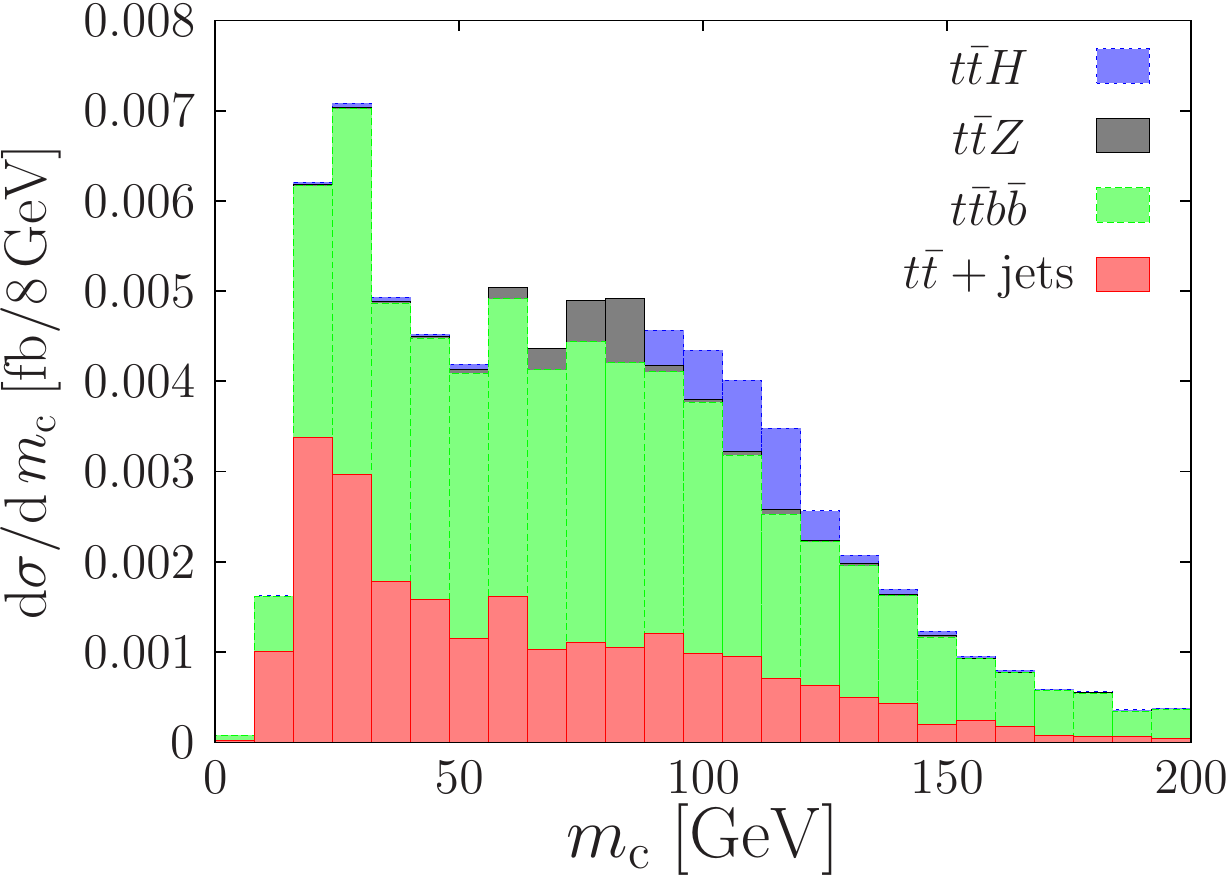}
  \end{subfigure}%
  \begin{subfigure}{\stew\textwidth}
    \centering
    \includegraphics[width=\stgw\linewidth]{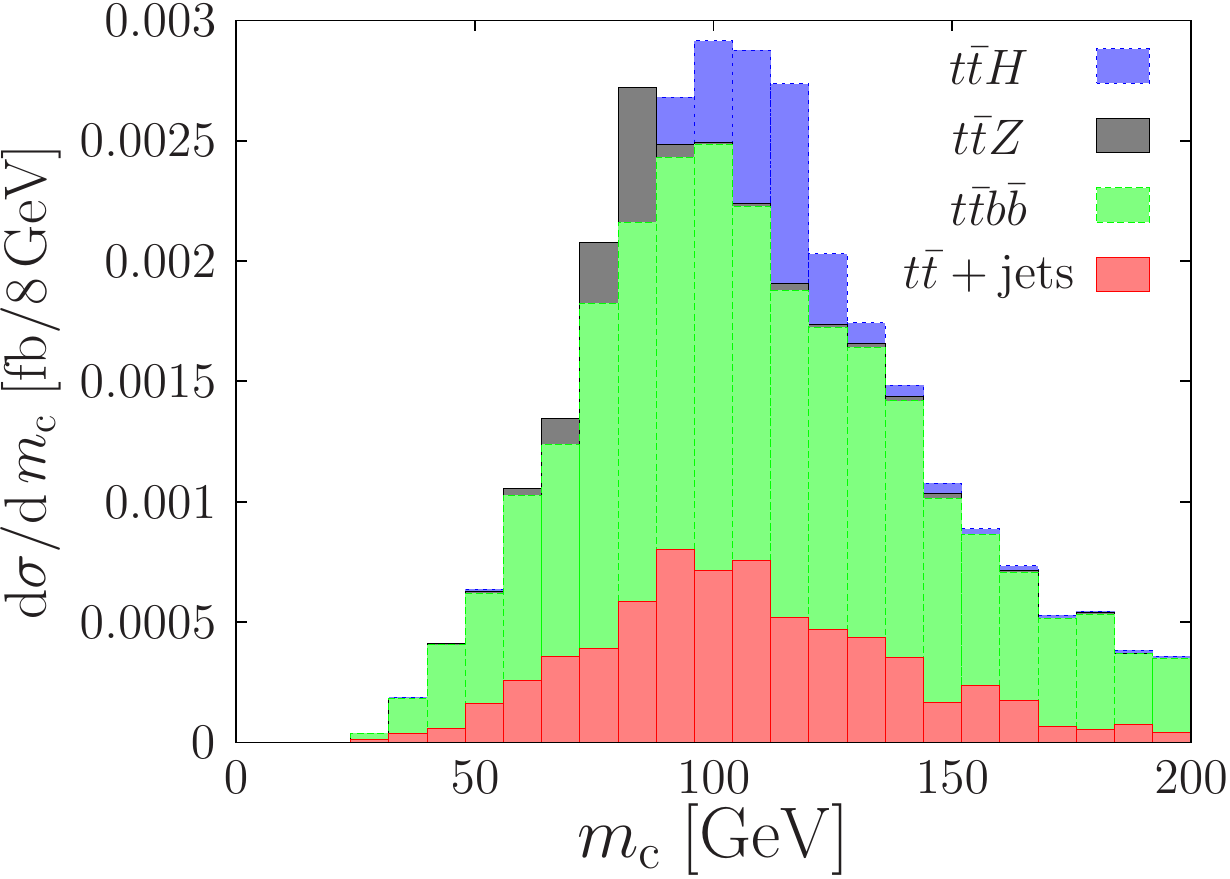}
  \end{subfigure}\\[1ex]
  \caption{
        $m_\mathrm{c}$ distribution from the selection channel with a single Higgs candidate in the fat jet and a tagged boosted hadronic top ({\bf T1}).
        The left(right) figure is without(with) a $\hat{t}$ cut on the Higgs candidate constituents.
          }
  \label{fig:2pr}
\end{figure}
Figure~\ref{fig:2pr} shows the different mass distributions of the Higgs
candidate in the {\bf T1} selection channel with and without a cut
$\hat{t}<0.2$.
As discussed in Sec.~\ref{sec:results},
the ellipticity cut allows one to achieve an appreciable improvement in $S/B$
with minor losses in terms of signal yield.  However, given its fairly small cross section,
the {\bf T1} channel alone does not provide substantial sensitivity to $t\bar tH$ at Run 2.

The complementary category {\bf T2}
has a four times higher rate. Thus, increasing $S/B$ in this channel, 
which is dominated  by the Higgs-jet topology B3 in Tab.~\ref{tab:topolH},
can boost the sensitivity of $t\bar{t}H(b\bar{b})$.
To this end, after step~\ref{enum:newana}
of the standard boosted analysis, we try to tag the leptonic top,
$t_\mathrm{lep}$, as well as the Higgs if the fat jet has more than one Higgs
candidate.  The reconstruction of $t_\mathrm{lep}$ helps to
uniquely identify the origin of the three $b$-tagged jets and ameliorates
the combinatorial smearing of the Higgs peak evident in Fig.~\ref{fig:topolH3b3pr}. 
The reconstruction is implemented by minimising a $\chi^2$ score computed for 
all combinations of final state objects
that can form a Higgs--$t_\mathrm{lep}$ pair.  
For each of the three Higgs
candidates there are a number of possible combinations, and the relevant physical objects are:
\begin{enumerate}
\item two subjets 
reconstructed from the hadrons of the filtered Higgs candidate using the exclusive-$k_T$ algorithm.
\item 
the inner and outer jets with respect to the current Higgs candidate (see definition
in Sec.\ref{sec:mspana});

\item the isolated lepton; 
\item the missing transverse momentum of the event $\mathbf{\slashed{E}_T}$.  
\end{enumerate}

The neutrino momentum can be reconstructed from the
lepton momentum and $\mathbf{\slashed{E}_T}$  imposing the
on-shell condition for the corresponding $W$ boson.
The ambiguity related to the two solutions of the quadratic equation 
is not resolved at this point, i.e.~both possibilities are 
taken into account in the following steps.
Since a leptonic top
consists of a $b$-quark, a charged lepton and a  neutrino, we call a
H$t_\mathrm{lep}$ configuration any unique choice of one out of $n$ inner
and outer jets, one of the two neutrino candidates, the isolated lepton and
the two exclusive Higgs candidate subjets.
Therefore, any 3-Higgs-candidate
fat jet has 
a number $2\sum_{i=1}^{3}n_i$  of $Ht_\mathrm{lep}$ configurations.
We define $\chi^2$ for a configuration in the
following way:

\begin{eqnarray} 
\label{eq:X2}
\chi^2 &=& \chi^2_\mathrm{top} + \chi^2_\mathrm{Higgs}, \nonumber \\
\chi^2_\mathrm{top} &=& \frac{(m_\mathrm{t_{lep},reco} - m_\mathrm{t_{had},max})^2}{\sigma^2_\mathrm{t_{had}}}, \nonumber \\
\chi^2_\mathrm{Higgs} &=&\frac{(m_\mathrm{H,reco} - m_\mathrm{H,max})^2}{\sigma^2_\mathrm{H+}}\Theta(m_\mathrm{H,reco} - m_\mathrm{H,max}) + \frac{(m_\mathrm{H,reco} - m_\mathrm{H,max})^2}{\sigma^2_\mathrm{H-}}\Theta(m_\mathrm{H,max} - m_\mathrm{H,reco}),
\end{eqnarray}
where $\Theta$
is the Heaviside step function. The errors
$\sigma_\mathrm{H\pm}$ are the standard deviations of Gaussian fits to the
data to the right (+) and left (-) of the peak in {\bf T1}
(Fig.~\ref{fig:topolH2b1pr}).  We make this choice because the reconstructed
Higgs mass distribution is heavily skewed to lower values, thus a single
Gaussian fit will overestimate one and underestimate the other deviation. 
We take $m_\mathrm{H,max}$ as the position of the peak.  The
$t_\mathrm{had}$ mass distribution from the topology 
$A_1$ in
Table~\ref{tab:topolT} is much more symmetric.  Therefore, a single Gaussian
fit suffices to extract $\sigma_\mathrm{t_{had}}$ and
$m_\mathrm{t_{had},max}$.
We order all configurations in ascending $\chi^2$ and choose the first
quarter of unique configurations.  Then we keep the configurations with two
successful $b$-tags in the Higgs candidate and another for the inner or
outer jet (there is only one such object per configuration).  We record the
Higgs candidate mass $m_\mathrm{c}$ of each of the configurations remaining
after the $\chi^2$ and $b$-tag cuts.

\begin{figure} 
\centering \begin{subfigure}{\stew\textwidth}
    \centering
    \includegraphics[width=\stgw\linewidth]{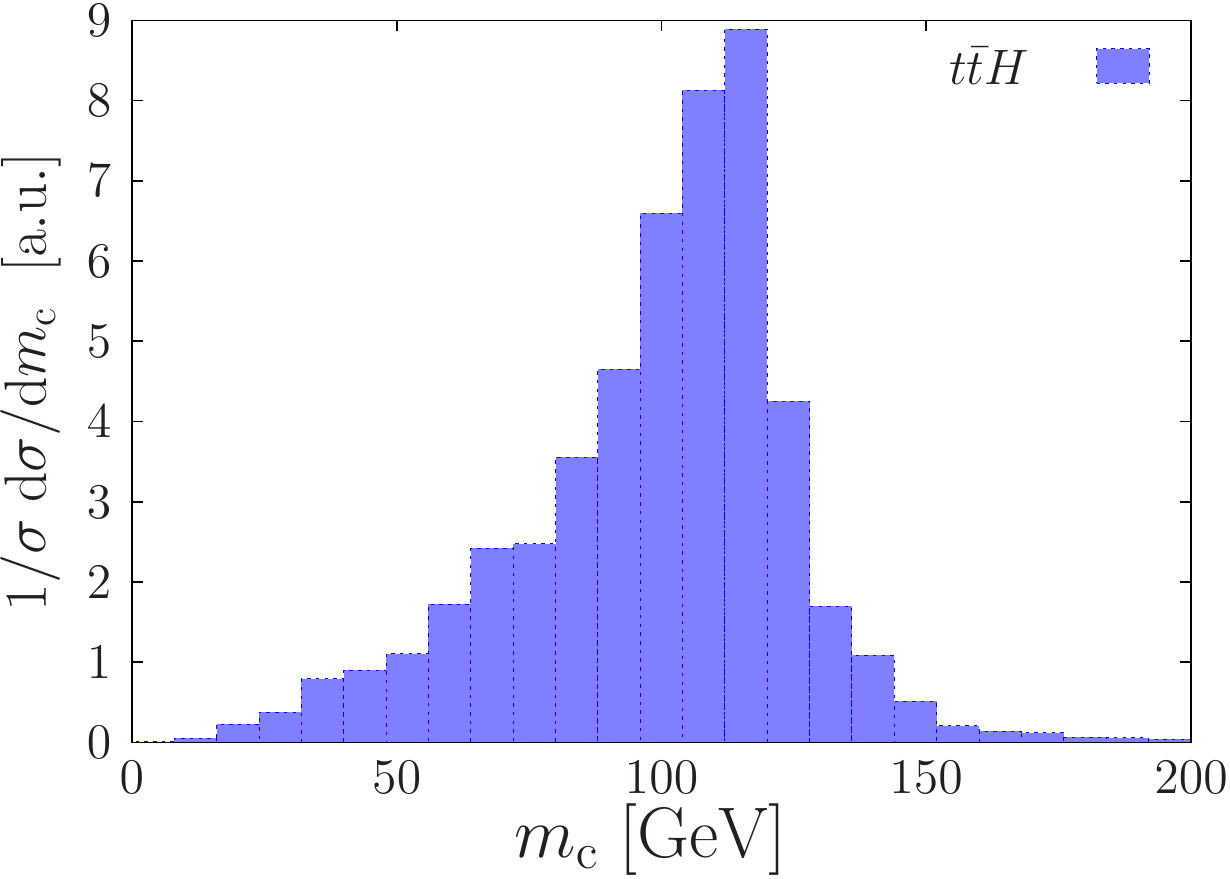}
    \label{fig:new3b3pr}
  \end{subfigure}%
  \begin{subfigure}{\stew\textwidth}
    \centering
    \includegraphics[width=\stgw\linewidth]{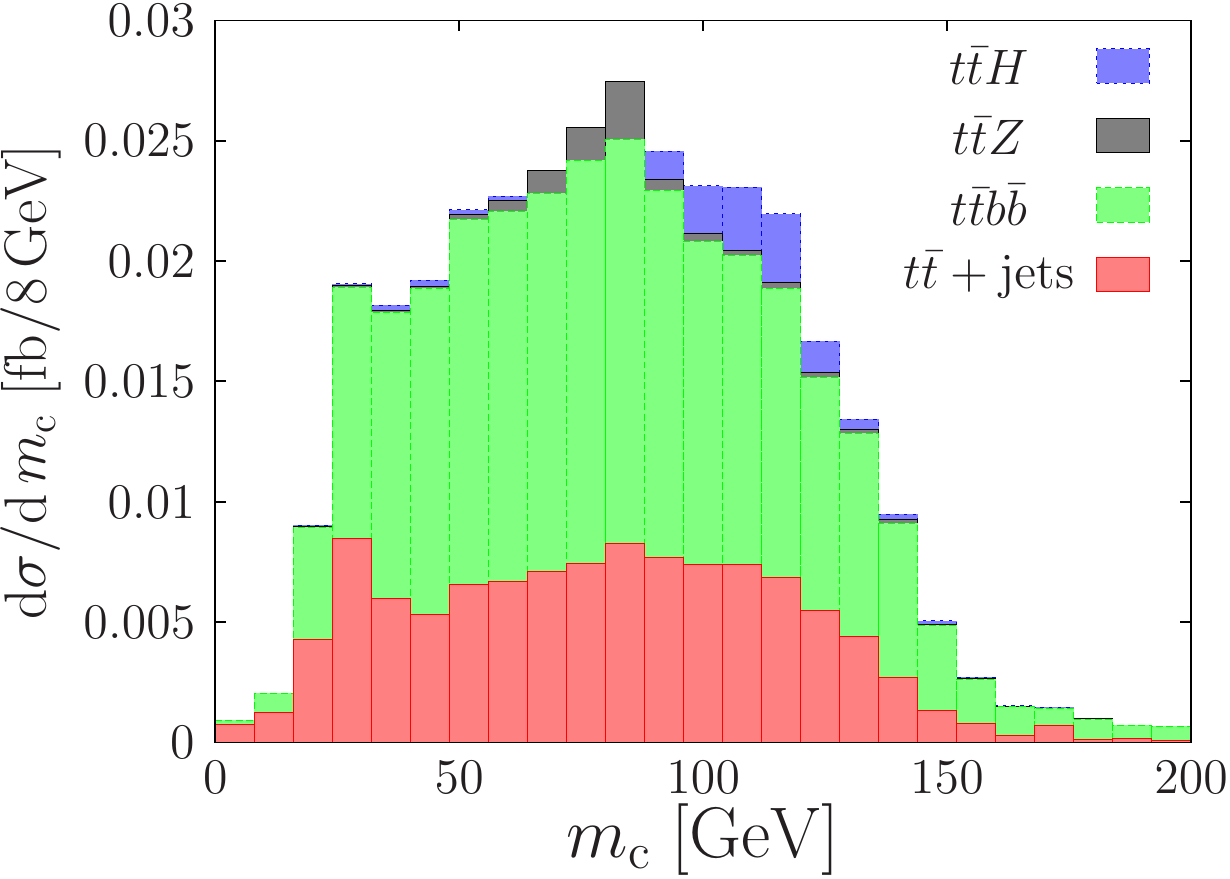}
  \end{subfigure}\\[1ex]
  \caption{
        $m_\mathrm{c}$ distribution obtained from the 25\% of configurations with lowest $\chi^2$ score in the 3-Higgs-candidate selection channel ({\bf T2}).  The left figure is the signal $t\bar{t}H$ and the
        figure to the right contains signal and background.
          } 
\label{fig:2boost} 
\end{figure}

The $\chi^2$ ordering complements the $b$-tagging in the following way.
Assume the Higgs fat jet contains both the $b$-quark of $t_\mathrm{lep}$ and
the two Higgs decay products.  Without ranking the different configurations,
all three Higgs candidates will certainly contribute to the $m_\mathrm{c}$
distribution.  We ameliorate combinatorial issues by removing configurations
with large $\chi^2$ scores, i.e.~we only keep the $25\%$ of combinations
with lowest $\chi^2$ score.  By requiring 
three
 $b$-tags, we remove the
configurations that contain non-b-tagged inner/outer jets.  Consequently, we
veto the correct configuration less often than the one where the
$b$-quark from the leptonic top
fakes one of the Higgs decays, before $b$-tagging
of the Higgs candidate is even implemented.  Figure~\ref{fig:2boost} shows
the change of $m_\mathrm{c}$ in the {\bf T2} topology (compare with
Figure~\ref{fig:topolH3b3pr}).

There are additional steps that one can study to remove fake configurations
before applying the $b$-tags.  We studied the impact of including the jet
shape observable $\hat{t}$ and the helicity angle \cite{Kaplan:2008ie}
between the bottom quark and the charged lepton.  However, we found them not
to be efficient in increasing $S/B$ at this stage of the analysis.

\begin{figure} 
  \centering
  \begin{subfigure}{\stew\textwidth}
    \centering
    \includegraphics[width=\stgw\linewidth]{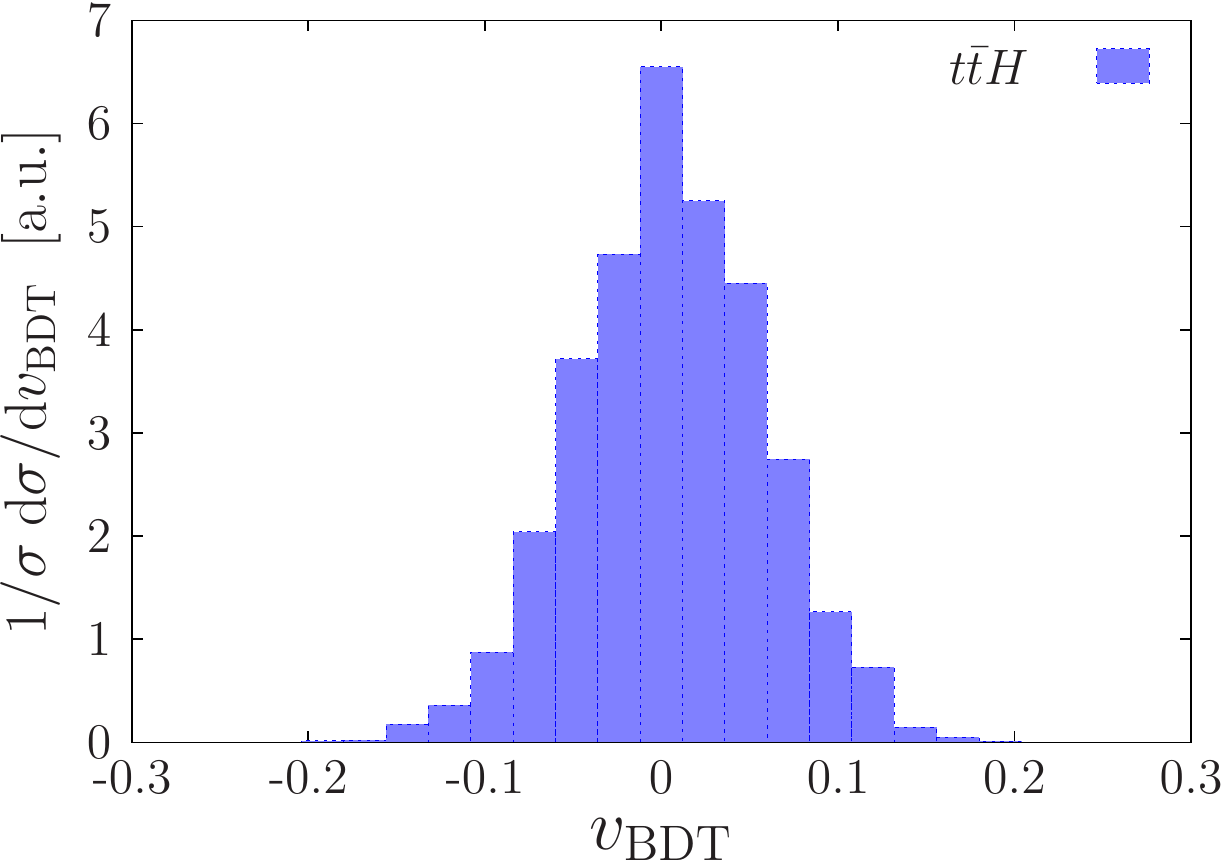}
  \end{subfigure}%
  \begin{subfigure}{\stew\textwidth}
    \centering
    \includegraphics[width=\stgw\linewidth]{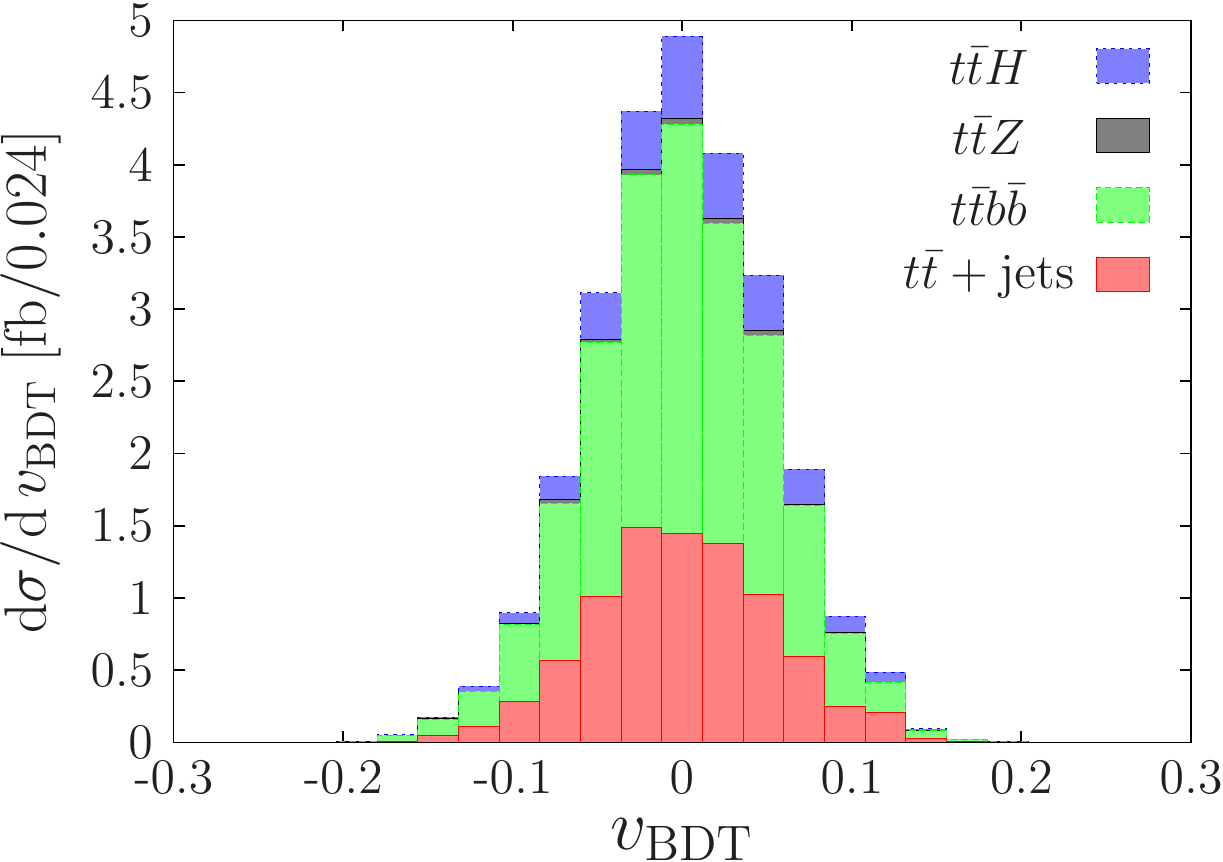}
  \end{subfigure}\\[1ex]
  \caption{
        Boosted Decision Trees score distribution from 5 variables calculated with the reconstructed $t\bar{t}H$ objects after the mass cut in {\bf T2}.
            The left figure is the signal $t\bar{t}H$ and the figure to the right contains signal and background.
              }
  \label{fig:BDTboost}
\end{figure}

At this point we have reconstructed all three resonances in $t\bar{t}H$.
Therefore, it is conceivable to use them in a 
multi-variate analysis (MVA) 
to exploit any angular
dependencies between these fundamental objects.  In particular we select the
mass, $p_\mathrm{T}$ and rapidity of the $t\bar{t}H$ system as well as the
angles between the Higgs and the top/anti-top in the $t\bar{t}H$ centre of
mass frame to build a 
boosted decision tree (BDT) discriminant.
For the numerical evaluation we use the TMVA
\citep{TMVA} package of ROOT \citep{ROOT}.  We build a forest of 850 trees
each with 
three
 layers and require at least 5\% of signal in each leaf node to
explicitly avoid any overtraining.  There is a great freedom of choice for
the number and nature of variables in a MVA at this stage, depending on the
remaining statistics and the systematic uncertainties of the input
variables.  Particularly the latter requires a full detector simulation for
a reliable estimate.  The distribution of the BDT score is displayed in
Figure~\ref{fig:BDTboost}.  Despite the previously described attempts at
improving the $S/B$ of {\bf T2}, 
the results presented in Sec.~\ref{sec:results} indicate
that the obtained improvement in $S/B$ is rather modest.

\subsubsection{Topologies
{\bf T3}--{\bf T5}:
boosted $\mathbf{t_{had}}$ or boosted $\mathbf{H}$} \label{sec:1boost}

So far we have only used the selection channels where a fat jet has been
tagged as a top, and there is at least one more fat jet to be tagged as a
Higgs ({\bf T1} and {\bf T2}).  
In the following we will consider two additional 
types of channels.
If there is one or more fat jets in the event but
neither is top-tagged, we will test for a boosted Higgs among them and aim to
reconstruct a top using the radiation outside the Higgs candidate (channel {\bf T3}--{\bf T4}).
Vice versa, if there is only one fat jet and it has been top-tagged, we
will look for a non-boosted Higgs among the remaining particles in the event
(channel {\bf T5}).
In the first case
({\bf T3} and  {\bf T4}) we follow
Sec.~\ref{sec:boostTHana} for the reconstruction of the Higgs candidate. 
For each fat jet we find the mass drop subjets and we group them into Higgs
candidates.  Then we keep fat jets with up to three candidates, but we
separate the 1-candidate ({\bf T3}) from the 3-candidate ({\bf T4}) fat jets.  We again construct
inner and outer jets after removing the Higgs candidates, see
step~\ref{enum:trippletag} of the boosted analysis
in Sec.~\ref{sec:mspana}.  As we would like to
reconstruct the unboosted hadronic top as well, we require at least four
inner or outer jets, accounting for the hadronic decay products of the
leptonically and hadronically decaying top quarks.

There is a three-fold  way to assign the $b$-quark within the hadronic top.
To remove this ambiguity, we first reconstruct the hadronic $W$ by
minimising $\Delta m_\mathrm{W} = \left|m_\mathrm{W_reco} - m_\mathrm{W}
\right|$.  Eventually, a $\chi^2$ value of every $t_\mathrm{had}$ and Higgs
candidate configuration is calculated, i.e.
\begin{eqnarray}
\chi^2 &=& \chi^2_\mathrm{top} + \chi^2_\mathrm{W} + \chi^2_\mathrm{Higgs}, \nonumber \\
\chi^2_\mathrm{top} &=& \frac{(m_\mathrm{t_{had},reco} - m_\mathrm{t_{had},max})^2}{\sigma^2_\mathrm{t_{had}}}, \nonumber \\
\chi^2_\mathrm{W} &=& \frac{(m_\mathrm{W_{had},reco} - m_\mathrm{W_{had},max})^2}{\sigma^2_\mathrm{W_{had}}}.
\end{eqnarray}

\begin{figure} 
  \centering
  \begin{subfigure}{\stew\textwidth}
    \centering
    \includegraphics[width=\stgw\linewidth]{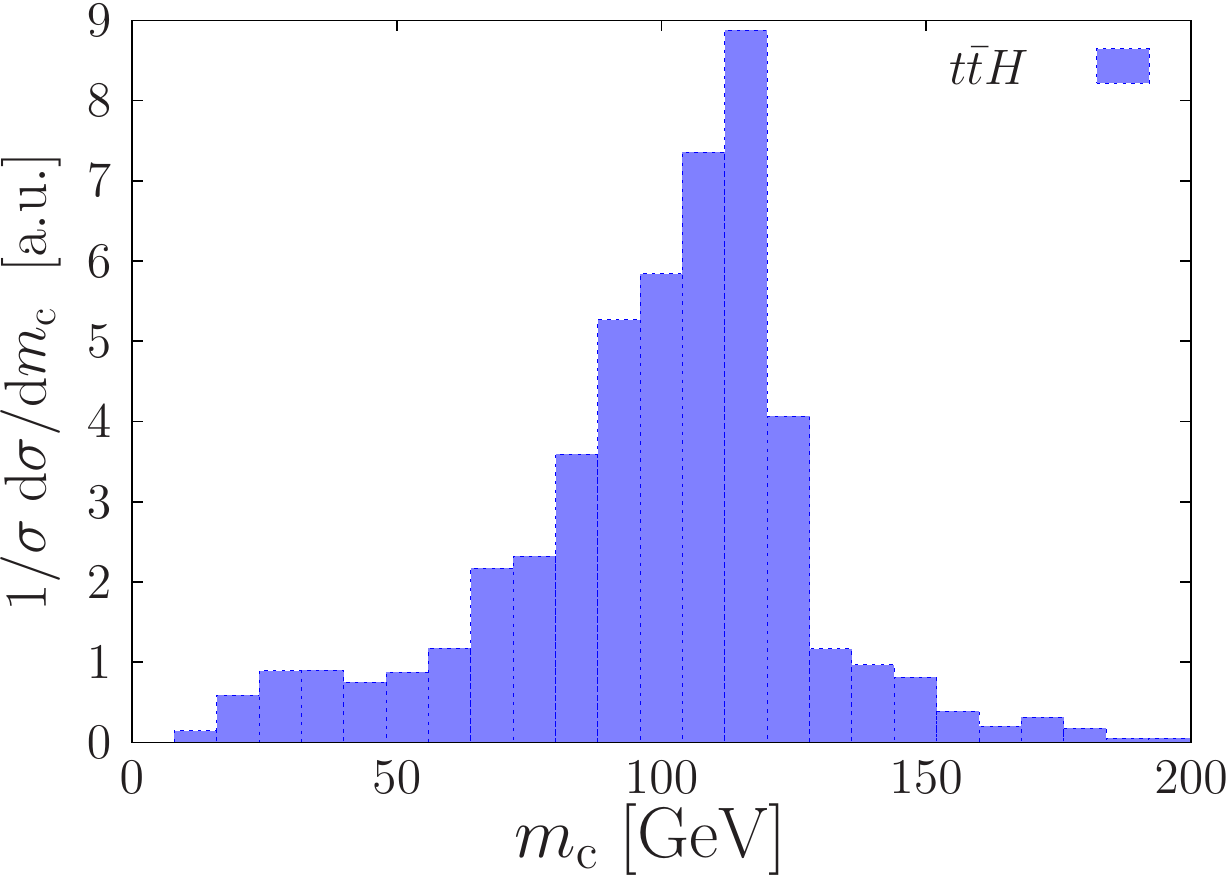}
  \end{subfigure}%
  \begin{subfigure}{\stew\textwidth}
    \centering
    \includegraphics[width=\stgw\linewidth]{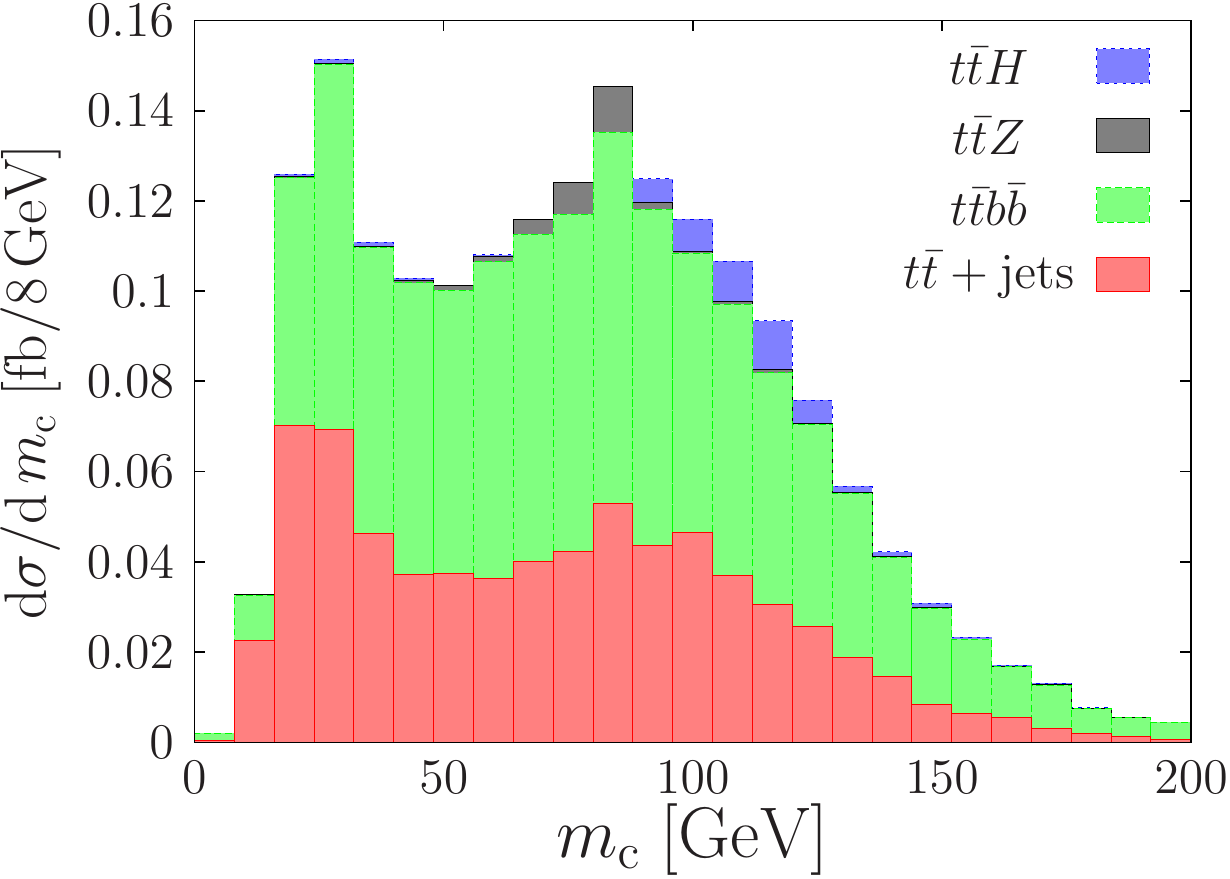}
  \end{subfigure}\\[1ex]
  \begin{subfigure}{\stew\textwidth}
    \centering
    \includegraphics[width=\stgw\linewidth]{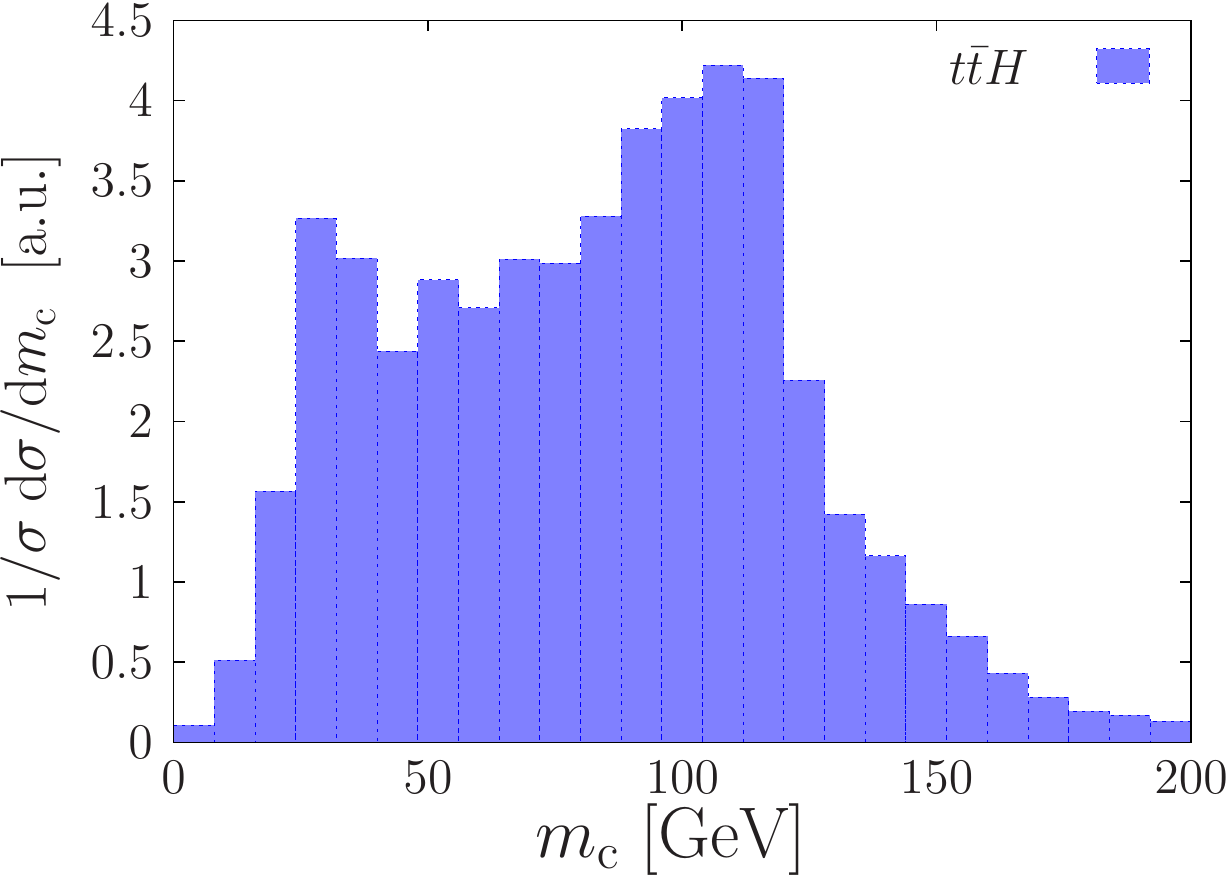}
  \end{subfigure}%
  \begin{subfigure}{\stew\textwidth}
    \centering
    \includegraphics[width=\stgw\linewidth]{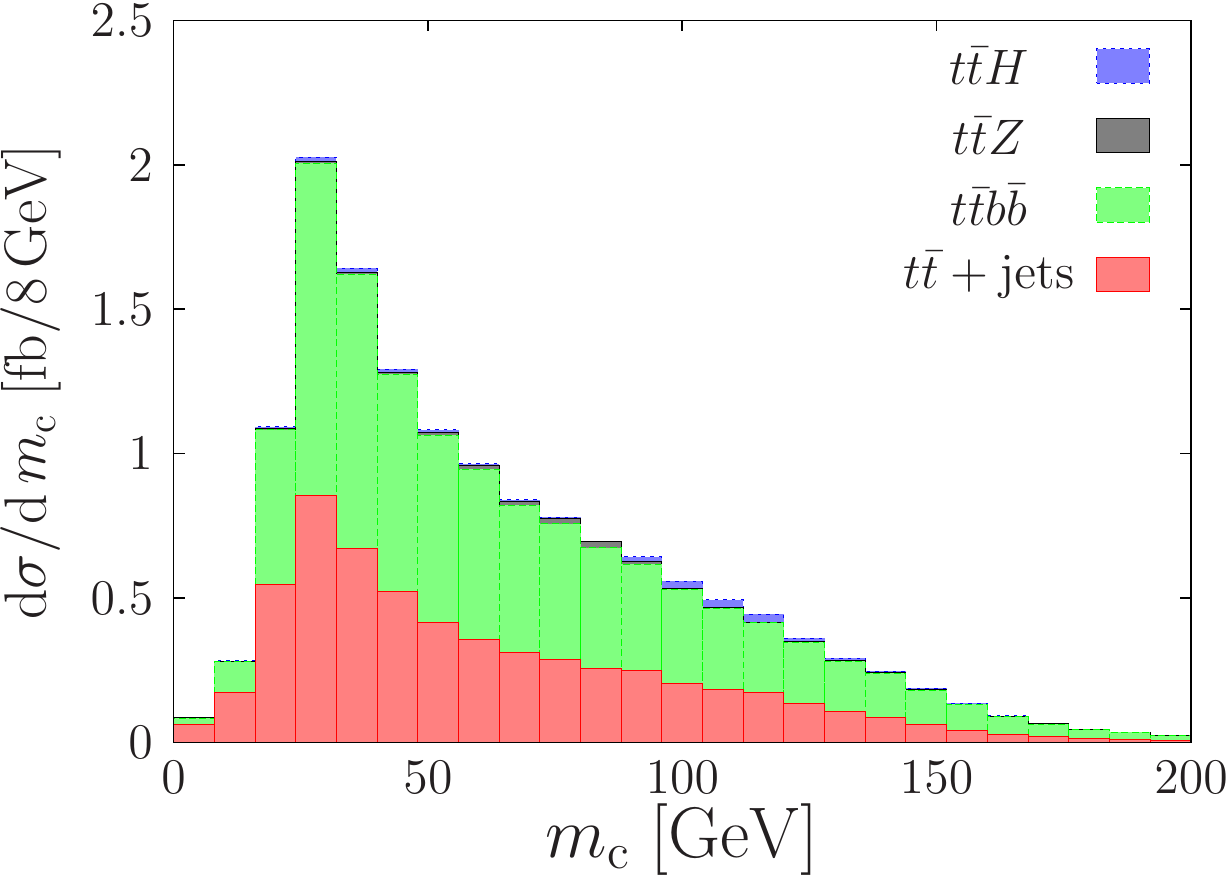}
  \end{subfigure}\\[1ex]
  \caption{
        $m_\mathrm{c}$ distribution obtained from the selection channels without any top tags - {\bf T3} (top) and {\bf T4} (bottom).
        The left figures show the $t\bar{t}H$ signal only and the figures to the right contain signal and background.}
  \label{fig:Hboost}
\end{figure}

The $\chi^2_\mathrm{Higgs}$ and $\chi^2_\mathrm{top}$ are identical to the
ones defined in Sec.~\ref{sec:boostTHana}.  The $\chi^2_\mathrm{W}$
parameters are extracted in the same way as the $\chi^2_\mathrm{top}$
parameters, i.e.~using a Gaussian fit to the mass distribution (right plot
in Fig.~\ref{fig:topolT}) of the two $W$ subjets in the reconstructed
hadronic top, in the case when it falls into the 
cleanest topology (A1)
in Table~\ref{tab:topolT}.  The configurations are ordered by $\chi^2$ and
the highest 75\% are rejected.  From each remaining configuration, three
successful $b$-tags are required---two among the Higgs candidate filtered
subjets and another for the leptonic top.  We do not require an additional
$b$-tag for the hadronic top candidate.  As before, the Higgs candidates'
masses of all surviving configurations are recorded, and the resulting
distributions plotted in Figure~\ref{fig:Hboost}.  
Results in Sec.~\ref{sec:results} confirm that, as expected, topology {\bf T3} features
a cleaner peak and therefore a better $S/B$ ratio than {\bf T4}.
Most importantly,
the signal yields of those two channels, at approximately
1\,fb, are an order of magnitude larger than 
for
{\bf T1} and {\bf T2}.

\begin{figure} 
  \centering
  \begin{subfigure}{\stew\textwidth}
    \centering
    \includegraphics[width=\stgw\linewidth]{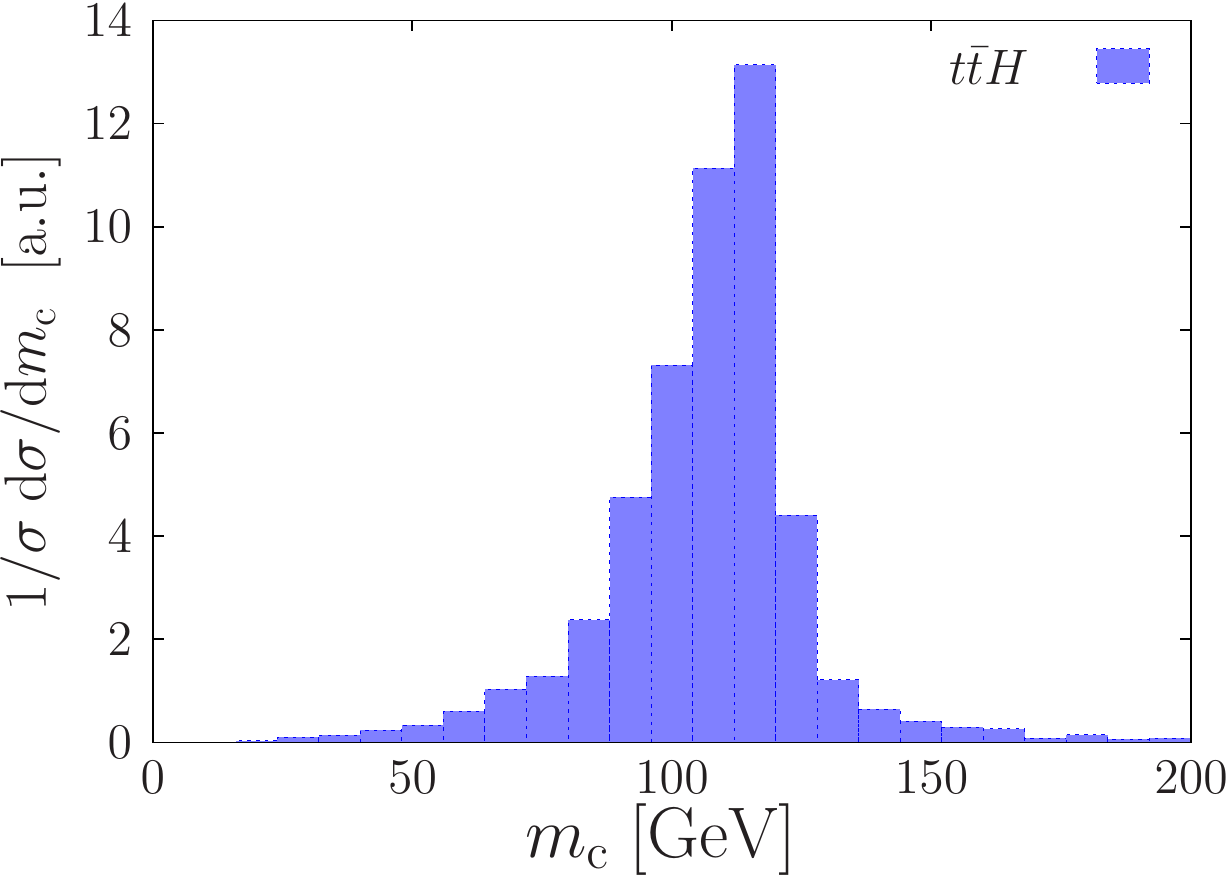}
  \end{subfigure}%
  \begin{subfigure}{\stew\textwidth}
    \centering
    \includegraphics[width=\stgw\linewidth]{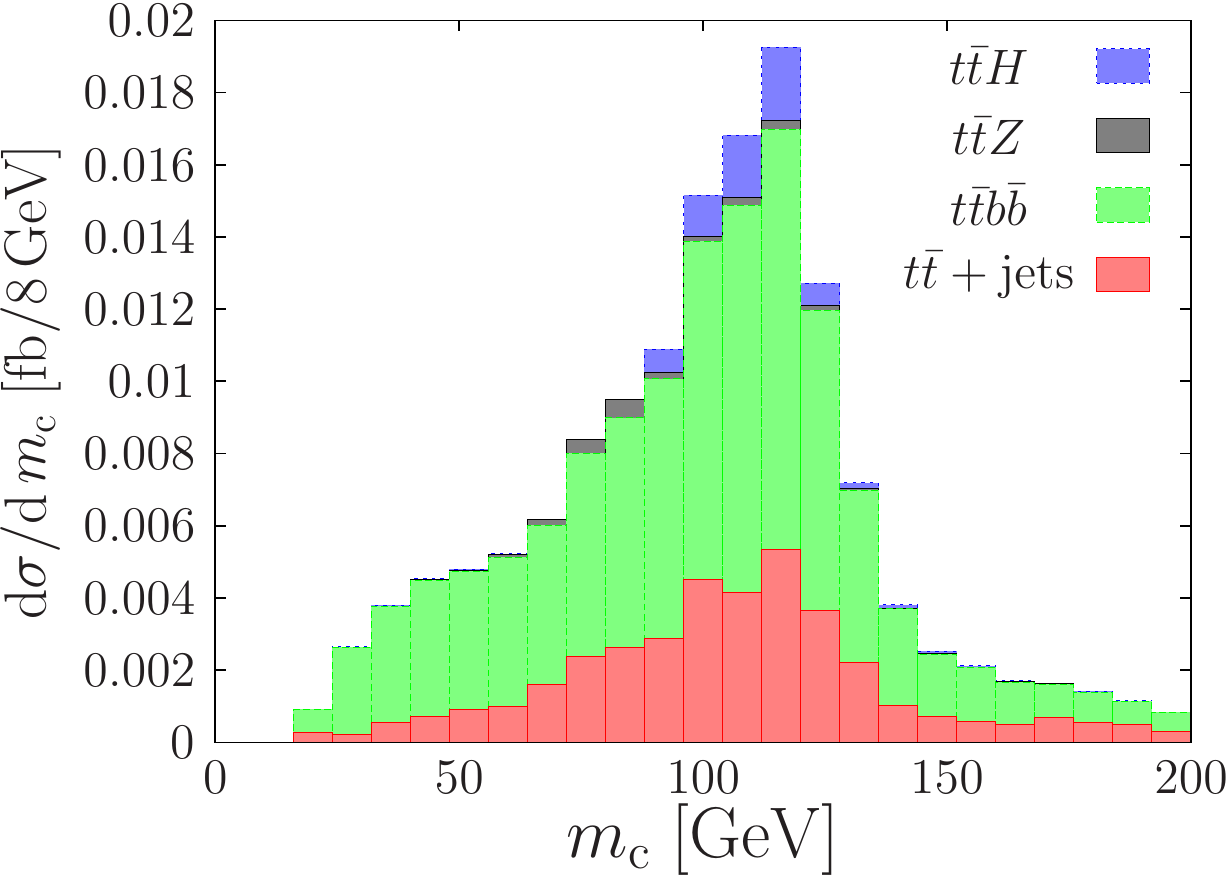}
  \end{subfigure}\\[1ex]
  \caption{
        $m_\mathrm{c}$ distribution obtained from the selection channel with only one fat jet that has been top-tagged ({\bf T5}).
        The left figure is the signal $t\bar{t}H$ and the figure to the right contains signal and background.
          }
  \label{fig:Tboost}
\end{figure}

Finally we focus on the channel with a single fat jet that is top tagged
({\bf T5}).  
In this case we cluster
all final state hadrons into C/A $R=0.4$ jets with
$p_T>30$ GeV and require 3 $b$-tagged jets.  We reconstruct a Higgs and a
leptonic top using the same $\chi^2$ as in the {\bf T2} scenario of
Sec.~\ref{sec:boostTHana}.  The $Ht_{lep}$ configuration consists of two
$b$-tagged jets (Higgs candidate), the other $b$-jet, and a reconstructed
neutrino ($t_{lep}$).  We calculate the $\chi^2$ defined in Eq.~\ref{eq:X2}
and choose the Higgs candidate with best $\chi^2$ value. 
Figure~\ref{fig:Tboost} shows this candidate's mass distribution.  
It turns out that the
{\bf T5} channel has a $S/B$ ratio similar to {\bf T3}, but the signal yield is
smaller by a factor of five and the background shape is more biased
(see Sec~\ref{sec:results}).

\subsection{MVA Without Boost}\label{sec:unboost}

As a generalisation of the boosted configurations discussed in
Secs.~\ref{sec:mspana}-\ref{sec:boostAna}, we perform a MVA with 
seven
observables that do not require any of the resonances to be boosted.  The
analysis may include parts of all phase space regions defined in
Fig.~\ref{fig:PhSp}, however the input objects to the observables are
different.  We start by asking for a single isolated lepton and at least six
C/A R=0.4 jets with $p_\mathrm{T}>30$ GeV, of which exactly four must be
$b$-tagged.  We will only use six jets for the reconstruction, thus from the
remaining jets without a $b$-tag we keep the two with largest
$p_\mathrm{T}$.

\begin{figure} 
  \centering
  \begin{subfigure}{\stew\textwidth}
    \centering
    \includegraphics[width=\stgw\linewidth]{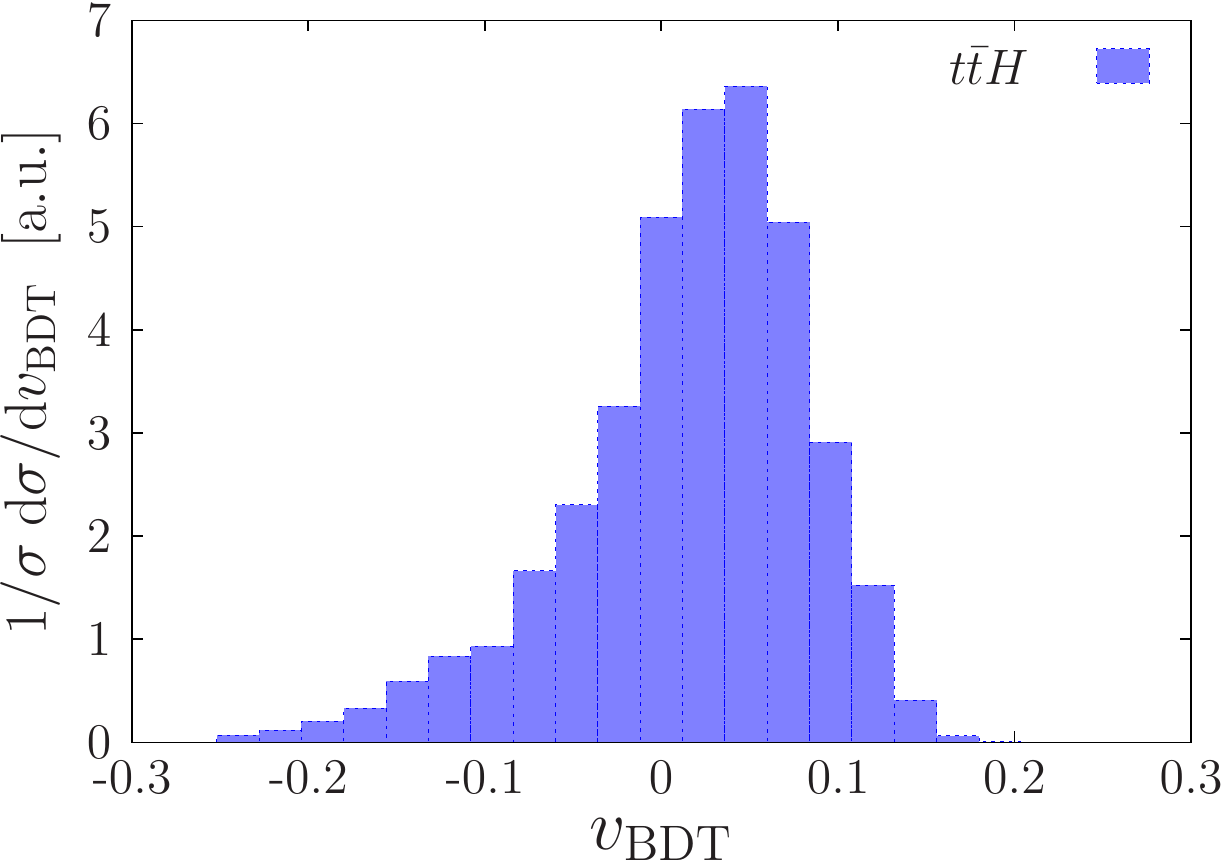}
    \caption{Sig}
  \end{subfigure}%
  \begin{subfigure}{\stew\textwidth}
    \centering
    \includegraphics[width=\stgw\linewidth]{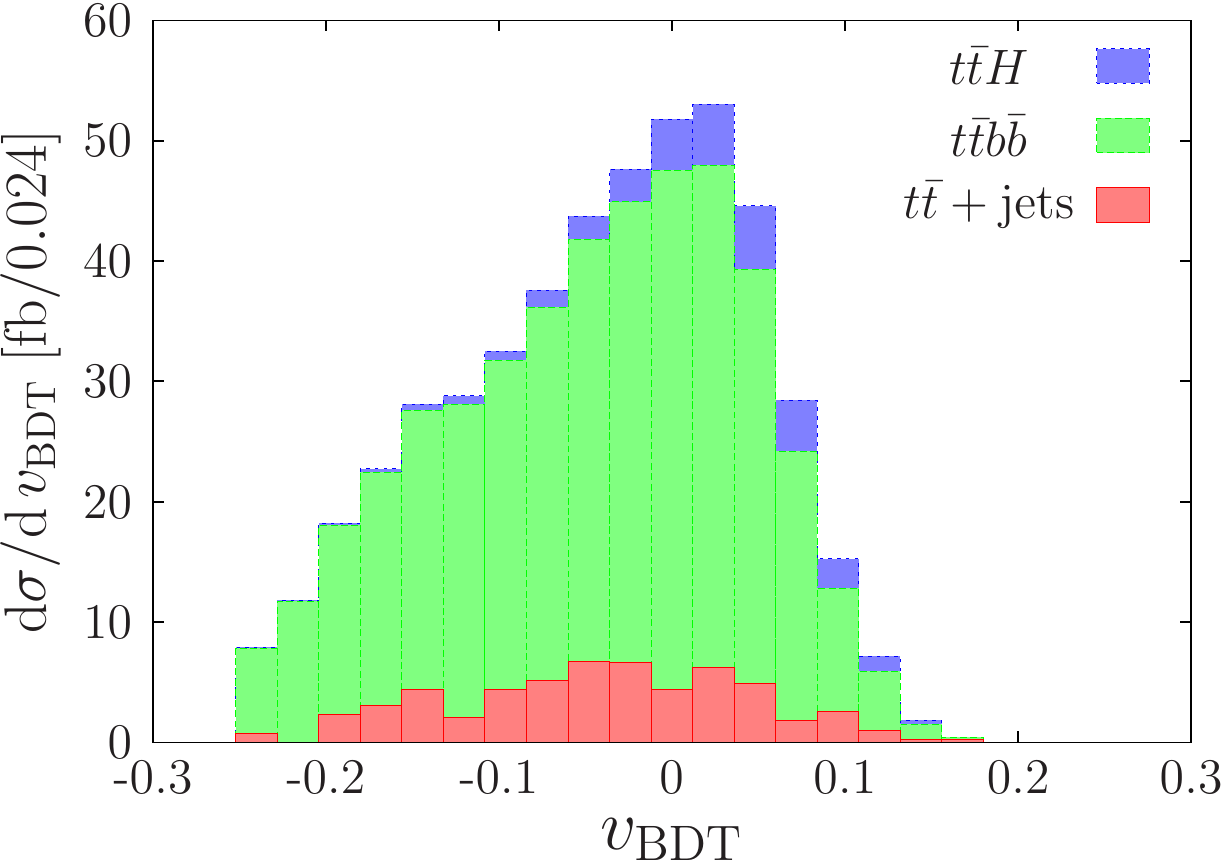}
    \caption{Sig+Bkg}
  \end{subfigure}\\[1ex]
  \caption{
        Boosted Decision Trees score distribution from 7 variables calculated from objects in the non-boosted analysis.
            The left figure is the signal $t\bar{t}H$ and the figure to the right contains signal and background.
              }
  \label{fig:BDTunboost}
\end{figure}

Using these six jets ($\mathrm{b_1,b_2,b_3,b_4,q_1,q_2}$)\footnote{The
numbering scheme signifies the $p_\mathrm{T}$ in descending order.}, as well
as the isolated lepton $\ell$ and missing energy
$\mathbf{\slashed{E}_T}$, we define simple kinematic variables: $\Delta
m_\mathrm{H} = \mathrm{min_{ij}}\left|m_\mathrm{H,max} -
m_\mathrm{b_ib_j}\right|$, $p_{\mathrm{Tq_2}}/p_{\mathrm{Tq_1}}$,
$\mathrm{max_{ij}}\Delta R_\mathrm{b_ib_j}$, $\mathrm{min}_i\Delta
R_{W,b_i}$, $\Delta \phi_\mathbf{\slashed{E}_T,b_3}$, $\Delta
R_\mathrm{\ell,b_3}$, $\Delta R_\mathrm{W,b_4}$.  We found these seven
variables to have highest rank, as defined in \cite{TMVA}, after running all
possible kinematic combinations of our input objects, using a BDT.  
Signal and background distributions in the 
BDT discriminant are plotted in Figure~\ref{fig:BDTunboost}, 
and detailed results of this analysis are presented 
in Sec.~\ref{sec:results}.

\section{Effects from b-jet energy correction}
\label{sec:bjet}

Throughout 
Secs.~\ref{sec:mspana} and~\ref{sec:analysis}
we have neglected energy
corrections of b-tagged (sub)jets.  As a result, the mass of the
reconstructed Higgs candidate and the top quark show a broad, smeared-out
distribution.  ATLAS and CMS apply jet-energy corrections to 
compensate for energy losses from unobserved neutrinos
in the decay of $B$-mesons.  While the correct inclusion of these
corrections requires a full detector simulation and is beyond the scope of
this analysis, 
at the end of Sect.~\ref{sec:results} we will present
the most optimistic results for the $t
\bar{t} H$ reconstruction by including the neutrino momenta in the jet
finding.

\def\clwr{1.5}
 
  \begin{figure} 
  \centering
  \begin{subfigure}{.5\textwidth}
    \centering
    \includegraphics[width=\stgw\linewidth]{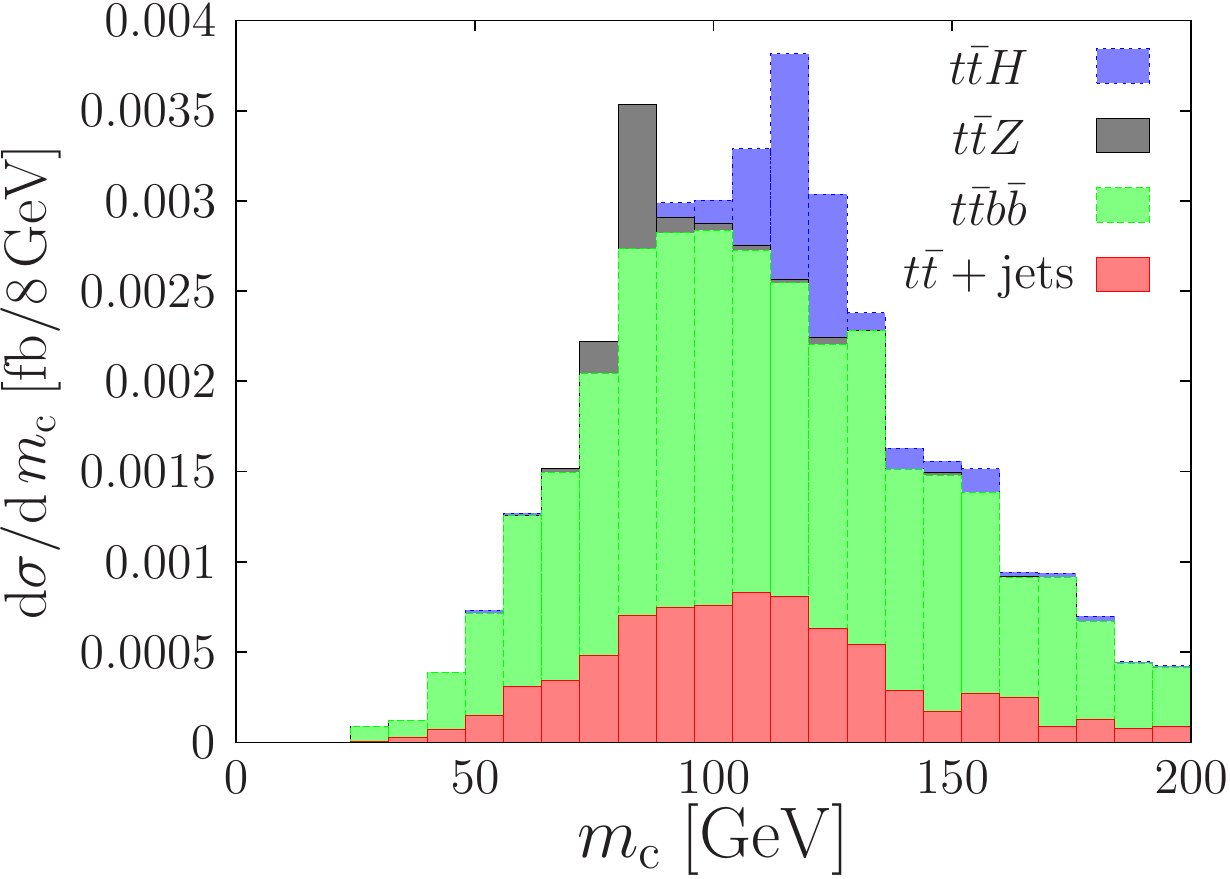}
    \caption{{\bf T1}}
  \end{subfigure}%
  \begin{subfigure}{.5\textwidth}
    \centering
    \includegraphics[width=\stgw\linewidth]{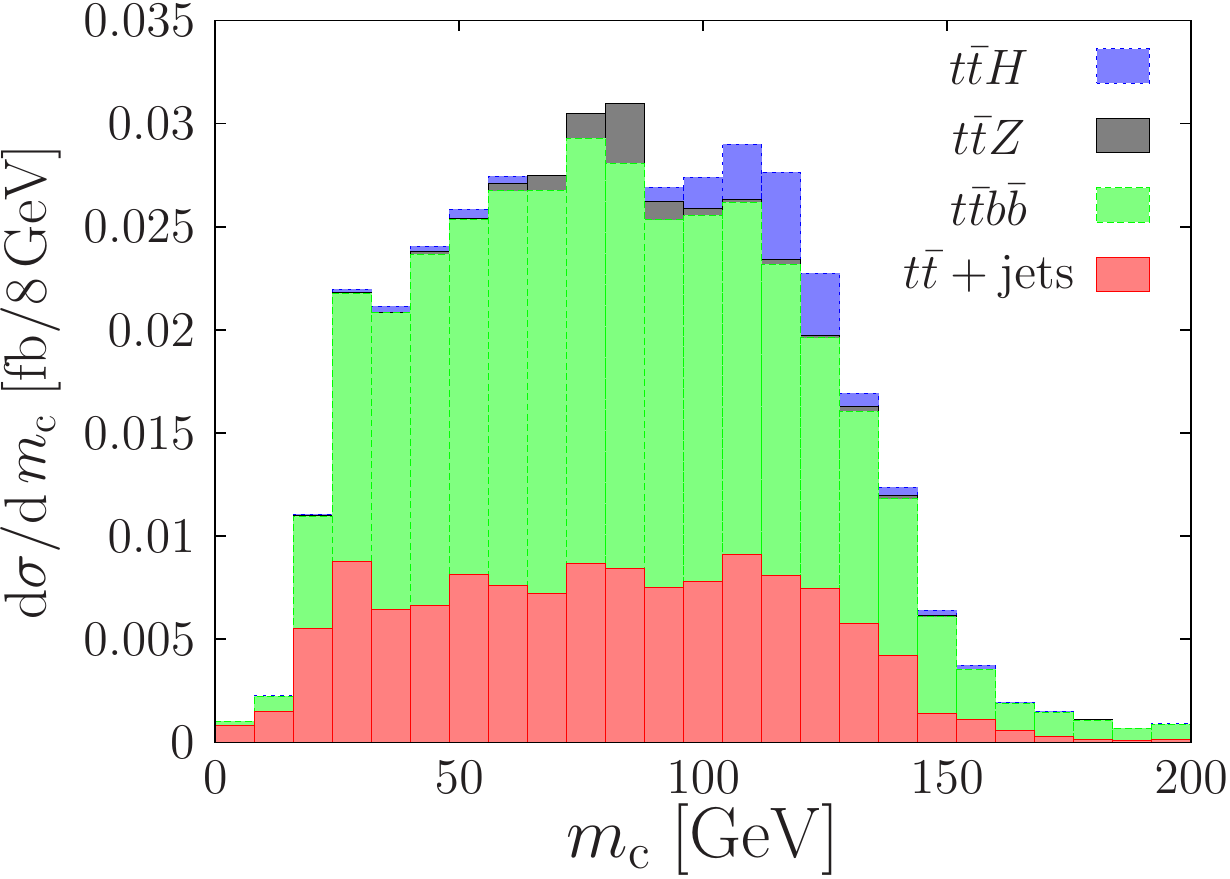}
    \caption{{\bf T2}}
  \end{subfigure}\\[3ex]
  \begin{subfigure}{.5\textwidth}
    \centering
    \includegraphics[width=\stgw\linewidth]{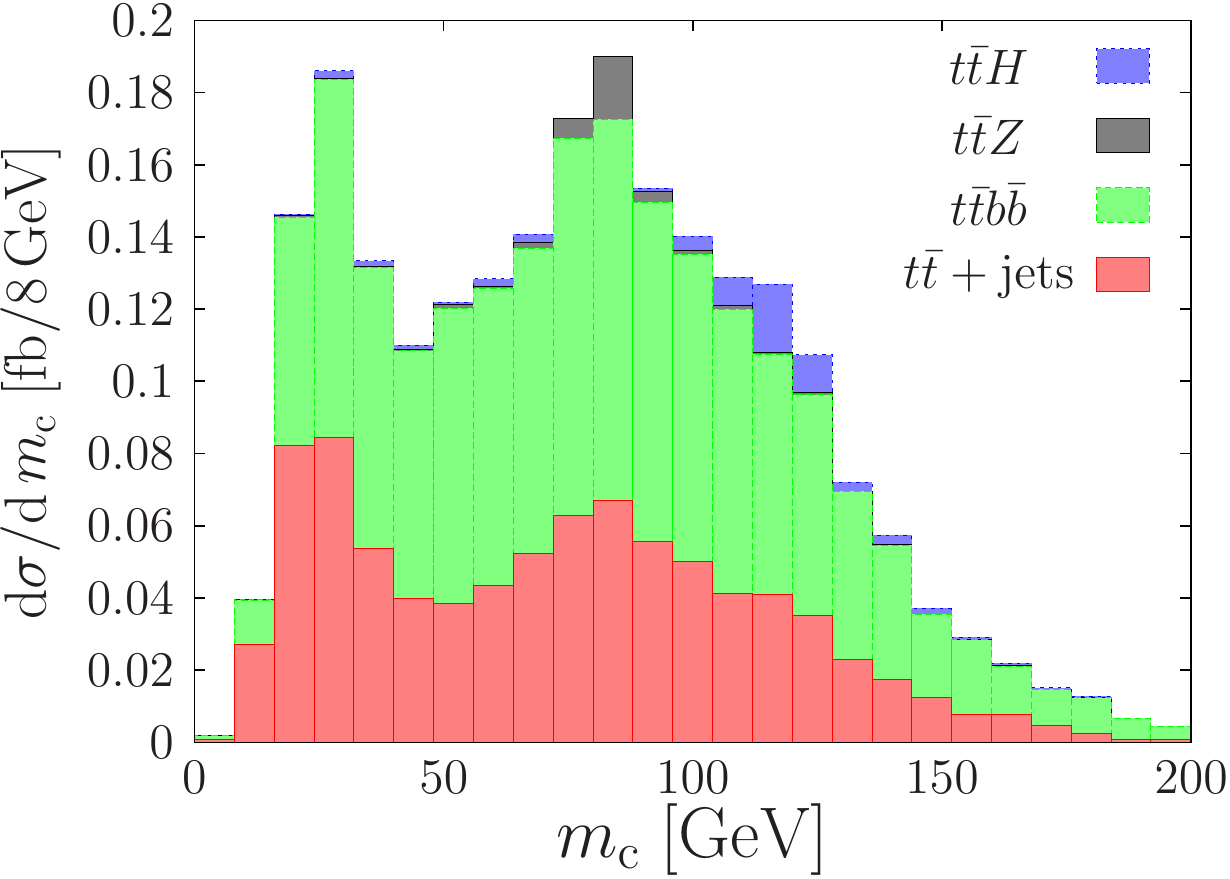}
    \caption{{\bf T3}}
  \end{subfigure}%
  \begin{subfigure}{.5\textwidth}
    \centering
    \includegraphics[width=\stgw\linewidth]{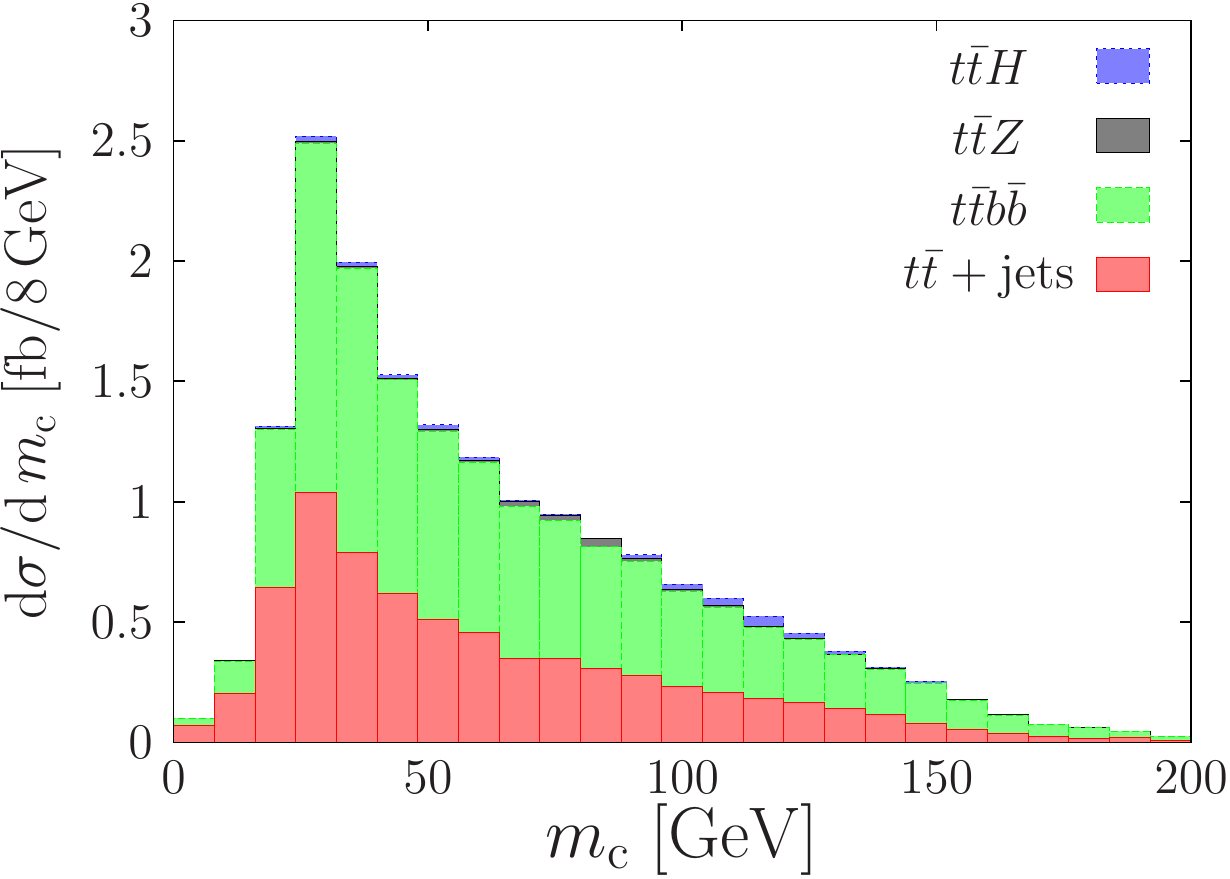}
    \caption{{\bf T4}}
  \end{subfigure}\\[1ex]
  \begin{subfigure}{.5\textwidth}
    \centering
    \includegraphics[width=\stgw\linewidth]{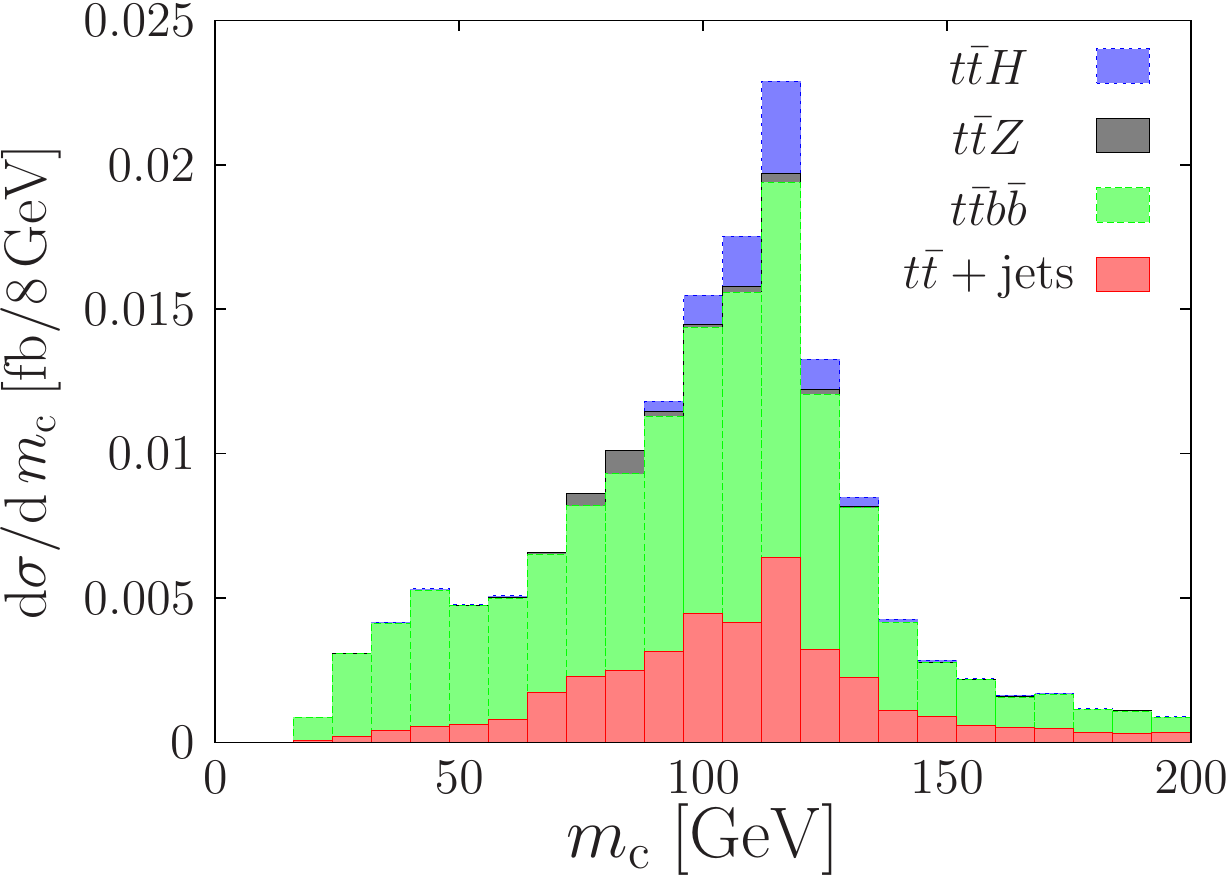}
    \caption{{\bf T5}}
  \end{subfigure}\\[1ex]
  \caption{
Distributions in the Higgs-candidate mass $m_\mathrm{c}$ after three $b$-tags for
the various selection topologies as in
Figs.~\ref{fig:2pr}--\ref{fig:Tboost},
but including neutrinos in the reconstructed $B$-hadrons.
}
  \label{fig:mHiggsNU}
\end{figure}

Distributions in the Higgs-candidate invariant mass with 
full $B$-reconstruction are illustrated in Figure~\ref{fig:mHiggsNU} 
for all analysis channels ({\bf T1}--{\bf T5}). As compared to Figs.~\ref{fig:2pr}--\ref{fig:Tboost},
including the neutrino momenta results in a narrower and
more pronounced mass peak at $m_H=125$ GeV.  
This effect is especially pronounced in the {\bf T1} channel. In this case, 
using an optimal mass window
can increase the $S/B$ 
ratio up 
to 40\% without losing signal yield.  
Moreover, for the {\bf T1} channel
a side-band analysis appears to be possible where
the signal strength can be estimated by comparing the $Z$ boson peak with
the adjacent Higgs peak,
while the signal depleted regions can be exploited for a data driven background 
determination.
However, the
{\bf T1} channel collects only a modest fraction of the $t\bar tH$ 
signal (see Sect.~\ref{sec:results}), while it is evident from 
Fig.~\ref{fig:mHiggsNU} that the other channels do not benefit 
in a similar way from an improved reconstruction of $B$-decays.

\section{Results}
\label{sec:results}

We present the results of the 
analyses described in the previous three sections 
in terms of $S/B$ ratios in signal enriched regions and in the form of $95\%$ CL limits
on the signal strength $\mu$.
We define $\mu$ as the 
observed deviation from the signal plus background SM
hypothesis as a fraction of the 
SM $t\bar{t}H$ cross section,
$\mu=\frac{\sigma^\mathrm{obs}-\sigma^\mathrm{SM}_\mathrm{S+B}}{\sigma^\mathrm{SM}_\mathrm{S}}$. 
Therefore, $\mu=0$ represents no deviation 
from the SM\footnote{
Note that the usual $\ttbar H$ signal strength corresponds to $1+\mu$.},
while coupling modifications due to new physics could result in $\mu < 0$ or $\mu > 0$. 
The limits are obtained from the final discriminating observables of the 
various selections, i.e.~the $m_c$ or $v_\mathrm{BDT}$ distributions. More precisely,
we perform a two-sided frequentist test with the profile
likelihood test statistic and the CLs variant of the p-value using the
RooStats framework \cite{roostats}.  
We use the expected 
number of signal plus background SM events in each bin of the 
relevant distribution
as the null hypothesis and we look for the limits this
analysis could impose on BSM contributions to the signal strength (both
positive and negative).  The results from the statistical analysis 
are presented in Fig.~\ref{fig:15err} 
under the assumption of a constant normalisation uncertainty
of 15\% for the SM background.
In Fig.~\ref{fig:muVlum} a more optimistic scenario is presented, where
the background uncertainty starts decreasing as the inverse square root of the 
integrated luminosity above 300\,fb$^{-1}$.  
The green bands in Figs.~\ref{fig:15err}--\ref{fig:muVlum} cover the $\mu$ values that
cannot be excluded at 95\% CL assuming the
data is exactly as predicted by the SM.  The yellow bands extend this region
to include an upward (downward) fluctuation by $1\sigma$ of the SM median
when calculating the upper (lower) $95\%$ confidence limit.

\def\clwr{.99}
\def\clws{.49}

\begin{figure} 
\begin{subfigure}{\clws\textwidth}
  \centering
  \includegraphics[width=\clwr\textwidth]{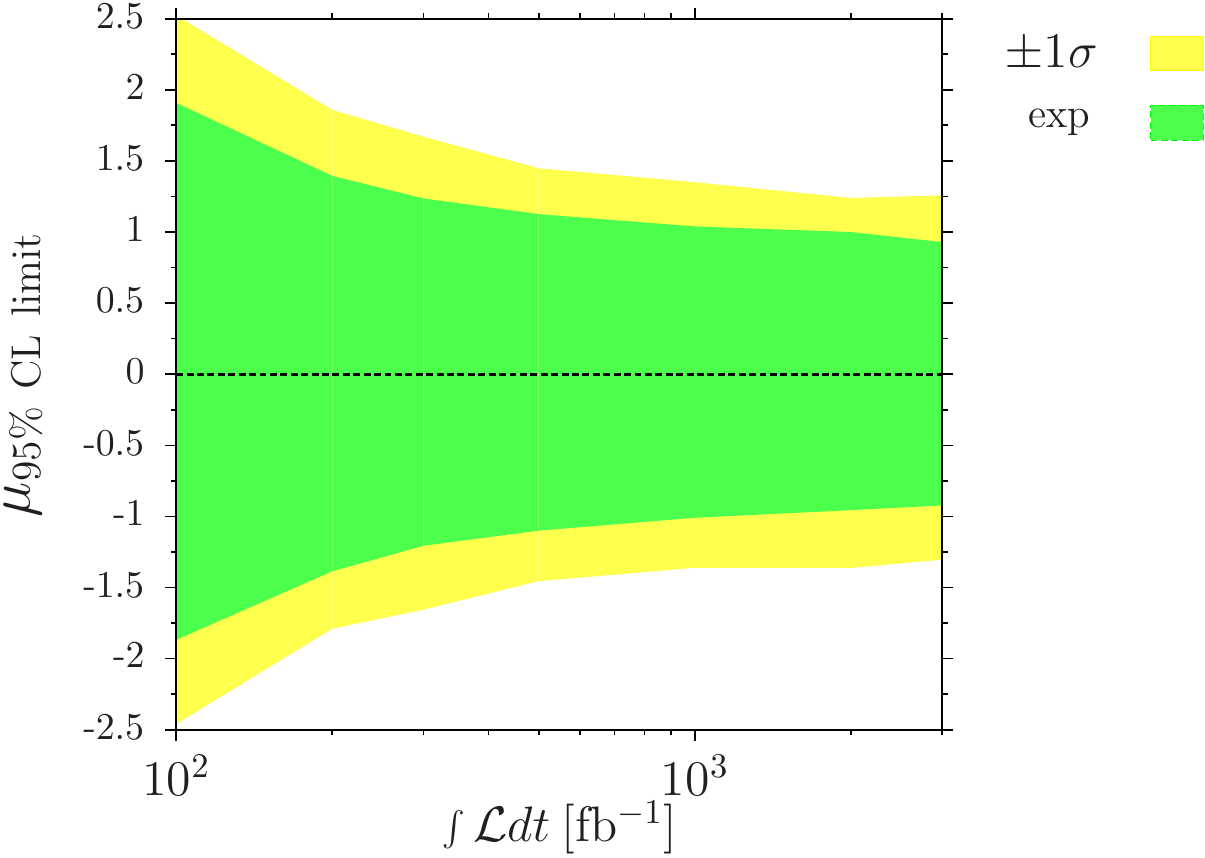}
  \caption{
Analysis of Sec.~\ref{sec:mspana} including all relevant topologies ({\bf T1} and {\bf T2}).
}
\label{fig:clsOLD}
\end{subfigure}
\begin{subfigure}{\clws\textwidth}
  \centering
  \includegraphics[width=\clwr\textwidth]{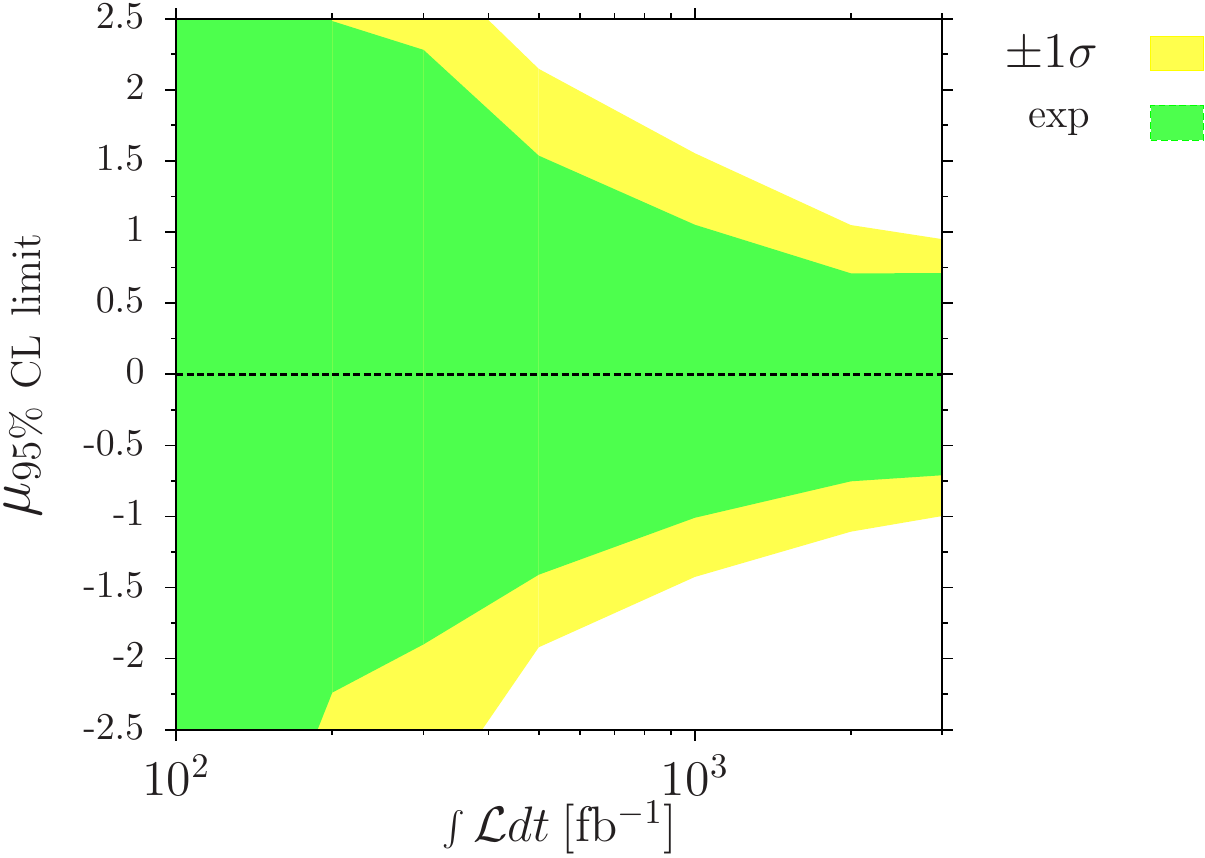}
  \caption{
Analysis of Sec.~\ref{sec:analysis} limited to topology {\bf T1}.
}
\end{subfigure}\\[1ex]
\begin{subfigure}{\clws\textwidth}
  \centering
  \includegraphics[width=\clwr\textwidth]{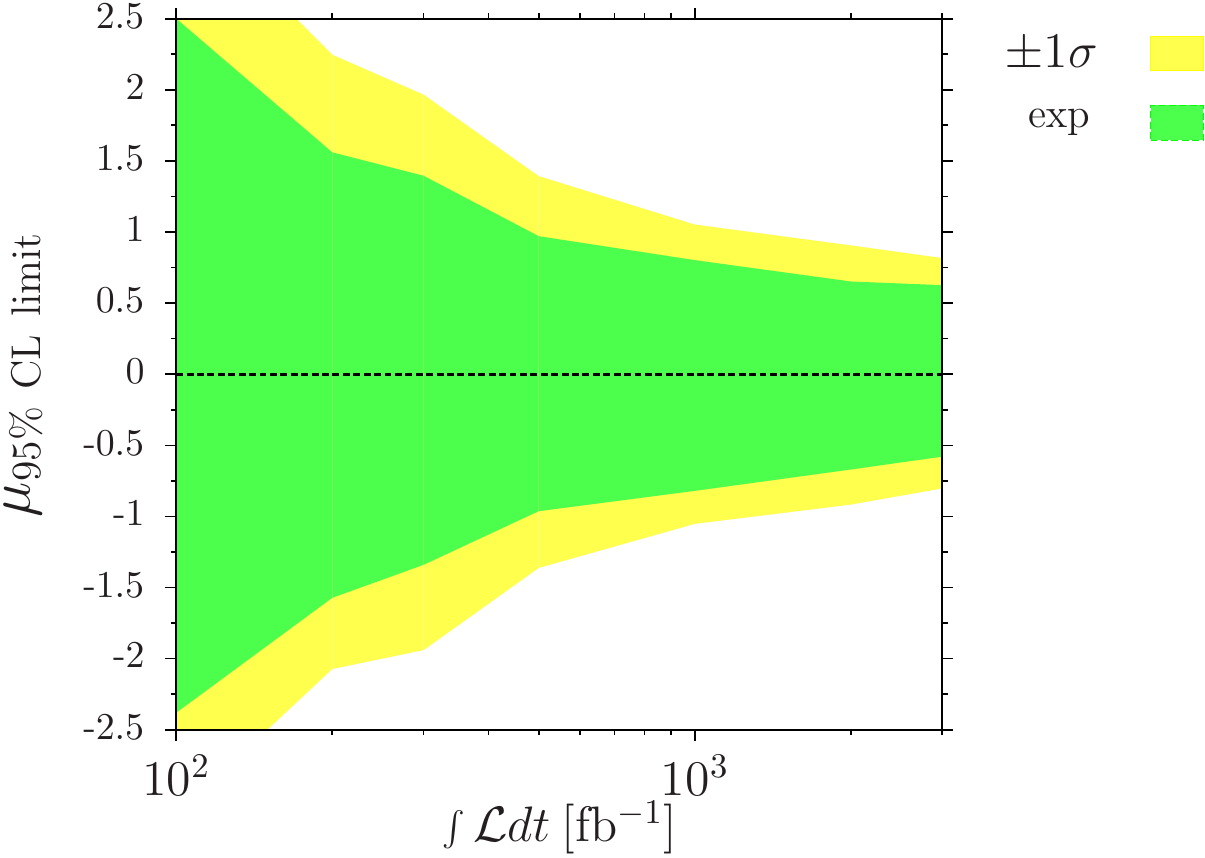}
  \caption{
Analysis of Sec.~\ref{sec:analysis} including topologies {\bf T1} and {\bf T2}.
}
\end{subfigure}
\begin{subfigure}{\clws\textwidth}
  \centering
  \includegraphics[width=\clwr\textwidth]{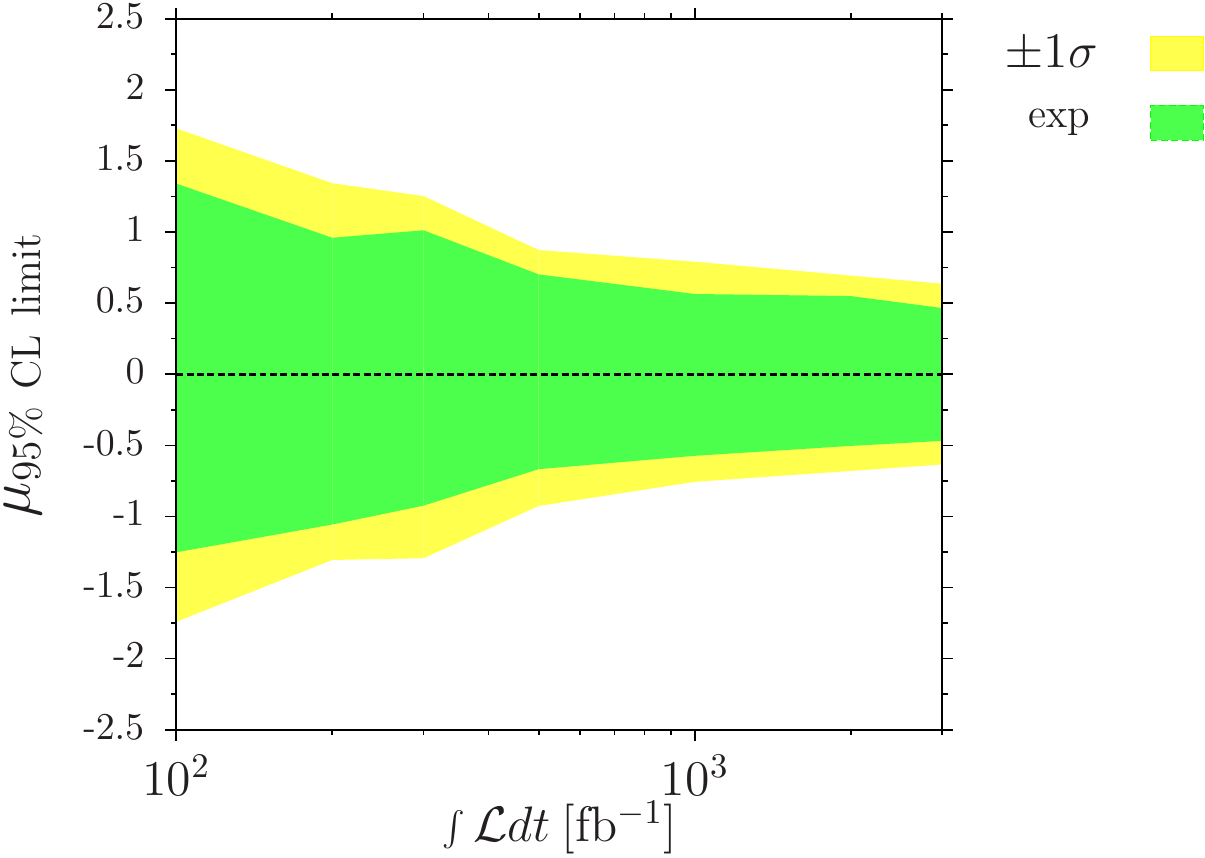}
  \caption{
Analysis of Sec.~\ref{sec:analysis} including all topologies ({\bf T1}--{\bf T5}).}
\end{subfigure}\\[1ex]
\begin{subfigure}{\clws\textwidth}
  \centering
  \includegraphics[width=\clwr\textwidth]{plots/CLs/truedata_err15/FS_CLs.pdf}
  \caption{
Analysis of Sec.~\ref{sec:analysis} including all topologies ({\bf T1}--{\bf T5})
and neutrinos in $B$-decay reconstruction.
}
\end{subfigure}
\begin{subfigure}{\clws\textwidth}
  \centering
  \includegraphics[width=\clwr\textwidth]{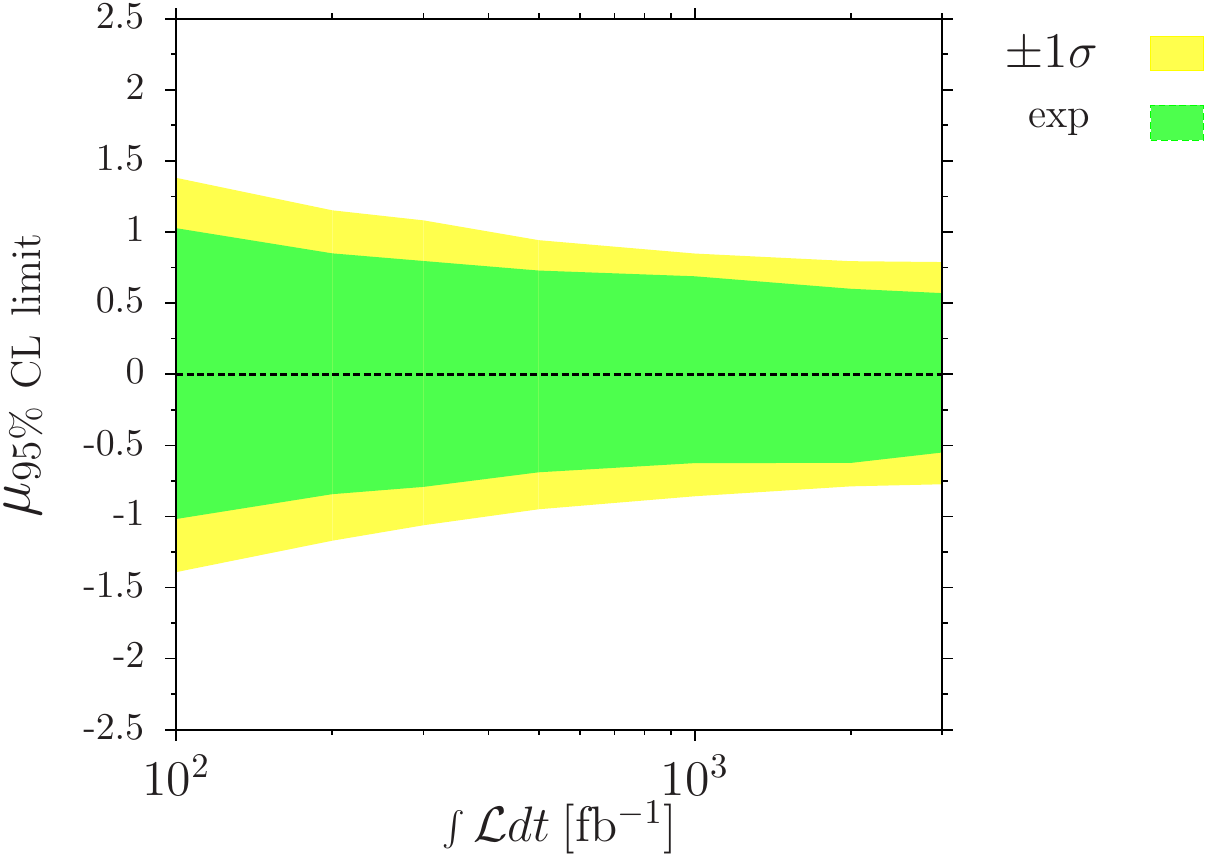}
  \caption{
Unboosted BDT analysis of Sec.~\ref{sec:unboost}.}
\end{subfigure}\\[1ex]
\caption{
Two-sided 95\% CL limit of the signal strength $\mu$ as a function of the
integrated luminosity  
assuming a constant 15\% normalisation uncertainty for the SM background.}
\label{fig:15err}
\end{figure}

\begin{figure} 
\begin{subfigure}{\clws\textwidth}
  \centering
  \includegraphics[width=\clwr\textwidth]{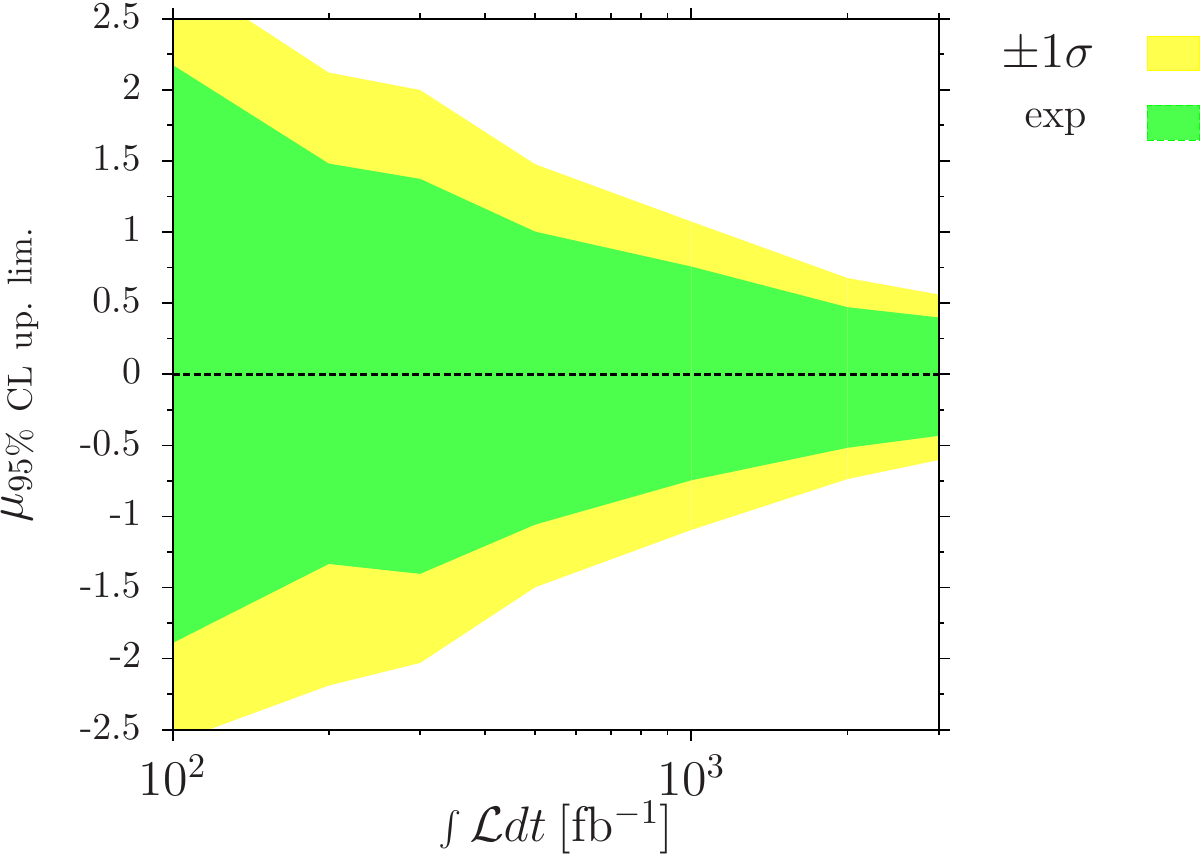}
  \caption{
Analysis of Sec.~\ref{sec:mspana} including all relevant topologies ({\bf T1} and {\bf T2}).}
\end{subfigure}
\begin{subfigure}{\clws\textwidth}
  \centering
  \includegraphics[width=\clwr\textwidth]{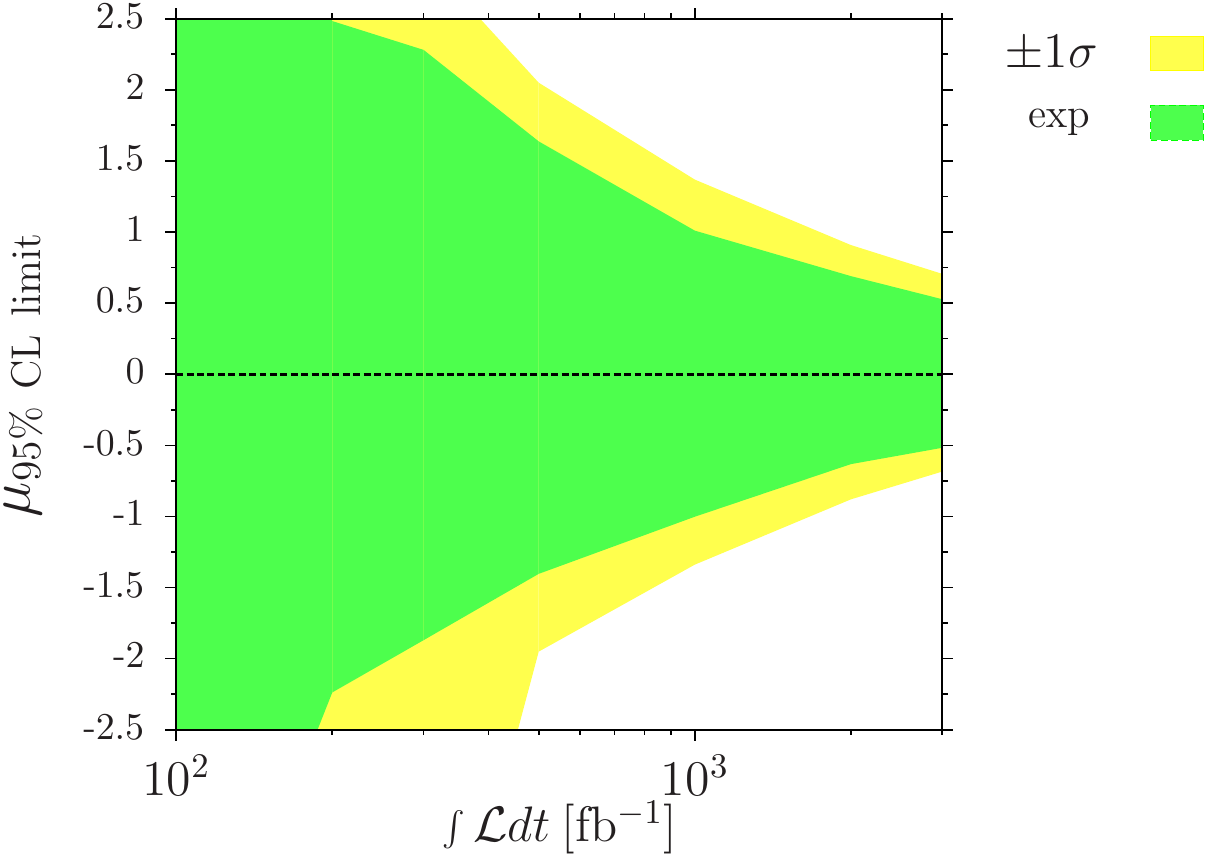}
  \caption{
Analysis of Sec.~\ref{sec:analysis} limited to topology {\bf T1}.}
\end{subfigure}\\[1ex]
\begin{subfigure}{\clws\textwidth}
  \centering
  \includegraphics[width=\clwr\textwidth]{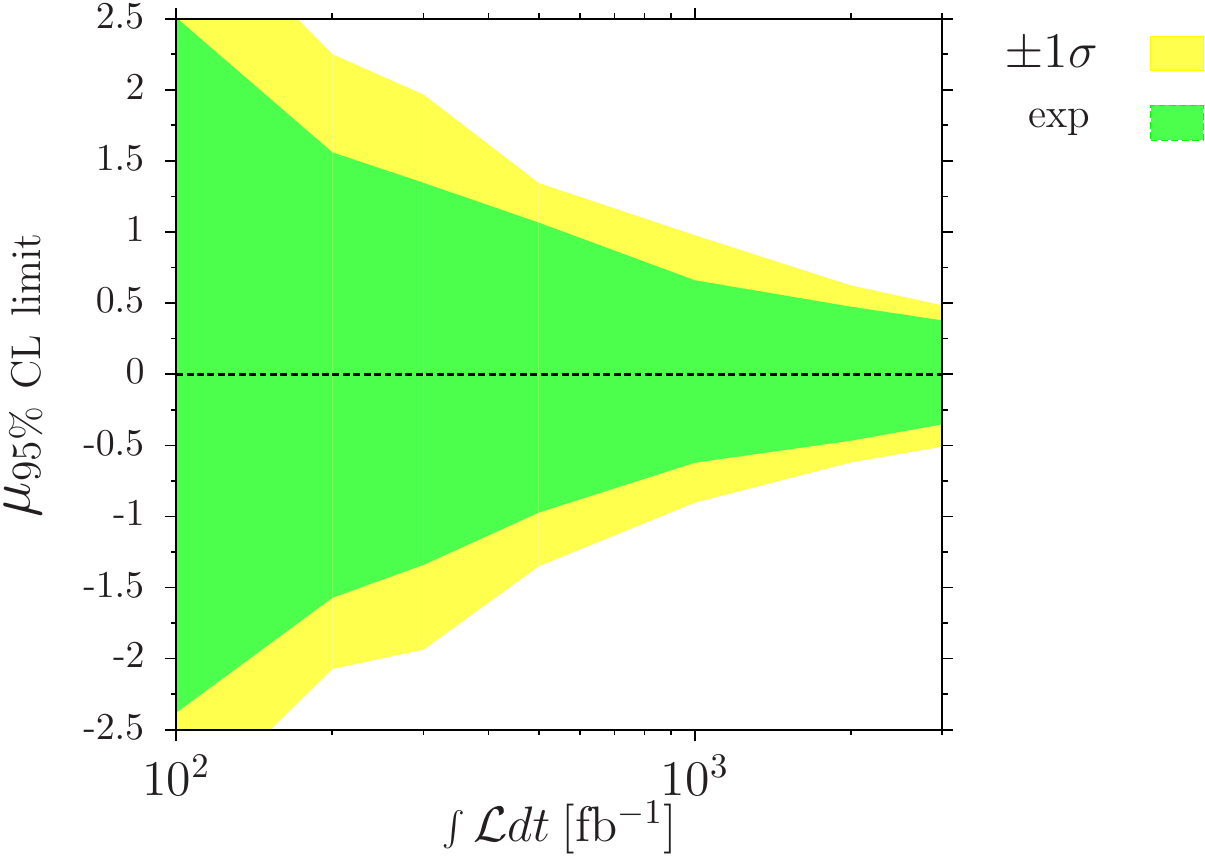}
  \caption{
Analysis of Sec.~\ref{sec:analysis} including topologies {\bf T1} and {\bf T2}.}
\end{subfigure}
\begin{subfigure}{\clws\textwidth}
  \centering
  \includegraphics[width=\clwr\textwidth]{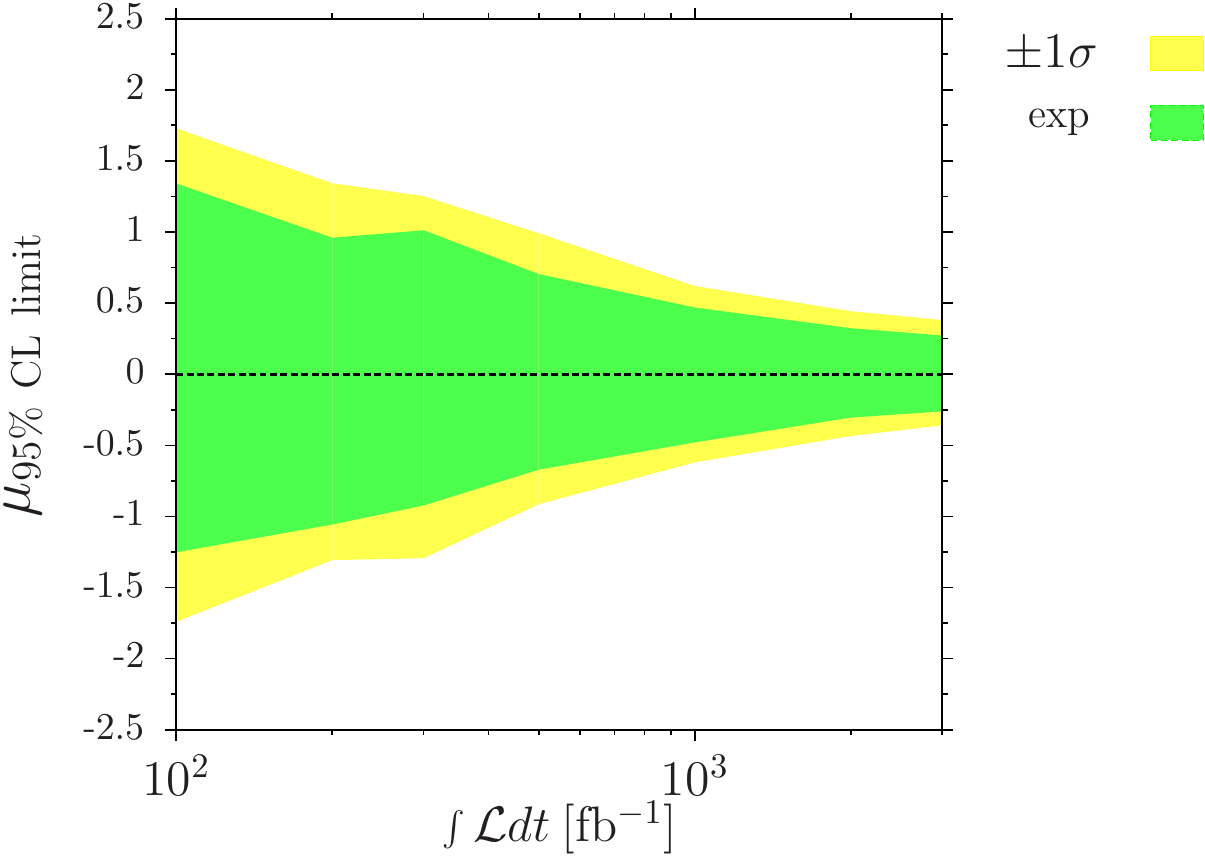}
  \caption{
Analysis of Sec.~\ref{sec:analysis} including all topologies ({\bf T1}--{\bf T5}).}
\end{subfigure}\\[1ex]
\begin{subfigure}{\clws\textwidth}
  \centering
  \includegraphics[width=\clwr\textwidth]{plots/CLs/truedata_scaleERR/FS_CLs.pdf}
  \caption{
Analysis of Sec.~\ref{sec:analysis} including all topologies ({\bf T1}--{\bf T5})
and neutrinos in $B$-decay reconstruction.}
\end{subfigure}
\begin{subfigure}{\clws\textwidth}
  \centering
  \includegraphics[width=\clwr\textwidth]{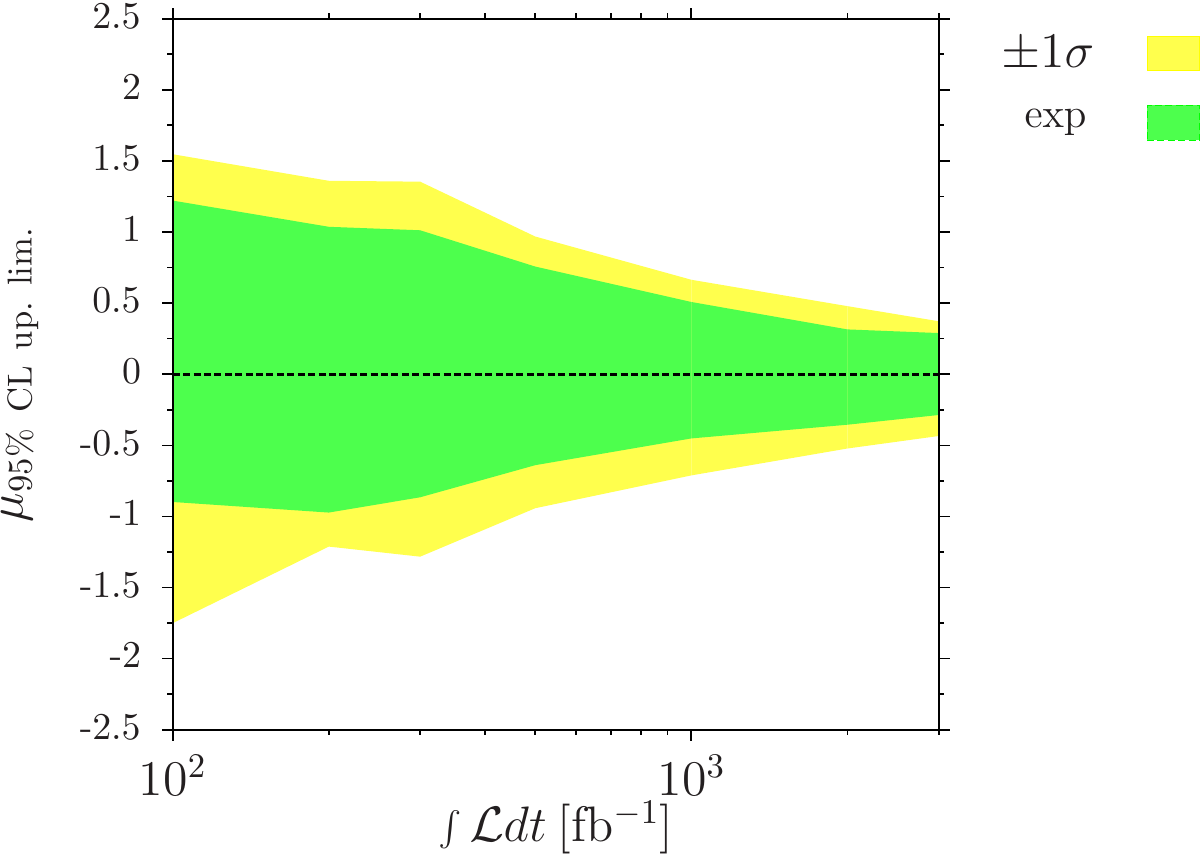}
  \caption{
Unboosted BDT analysis of Sec.~\ref{sec:unboost}.}
\end{subfigure}\\[1ex]
\caption{
Two-sided 95\% CL limit of the signal strength $\mu$ as a function of the integrated luminosity
assuming a  normalisation uncertainty for the SM background that remains constant at 15\% level up to
$300~\mathrm{fb}^{-1}$ and scales as $1/\sqrt{\mathcal{L}}$ for higher integrated luminosities.}
\label{fig:muVlum}
\end{figure}

\subsection{Standard boosted analysis}

\setlength\extrarowheight{2pt}
\def\shift{\hspace{3mm}}
\begin{table}
\begin{tabular}{@{\shift}c@{\shift}|@{\shift}c@{\shift} @{\shift}c@{\shift} @{\shift}c@{\shift} @{\shift}c@{\shift}|@{\shift}c@{\shift}}
 stage              	  &  ${t\bar{t}H}$  &  $t\bar{t}b\bar{b}$ &  $t\bar{t}+$jets  & $t\bar{t}Z$ & $S/B$ 		      \\\hline
 MC level     	    	  &  94           &  7.3$\times 10^3$   &  2.6$\times 10^5$ &  50         &  3.5$\times 10^{-4}$      \\
 1 lepton           	  &  60           &  4.7$\times 10^3$   &  1.6$\times 10^5$ &  22         &  3.6$\times 10^{-4}$      \\
 \textgreater1 fat jets   &  15           &  400                &  9.5$\times 10^3$ &  5.9        &  1.5$\times 10^{-3}$       \\ 
 1 top tag       	  &  4.8          &  110                &  2.6$\times 10^3$ &  1.9        &  1.8$\times 10^{-3}$       \\
 3 $b$-tags     	  &  0.59         &  7.6                &  4.2              &  0.25       &  0.049     			\\
 $m_\mathrm{c}$ cut       &  0.2          &  0.9                &  0.48             &  0.023      &  0.14    			\\\hline
\end{tabular}
\caption{
Signal and background cross sections in femtobarn and $S/B$ ratios
at different stages of the boosted analysis of Section~\ref{sec:mspana}.
}
\label{tab:Told}
\end{table}

Let us start discussing the results of the analysis of
Sec.~\ref{sec:mspana}, which represents a standard boosted selection  along
the lines of~\cite{MSp2009}.  The
signal and background contributions at various steps of the selection are
presented in Table~\ref{tab:Told}, and the 
overall picture is qualitatively similar to the original boosted
analysis~\cite{MSp2009}. However the quantitative differences are quite notable. 
In particular, for the $S/B$ ratio after the $m_\mathrm{c}$-cut we
observe a reduction from about\footnote{
To be precise, 
the $S/B$ ratios reported in~\cite{MSp2009} 
are 42\% and 28\% for $m_H=120$\,GeV and 130\,GeV, respectively.}~35\%
in~\cite{MSp2009} to 14\%. 
To a very large extent, this loss of discriminating power can be attributed 
to the differences in signal and background 
rates between the two analyses.
In particular, the dominant effects are a 35\% 
increase---driven by $t\bar t$+jets---of the 
overall background level, and a 30\% reduction of the $t\bar t H$ 
signal within 
final
selection cuts. As for the modified top taggers and the inclusion of $B$-meson decays with 
related neutrino-energy losses (which require a  modified Hiss-mass window),
we checked that the impact on $S/B$ is relatively small.

\def\epsmis{\epsilon_\mathrm{mistag}}
\def\epsb{\epsilon_b}

The NLO tools used in the present study (see Sect.~\ref{se:MC}) provide more
reliable signal and background simulations as compared to the LO+PS samples
employed in~\cite{MSp2009}.  In the case of the $t\bar t+$jets background
we observe a very large enhancement---close to one order of magnitude in the 
 signal region---that can be in part attributed to the
usage of a rather crude approximation based on $t\bar t+1$\,jet LO matrix
elements matched to Herwig++ in~\cite{MSp2009}.
Moreover,
the large relative increase of 
the $t\bar{t}$+jets contribution
can be explained in part by
the way we simulate 
$b$-tagging and top-tagging.
On the one hand, Table~\ref{tab:Told} shows that imposing three
$b$-tags results in 
a reduction of the signal by about a factor ten,
which is
well beyond the naive expectation of a 
suppression factor $\epsilon_b^{-3}\simeq 3$
for a constant $b$-tagging efficiency 
$\epsilon_b=0.7$.  This
observation can be explained by the fact that, in the present analysis, each $b$-tag is
accompanied by a \mbox{$B$-meson} acceptance cut $p_{T,B}>10$\,GeV 
and by the requirement that
the $B$-meson is matched to the actual jet. In particular, this latter condition can be rather
restrictive in case of the two narrow $b$-jets that form the Higgs candidate substructure.  
On the other hand, it turns out that applying three $b$-tags 
to the $t\bar t$+jets background results in a suppression factor 
$2600/4.2\simeq 620$,
while using a light-jet mistagging efficiency 
$\epsmis=0.01$ 
and excluding the $b$-jet from the reconstructed hadronic top
one would naively expect a 
much stronger
suppression factor 
$[n(n-1)/2\,\epsmis^2\,\epsb]^{-1}\simeq 2400$
for an average number of light jets 
$n= 4$.
However, there is a probability of about 7\% 
that one $b$-quark escapes the hadronic top reconstruction 
(see topology $A_3$ in Tab.~\ref{tab:topolT}). In this case a single mistag 
is sufficient in order to arrive at three $b$-tags, 
and the corresponding suppression factor $(0.07\,n\,\epsmis\,\epsb^2)^{-1}\simeq 730$ for $n=4$ is well consistent with the observed drop in $t\bar t+$\,jets.
These considerations indicate that
the signal and background cross sections
after three $b$-tags are very sensitive
to the details of  $b$- and top-tagging,
and to multi-jet emissions in the $t\bar t+X$ background.
The changes in $S/B$ with respect to 
the original analysis of~\cite{MSp2009} can thus be attributed to 
the various modifications relative to these three aspects.
We also note that an optimal suppression of the $t\bar t+$\,jets background 
could be easily achieved by requiring a fourth $b$-tag.

Concerning the differences between the HEPTopTagger used in the
present analysis and the top tagging method employed in~\cite{MSp2009},
the results presented in
Sec.~\ref{sec:topreco} demonstrate that the HEPTopTagger efficiency 
is very robust irrespective of the pollution within the fat jet.
In contrast, the tagger used in~\cite{MSp2009} has higher efficiency and
light QCD jet mistag rate when a fat jet contains more hard
radiation than only the top decay products.
On the one hand, in general this can reduce $S/B$. But on the other
hand the tagger used in~\cite{MSp2009} has a slightly higher overall efficiency 
as compared to the HEPTopTagger's efficiency of
about $40\%$. For top-rich backgrounds this has no direct influence on $S/B$, but it can
improve the statistical sensitivity of the analysis.

The sensitivity of the standard boosted analysis of
Sec.~\ref{sec:mspana} are displayed in~Fig.\ref{fig:clsOLD}.
For the scenario of a constant systematic uncertainty of 15\% 
we find no sensitivity to modifications of the signal strength of $|\mu|\lesssim 1$ at
$\mathcal{L}=3000~\mathrm{fb}^{-1}$.  Hence, one can only exclude deviations
from the Standard Model that are at least as large as the $t\bar{t}H$
contribution it predicts.  In Fig.~\ref{fig:15err} we see that this limit
remains almost constant for integrated luminosities expected at the end of
the following two LHC runs, suggesting the analysis will be dominated by the
systematic uncertainty.
In the more optimistic scenario presented in  Fig.~\ref{fig:muVlum}, where the
systematic uncertainty scales as $1/\sqrt{\mathcal{L}}$, 
a limit around $|\mu|=0.5$ could be achieved at $\mathcal{L}=3000~\mathrm{fb}^{-1}$.

\subsection{Improved boosted analyses {\bf T1}--{\bf T5} and unboosted MVA approach}

\begin{table}
\begin{tabular}{@{\shift}c@{\shift}@{\shift}c@{\shift}|@{\shift}c@{\shift} @{\shift}c@{\shift} @{\shift}c@{\shift} @{\shift}c @{\shift}c@{\shift}}
Analysis            & stage                &  $t\bar{t}H$  &  $t\bar{t}b\bar{b}$ &  $t\bar{t}+$jets   & $t\bar{t}Z$         & $S/B$   \\\hline
\multirow{4}{*}{{\bf T1}} & before $b$-tag       &  1.1          & 27                  & 690                & 0.43                & 1.5$\times 10^{-3}$         \\
                    & 3 $b$-tags           &  0.075        & 0.77                & 0.37               & 0.032               & 0.064       \\
                    & $m_\mathrm{c}$ cut   &  0.042        & 0.13                & 0.053              & 2.0$\times 10^{-3}$ & 0.23      \\ 
                    & $\hat{t}$ cut        &  0.035        & 0.089               & 0.038              & 9.5$\times 10^{-4}$ & 0.27         \\\hline
\multirow{4}{*}{{\bf T2}} & before $b$-tag       &  12           &  240                &   4.6$\times 10^3$ & 4.5                 &  2.5$\times 10^{-3}$       \\
                    & 3 $b$-tags           &   0.25        &   3.0               &    1.5             & 0.11                &    0.054     \\
                    & $m_\mathrm{c}$ cut   & 0.14          & 0.66                &   0.36             &    0.01             & 0.13        \\
                    & $v_\mathrm{BDT}$ cut & 0.044         &   0.18              & 0.1                &  0.0031             &  0.15       \\\hline
\multirow{3}{*}{{\bf T3}} & before $b$-tag       & 51            & 1.2$\times 10^3$    & 1.9$\times 10^{4}$ & 18                  & 3.0$\times 10^{-3}$        \\
                    & 3 $b$-tags           & 1.0           &  17                 & 11                 & 0.48                & 0.04        \\
                    & $m_\mathrm{c}$ cut   &  0.53         &  3.2                &  2.0               & 0.032               & 0.1        \\\hline
\multirow{3}{*}{{\bf T4}} & before $b$-tag       &  630          & 1.5$\times 10^{4}$  & 2.2$\times 10^{5}$ & 210                 &  3.0$\times 10^{-3}$       \\
                    & 3 $b$-tags           &  5.6          & 130                 & 92                 &  2.2                &  0.02       \\
                    & $m_\mathrm{c}$ cut   &  1.5          &   16                &  10                &   0.2               &  0.06       \\\hline
\multirow{3}{*}{{\bf T5}} & before $b$-tag       &  4.2          & 220                 &  5.7$\times 10^3$  & 1.5                 & 7$\times 10^{-4}$            \\
                    & 3 $b$-tags           &  0.14         &  1.6                &   0.65             &  0.036              & 0.06         \\
                    & $m_\mathrm{c}$ cut   &  0.094        &  0.6                & 0.28               & 0.011               & 0.11        \\\hline
\multirow{3}{*}{MVA} & \textgreater5 jets   &  14           &  420                & 6.0$\times 10^3$   &  5.1                &  2.2$\times 10^{-3}$       \\
                    & 4 $b$-jets           &  1.5          & 19                  & 2.9                &  0.52               &  0.066        \\
                    & $v_\mathrm{BDT}$ cut &  0.041        & 0.16                &  0.033             &  2.4$\times 10^{-3}$&  0.21       \\\hline
\end{tabular}
\caption{Signal and background cross sections in femtobarn and $S/B$ ratios
at different stages of the various boosted analyses ({\bf T1}--{\bf T5})  of Section~\ref{sec:boostAna}
and for the unboosted MVA analysis of Section~\ref{sec:unboost}.}
\label{tab:summaryIV}
\end{table}

We now turn to the results of the boosted and unboosted analyses of
Section~\ref{sec:analysis}.  The evolution of signal and background cross
sections at subsequent stages of the improved boosted analyses {\bf T1}--{\bf T5} and of the
unboosted MVA analysis is presented in Table~\ref{tab:summaryIV}.  The {\bf T1}
selection provides the sharpest signal peak and allows one to increase
$S/B$ after $m_\mathrm{c}$-cut from $14\%$---in the case of a standard boosted analysis with 
{\bf T1} and {\bf T2} contributions---up to about $23\%$.  This
gain in sensitivity comes at a high price for the signal yield, 
as the {\bf T1} topology is suppressed by roughly a factor four with respect to the 
complementary {\bf T2} contribution.
However, an extra ellipticity cut, $\hat{t}<0.2$, can further 
increase $S/B$ to 27\% with only a 15\% loss in the {\bf T1} signal yield.
At 300\,fb$^{-1}$ of integrated luminosity the expected number of $t\bar{t}H$ events in the 
{\bf T1} signal region is only twelve, but the higher $S/B$ level for this topology 
results in an improved exclusion limit around $|\mu|=0.7$
at $\mathcal{L}=3000~\mathrm{fb}^{-1}$ (see Fig.~\ref{fig:15err}).
To increase the signal yield and accumulate sensitivity from other regions
of the 
phase space, the {\bf T1} selection can be combined
with four statistically independent
channels: {\bf T2}-{\bf T5}.

In the {\bf T2} channel, with three Higgs candidates in a fat jet, we are able to
improve $S/B$ from $13\%$ 
(after $m_\mathrm{c}$ cut) to $15\%$
by applying a BDT made of 5 variables that define the
reconstructed $t\bar{t}H$ system, 
as discussed in Sec.~\ref{sec:boostTHana}.
However, the $v_\mathrm{BDT}$ cut suppresses the signal by about a factor three. Moreover,
several of the attempted modifications, e.g.~cutting on
$\chi^2$, using the jet shape observable ellipticity, and exploiting the
angle between the leptonic top's $b$-quark and lepton, do not improve $S/B$. 
Nevertheless, combining {\bf T1} and {\bf T2} without the aforementioned modifications in the CLs does improve the limit from
$|\mu|\simeq 0.7$ to $|\mu|\simeq 0.6$.

The topologies {\bf T3}--{\bf T5},
which contain at least
one boosted resonance, feature $S/B$ ratios from 6\% to 11\%, and
a factor eleven more signal 
cross section than the {\bf T1}--{\bf T2} selections.
The inclusion of all five channels in the CL
calculation allows one to exclude $|\mu|\gtrsim 0.45$.

Finally, let us consider the analysis of
Section~\ref{sec:unboost}, which entirely relies on a BDT without requiring
the presence of a fat jet in the final state.  This channel is not
independent from {\bf T1}--{\bf T5} and is closer in spirit to analyses
performed by ATLAS and CMS during Run 1.  In Table~\ref{tab:summaryIV} we
present results for a fairly tight cut on the $v_\mathrm{BDT}$ discriminant,
which is chosen in such a way that the resulting signal yield is the same as for
the individual boosted selection with the highest discriminating power, i.e.~{\bf
T1}.  In this case the BDT analysis reaches $S/B\simeq 21\%$, which
lies slightly below the result of the {\bf T1} selection (23\% without $\hat
t$ cut).  With a looser $v_\mathrm{BDT}$ cut it is possible to
increase the signal yield by an order of magnitude while keeping $S/B\simeq
18\%$.
For the scenario of 15\% systematic background uncertainty 
we find that exploiting the full distribution of $v_\mathrm{BDT}$ in the profiled
likelihood method permits to exclude $|\mu|\gtrsim 0.55$ at $\mathcal{L}=3000~\mathrm{fb}^{-1}$.
Thus the BDT limit lies between the results of the {\bf T1} analysis
($|\mu|\simeq 0.7$) and the combination of {\bf T1}--{\bf T5}
($|\mu|\simeq 0.45$).  
However, when comparing boosted and BDT selections, one should keep in mind
that the emergence of a Higgs peak in the $m_c$ distribution represents a
key added value of the {\bf T1} analysis, as it permits to 
use a side-band approach in order to mitigate 
theoretical uncertainties in the shapes of the $t\bar{t}+X$ backgrounds. 
Moreover the presence of a measurable $Z\to b\bar b$ peak provides extra
opportunities for a further reduction of the systematic uncertainties.

It will take the LHC more than a decade to collect $3000~\mathrm{fb}^{-1}$,
thus we expect significant improvements in the systematic uncertainties of
$t\bar{t}+X$ final states.  
Thus in Fig.~\ref{fig:muVlum} we present a more optimistic scenario, where
above $\mathcal{L}=300$\,fb$^{-1}$ the background systematic uncertainty starts
decreasing below 15\% and scales inversely proportionally to the square root 
of the integrated luminosity.
For shrinking uncertainties, equally applied to all channels, the final
signal yield becomes of crucial importance.  Therefore individual boosted
channels cannot compete with the unboosted analysis in setting a limit for
$\mu$.  
The {\bf T1} topology alone
can exclude deviations larger than 50\% of the expected
null hypothesis, while the unboosted 
BDT
analysis achieves 29\%.  The
combination of all five boosted topologies however, still offers the best
exclusion limit at $|\mu|\gtrsim 0.26$.
This limit can be
further improved by including $b$-jet energy correction as discussed in
Sec.~\ref{sec:bjet}.  When neutrinos from hadronic decays are used in the
reconstruction, 
the combination of the {\bf T1}--{\bf T5} analyses yields
the limit $|\mu| = 0.2$.

\section{Summary and Conclusions}
\label{sec:conc}

We have re-evaluated LHC's potential to set a limit on the signal strength
$\mu$ in the semi-leptonic $t\bar{t}H(b\bar{b})$ channel.  We first focused
on events with simultaneously moderately boosted Higgs boson and hadronic
top quark.  This channel was proposed earlier to discover a light Higgs
boson \cite{MSp2009} and was expected to provide a good handle in measuring
$\mu$.  After improving on the simulation of signal and backgrounds and
applying a more conservative particle identification, we find that achieving
a sensitivity to small deviations in $\mu$ is more challenging than
anticipated.  Therefore, to increase the sensitivity, we split the
$t\bar{t}H$ phase space into several independent regions and combined their
individual contributions.  In particular, we extended the search into new
phase space regions, where only one of the hadronically decaying resonances
is boosted.  These improvements allow us to dramatically reduce the 95\% CL
limit from $|\mu|=1.0$ to $|\mu|=0.2$ 
for
$3\mathrm{ab}^{-1}$ of integrated luminosity.
We acknowledge the previous
statement is 
very dependent on the handling of both theoretical uncertainties
in the 
normalisation and shape
of the $t\bar{t}+X$ distributions
and systematic effects in the experimental reconstruction of particles.  Therefore, we
encourage theoretical and experimental studies of the boosted channels in
$t\bar{t}+X$ events in order to make use of the full phase space accessible
at the LHC.

\acknowledgements 

We thank A.~Denner, S.~Dittmaier and L.~Hofer for providing us
with the one-loop tensor-integral library~\Collier and F.~Krauss for collaboration during early stages of this work. 
This research was supported in part by the Swiss
National Science Foundation (SNF) under contract PP00P2-153027  
and by the Research Executive Agency (REA) of the European Union
under the Grant Agreements PITN--GA--2012--315877 ({\it MCnet}) and 
PITN--GA--2012--316704 ({\it HiggsTools}).

\bibliography{tthbb}

\end{document}